\documentclass[aps,prd,groupedaddress,showpacs,tighten,floats,nofootinbib]{revtex4}
\usepackage{graphicx}
\usepackage{latexsym}
\def\beq{\begin{equation}}
\def\eeq{\end{equation}}
\def\bey{\begin{eqnarray}}
\def\eey{\end{eqnarray}}

\def\lsim{\mathrel{\raise.3ex\hbox{$<$\kern-.75em\lower1ex\hbox{$\sim$}}}}
\def\gsim{\mathrel{\raise.3ex\hbox{$>$\kern-.75em\lower1ex\hbox{$\sim$}}}}

\begin{document}

\title{Neutralino Dark Matter as the Source of the WMAP Haze}  
\author{Gabriel Caceres$^{1,2}$ and Dan Hooper$^{1,3}$}
\address{$^{1}$ Fermi National Accelerator Laboratory, Theoretical Astrophysics, Batavia, IL  60510 \\$^{2}$ Pennsylvania State University, Department of Astronomy and Astrophysics, University Park, PA 16802 \\$^{3}$  The University of Chicago, Department of Astronomy and Astrophysics, Chicago, IL 60637\\
}

\date{\today}

\begin{abstract}

Previously, it has been argued that the anomalous emission from the region around the Galactic Center observed by WMAP, known as the ``WMAP Haze'', may be the synchrotron emission from relativistic electrons and positrons produced in dark matter annihilations. In particular, the angular distribution, spectrum, and intensity of the observed emission are consistent with the signal expected to result from a WIMP with an electroweak-scale mass and an annihilation cross section near the value predicted for a thermal relic. In this article, we revisit this signal within the context of supersymmetry, and evaluate the parameter space of the Constrained Minimal Supersymmetric Standard Model (CMSSM). We find that, over much of the supersymmetric parameter space, the lightest neutralino is predicted to possess the properties required to generate the WMAP Haze. In particular, the focus point, $A$-funnel, and bulk regions typically predict a neutralino with a mass, annihilation cross section, and dominant annihilation modes which are within the range required to produce the observed features of the WMAP Haze. The stau-coannihilation region, in contrast, is disfavored as an explanation for the origin of this signal.

\end{abstract}
\pacs{95.35.+d;95.30.Cq,95.55.Ka; FERMILAB-PUB-08-272-A}
\maketitle

\section{Introduction}

Although the primary mission of the Wilkinson Microwave Anisotropy Probe (WMAP) is to study the anisotropies in the cosmic microwave background (CMB) with the intention of better determining cosmological parameters~\cite{spergel}, it is also capable of measuring the properties of a number of astrophysical foregrounds, including synchrotron emission from supernova shock acceleration, and emission from thermal dust, spinning dust, and ionized gas~\cite{Gold:2008kp}. Surprisingly, WMAP's observations of the inner $20^{\circ}$ around the Galactic Center have revealed an excess of microwave emission which does not appear to be the result of any of the standard foreground mechanisms~\cite{haze1}. 

When first identified, the WMAP haze was interpreted as thermal bremsstrahlung (free-free) emission from hot gas. Such emission, however, in addition to being inconsistent with the the measured spectrum, would be accompanied by an H$\alpha$ recombination line or X-ray emission which have not been observed at the required level~\cite{haze1}. In light of this, the spectrum of the WMAP Haze leads one to interpret it as synchrotron emission with a very hard spectral index~\cite{haze1}. Although supernovae shocks generate relativistic electrons which produce synchrotron, the resulting spectrum after diffusion and energy losses are taken into account is expected to be much softer ({\it ie.} have a steeper spectral index) than is observed in the WMAP Haze. A singular explosive event (perhaps a recent gamma ray burst) could potentially produce a harder electron spectrum, but would likely be unable to generate enough electrons. Furthermore, propagation effects would lead to a softer spectrum by the time that the electrons had diffused far enough to fill the region over which the WMAP Haze is observed~\cite{haze1,greganddoug-privatecomm}.

More recently, the WMAP Haze has been interpreted as the synchrotron emission from relativistic electrons and positrons produced in dark matter annihilations~\cite{haze2,haze3} (for other work on the subject of the indirect detection of dark matter with synchrotron, see Ref.~\cite{syn}). In particular, it has been shown that dark matter in the form of weakly interactive massive particles (WIMPs) produced thermally in the early universe could naturally generate the observed emission. The angular distribution, spectrum and intensity of the WMAP Haze favor a WIMP with a mass in the range of 80 GeV to several TeV, a cusped halo profile within the several kiloparsecs around the Galactic Center ($\rho \propto r^{-1.2}$, which is slightly steeper than found by Navarro, Frenk and White (NFW)~\cite{nfw}, but less steep than predicted by Moore {\it et al.}~\cite{moore} or NFW if strongly adiabatically contracted~\cite{ac}), and an annihilation cross section within a factor of a few of $3 \times 10^{-26}$ cm$^3$/s (the preferred value for a thermal relic)~\cite{haze2}. 

Although many candidates for dark matter have been proposed~\cite{review}, among the best motivated are those which appear in supersymmetric extensions of the Standard Model. In particular, the lightest neutralino~\cite{neutralinodm} is a very attractive and well studied dark matter candidate. In this article, we revisit the dark matter interpretation of the WMAP Haze and study the supersymmetric parameter space which leads to a lightest neutralino with the properties required to generate the observed emission. We find that a large fraction of the phenomenologically viable parameter space in the Constrained Minimal Supersymmetric Standard Model (CMSSM) predicts a lightest neutralino with properties consistent with the observed properties of the WMAP Haze. If annihilating neutralino are, in fact, the source of this anomalous emission, we find the prospects for both direct and indirect detection of neutralinos to be quite promising.

\section{The Characteristics of Dark Matter Required To Generate The WMAP Haze}

In this section, we briefly summarize the results of Ref.~\cite{haze2}, which determined the range of masses, dominant annihilation channels and annihilation cross sections required of a WIMP if it is to generate the observed features of the WMAP Haze. 

Neutralinos (or other WIMP species) annihilating in the halo of the Milky Way produce a combination of gamma-rays, neutrinos, protons, antiprotons, electrons and positrons. The electrons and positrons which are produced move under the influence of the Galactic Magnetic Field, losing energy through inverse Compton scattering with starlight, emission from dust, and the CMB, and through synchrotron emission. To determine the resulting electron/positron spectrum in the inner Galaxy, one solves the diffusion-loss equation~\cite{prop}:
\begin{eqnarray}
\frac{\partial}{\partial t}\frac{dn_{e}}{dE_{e}} = \vec{\bigtriangledown} \cdot \bigg[K(E_{e},\vec{x})  \vec{\bigtriangledown} \frac{dn_{e}}{dE_{e}} \bigg]
+ \frac{\partial}{\partial E_{e}} \bigg[b(E_{e},\vec{x})\frac{dn_{e}}{dE_{e}}  \bigg] + Q(E_{e},\vec{x}),
\label{dif}
\end{eqnarray}
where $dn_{e}/dE_{e}$ is the number density of positrons per unit energy, $K(E_{e},\vec{x})$ is the diffusion constant, and $b(E_{e},\vec{x})$ is the rate of energy loss. The source term in the diffusion-loss equation, $Q(E_e, \vec{x})$, reflects both the distribution of dark matter in the Galaxy, and the mass, annihilation cross section, and dominant annihilation channels of the neutralino. Following Ref.~\cite{haze2}, we adopt the following diffusion parameters: $K(E_e) \approx 10^{28} \, (E_{e} / 1 \, \rm{GeV})^{0.33} \,\rm{cm}^2 \, \rm{s}^{-1}$, and $b(E_e) = 5 \times 10^{-16} \, ({E_e} / 1 \, \rm{GeV})^2 \,\, \rm{s}^{-1}$. We also select boundary conditions corresponding to a slab of half-thickness 3 kiloparsecs, beyond which cosmic ray electrons/positrons are allowed to freely escape the Galactic Magnetic Field.

Once the steady-state spatial and energy distributions of electrons and positrons in the inner Galaxy have been determined, we can proceed to calculate the resulting synchrotron spectrum. For the purposes of the spectral shape of the synchrotron, we adopt an average magnetic field strength of 10 $\mu$G within the inner few kiloparsecs of the Milky Way. To determine the fraction of the energy in electrons and positrons which is transferred into synchrotron emission, we compare the energy loss rates to synchrotron and inverse Compton scattering, which scale as the energy density in magnetic fields and radiation fields, respectively. The flux of synchrotron emission, therefore, scales as $U_{B}/(U_B+U_{\rm rad})$, where $U_B$ and $U_{\rm rad}$ are the average magnetic field and radiation densities. As our central estimate, we adopt  an average ratio of: $U_{B}/(U_B+U_{\rm rad}) = 0.25$, although the true value could depart from this significantly. In order to account for this uncertainty, we will allow in our analysis values of $U_{B}/(U_B+U_{\rm rad})$ which are within the range of $0.1-1.0$.

\begin{figure}[t]
\centering\leavevmode
\mbox{
\includegraphics[width=3.5in,angle=0]{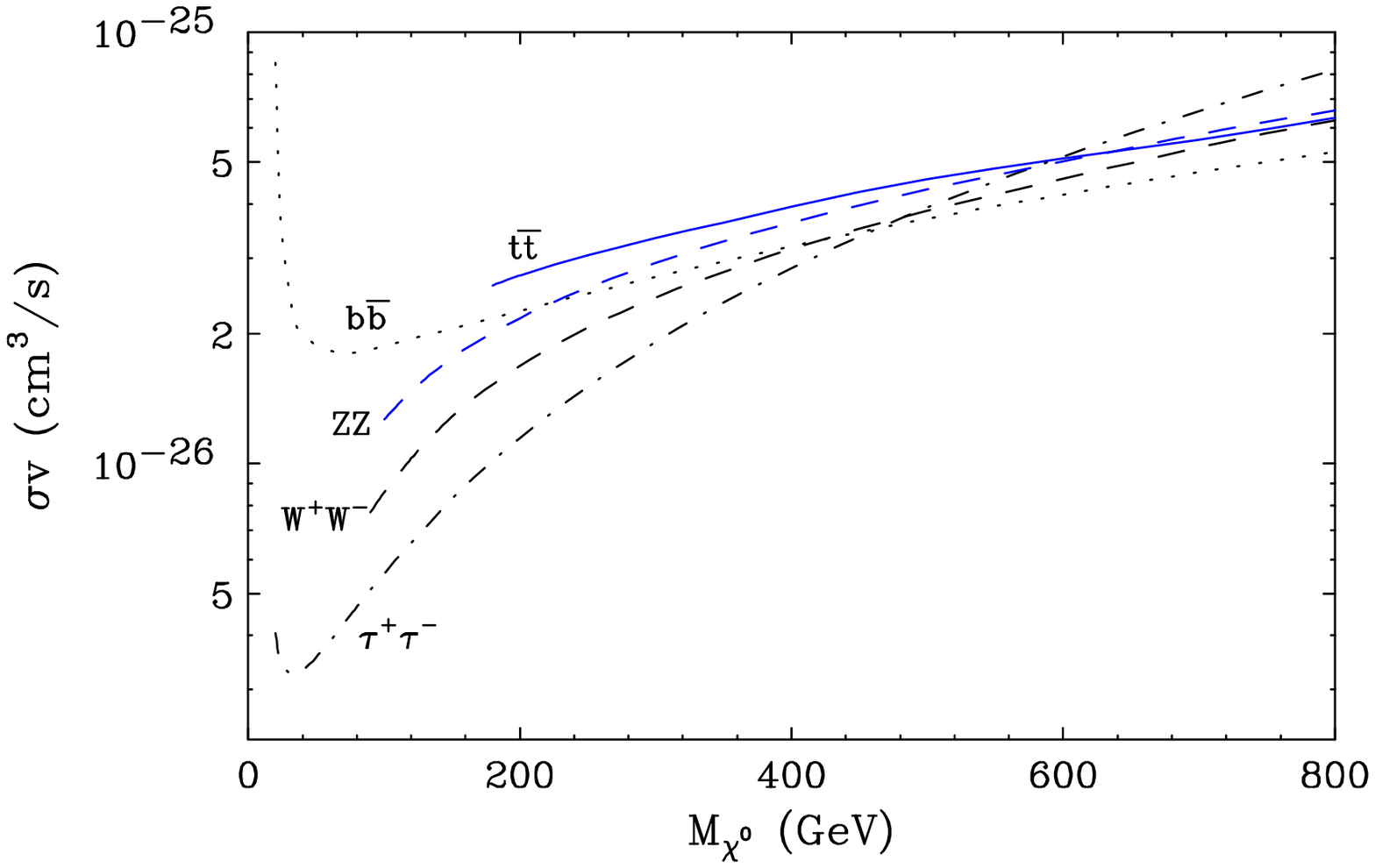}
\includegraphics[width=3.5in,angle=0]{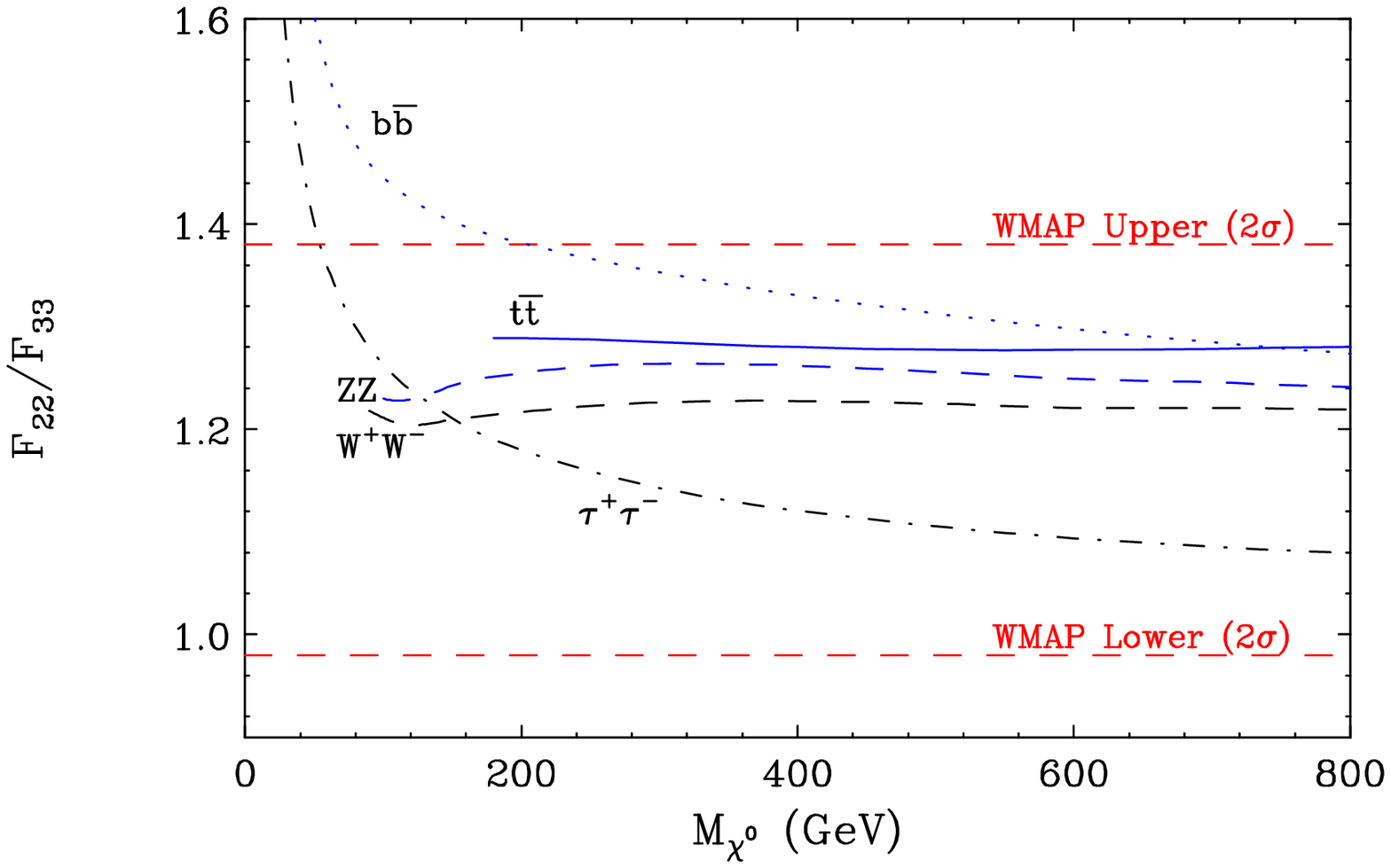}
}
\caption{Left: The neutralino annihilation cross section required to produce the observed intensity of the WMAP Haze as a function of mass, for several dominant annihilation channels. These results were calculated using $U_{B}/(U_B+U_{\rm rad}) = 0.25$ and will shift upward or downward with the inverse of this quantity. Right: The ratio of synchrotron intensities in WMAP's 22 GHz and 33 GHz frequency bands as a function of the neutralino's mass, for several dominant annihilation channels. The spectrum produced from a neutralino with any of the annihilation channels shown is consistent with the measured spectrum of the WMAP Haze with the exception of light neutralinos annihilating mostly to $b\bar{b}$ or $\tau^+ \tau^-$. Adapted from Ref.~\cite{haze2}.}
\label{sigma}
\end{figure}

In Ref.~\cite{haze2}, the WIMP mass and annihilation cross section required to generate the WMAP Haze was calculated for various dominant annihilation channels (after fixing the halo profile by matching to the angular distribution of the emission). In Fig.~\ref{sigma}, we plot this result, adapted slightly to the case of neutralino dark matter. In the left frame, the annihilation cross section (in the low velocity limit) required to normalize the synchrotron emission from dark matter annihilation products to the observed intensity of the WMAP Haze is shown, as a function of the neutralino mass, for several annihilation channels. These results were calculated using $U_{B}/(U_B+U_{\rm rad}) = 0.25$ and will shift upward or downward with the inverse of this quantity. From this frame, we see that an annihilation cross section of approximately $(1-5) \times 10^{-26}$ cm$^3$/s is required to produce the observed intensity of the WMAP Haze. This is remarkably similar to the value of approximately $3 \times 10^{-26}$ cm$^3$/s which is required of a thermal relic to be produced in the early universe with the measured dark matter abundance (in the absence of coannihilations, resonances or s-wave suppression). 

In the right frame of Fig.~\ref{sigma}, we turn our attention to the spectrum of the WMAP Haze. In particular, we show the ratio of synchrotron intensities produced in the 22 GHz and 33 GHz frequency bands of WMAP, again as a function of the neutralino's mass and dominant annihilation channel. The horizontal dashed lines are the (2$\sigma$) upper and lower limits of this ratio from measurements of the WMAP Haze~\cite{haze2}. From this frame, we see that for each of the annihilation channels shown, the resulting spectrum is consistent with that of the WMAP Haze, with the exception of light neutralinos ($m_{\chi^0} \lsim 200$ GeV) annihilating mostly to b quarks (or a very light neutralino annihilating to tau leptons).

\section{Neutralino Dark Matter in the Constrained Minimal Supersymmetric Standard Model (CMSSM)}

In this section, we calculate the mass, annihilation cross section and dominant annihilation modes of the lightest neutralino over the phenomenologically acceptable regions of the CMSSM parameter space, and compare those results to the values required to generate the observed properties of the WMAP Haze, as discussed in the previous section. 

The parameter space of the CMSSM consists of four continuous parameters: the universal scalar mass $m_0$, the universal gaugino mass $m_{1/2}$, the universal trilinear scalar coupling $A_0$, and the ratio of the vacuum expectation values of the two Higgs doublets $\tan \beta$, and one discrete parameter: the sign of the higgsino mass parameter $\mu$. Under the assumptions implicit in the CMSSM, the masses and couplings of the entire MSSM can be calculated from these five quantities. 

\begin{figure}[t]
\centering\leavevmode
\includegraphics[width=3.0in,angle=-90]{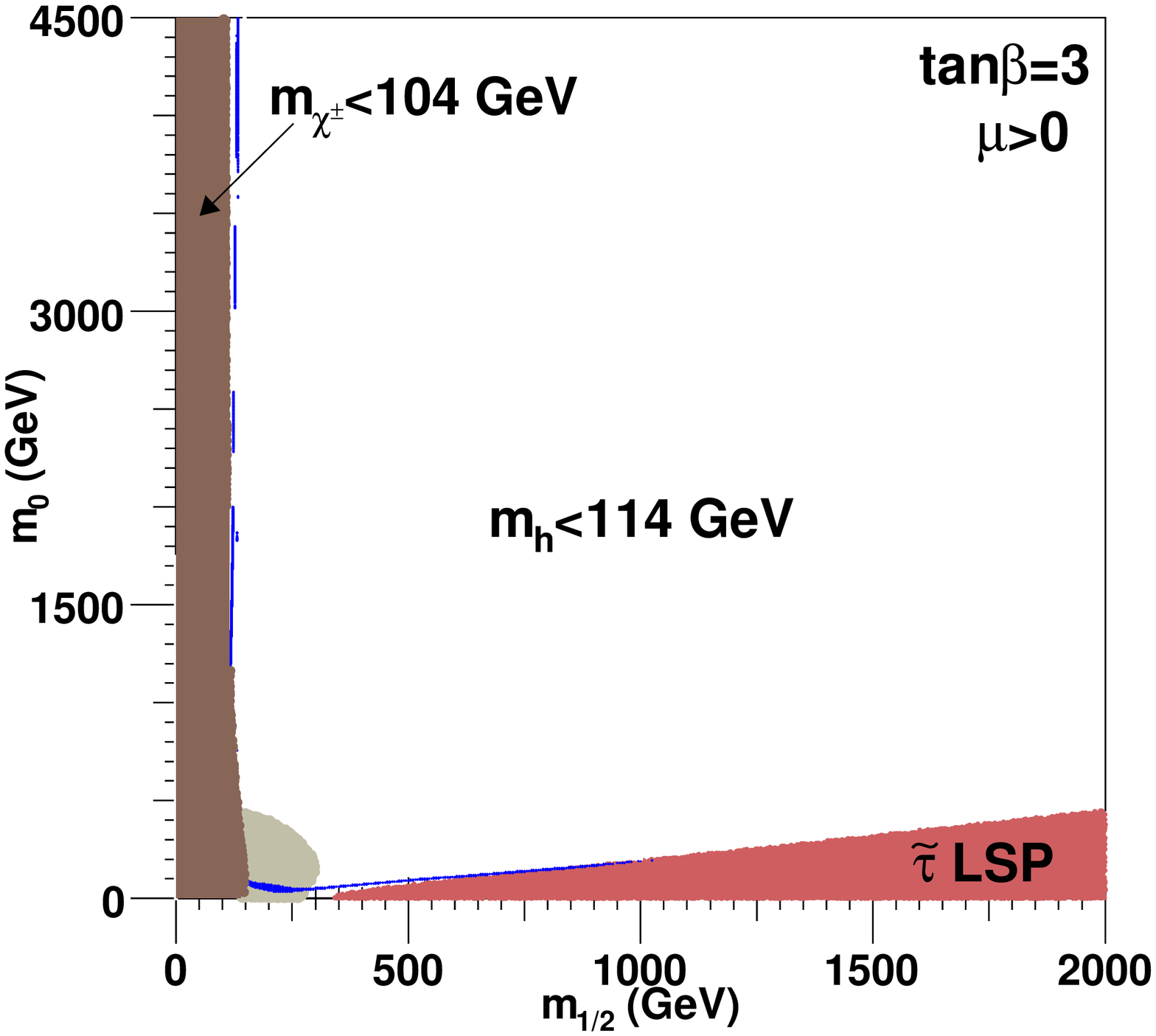}
\includegraphics[width=3.0in,angle=-90]{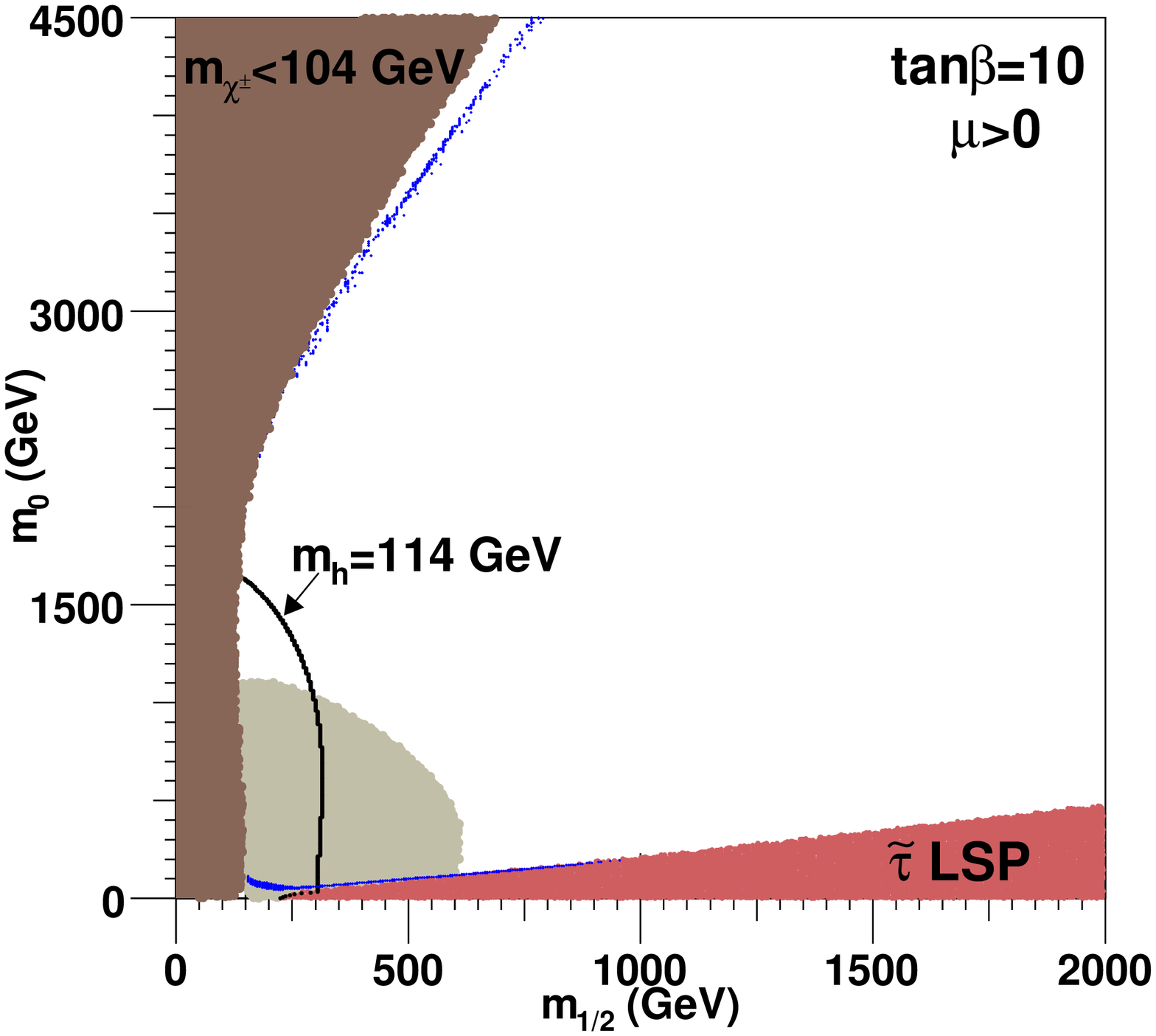}\\
\vspace{0.7cm}
\includegraphics[width=3.0in,angle=-90]{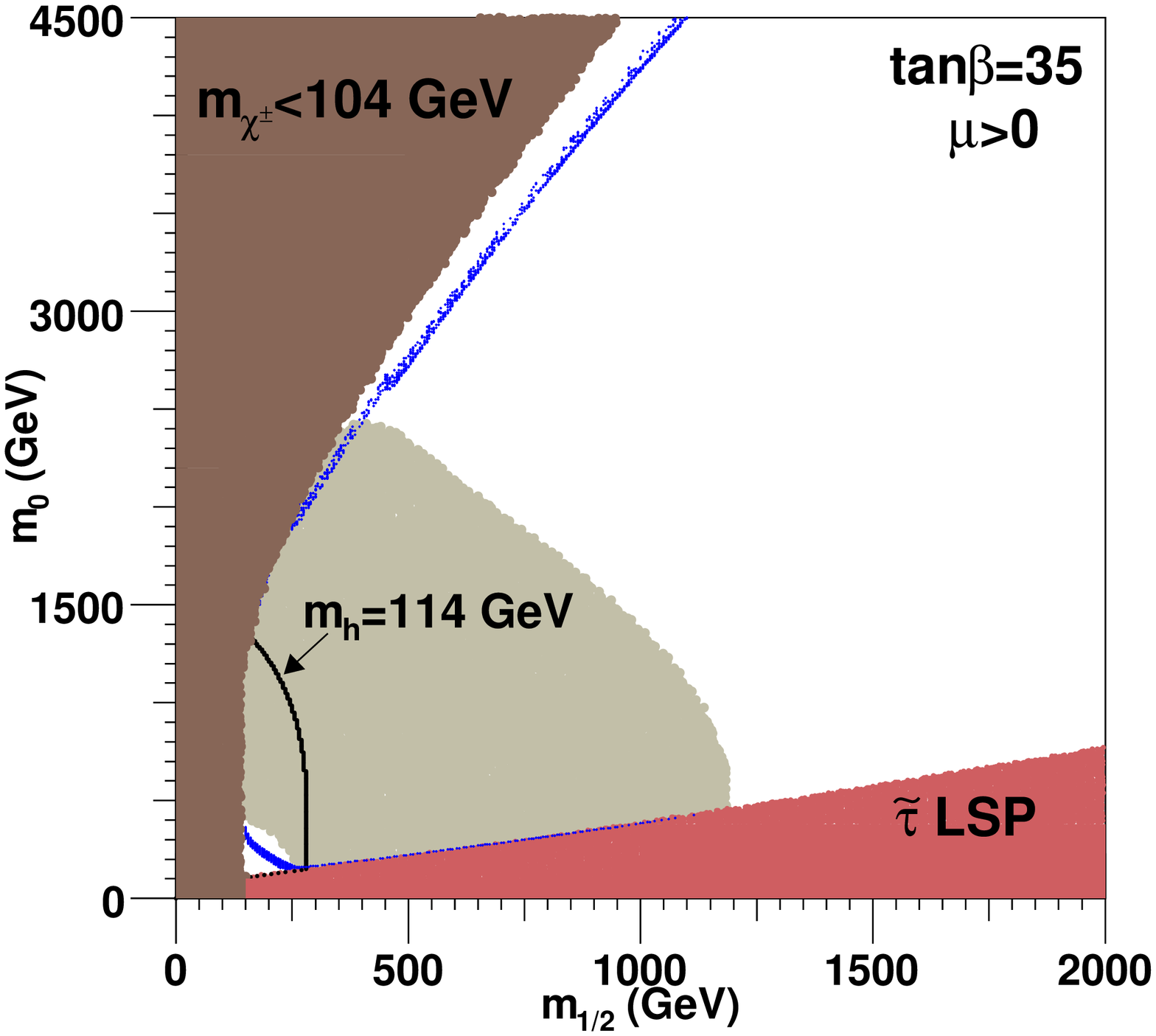}
\includegraphics[width=3.0in,angle=-90]{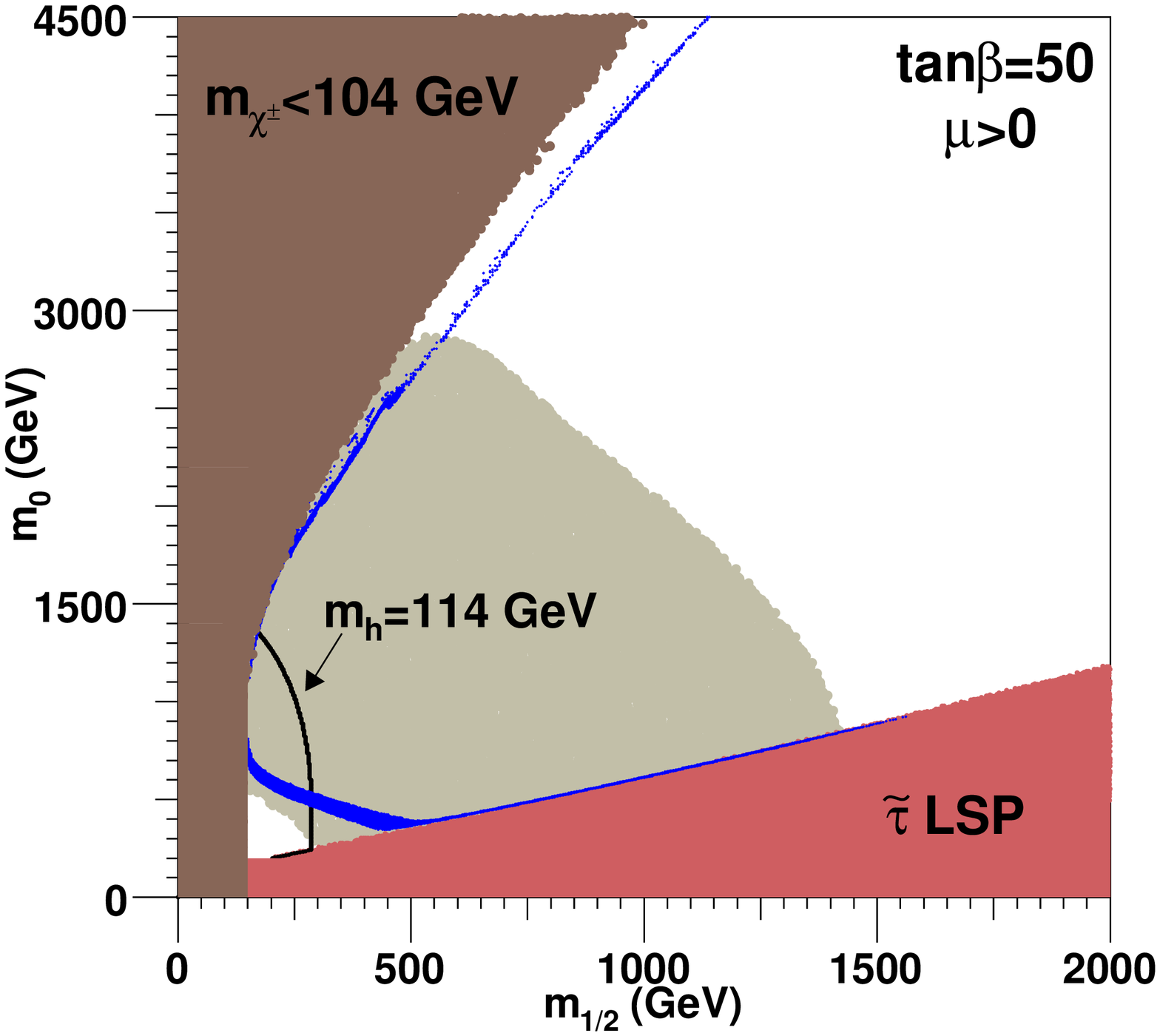}
\caption{Constraints on the CMSSM parameter space, including the region with an acceptable neutralino relic density, $0.0913 < \Omega_{\chi^0} h^2 < 0.1285$ (blue)~\cite{spergel}. The shaded regions to the upper left and lower right are disfavored by the LEP chargino bound and as a result of containing a stau LSP, respectively. The LEP bound on the light Higgs mass is shown as a solid line ($m_h=114$ GeV). The region favored by measurements of the muon's magnetic moment are shown as a light shaded region (at the 3$\sigma$ confidence level)~\cite{gm2SM}. In each frame, we have used $A_0=0$ and $\mu > 0$.}
\label{mzeromhalf}
\end{figure}

\begin{figure}[t]
\centering\leavevmode
\includegraphics[width=3.0in,angle=-90]{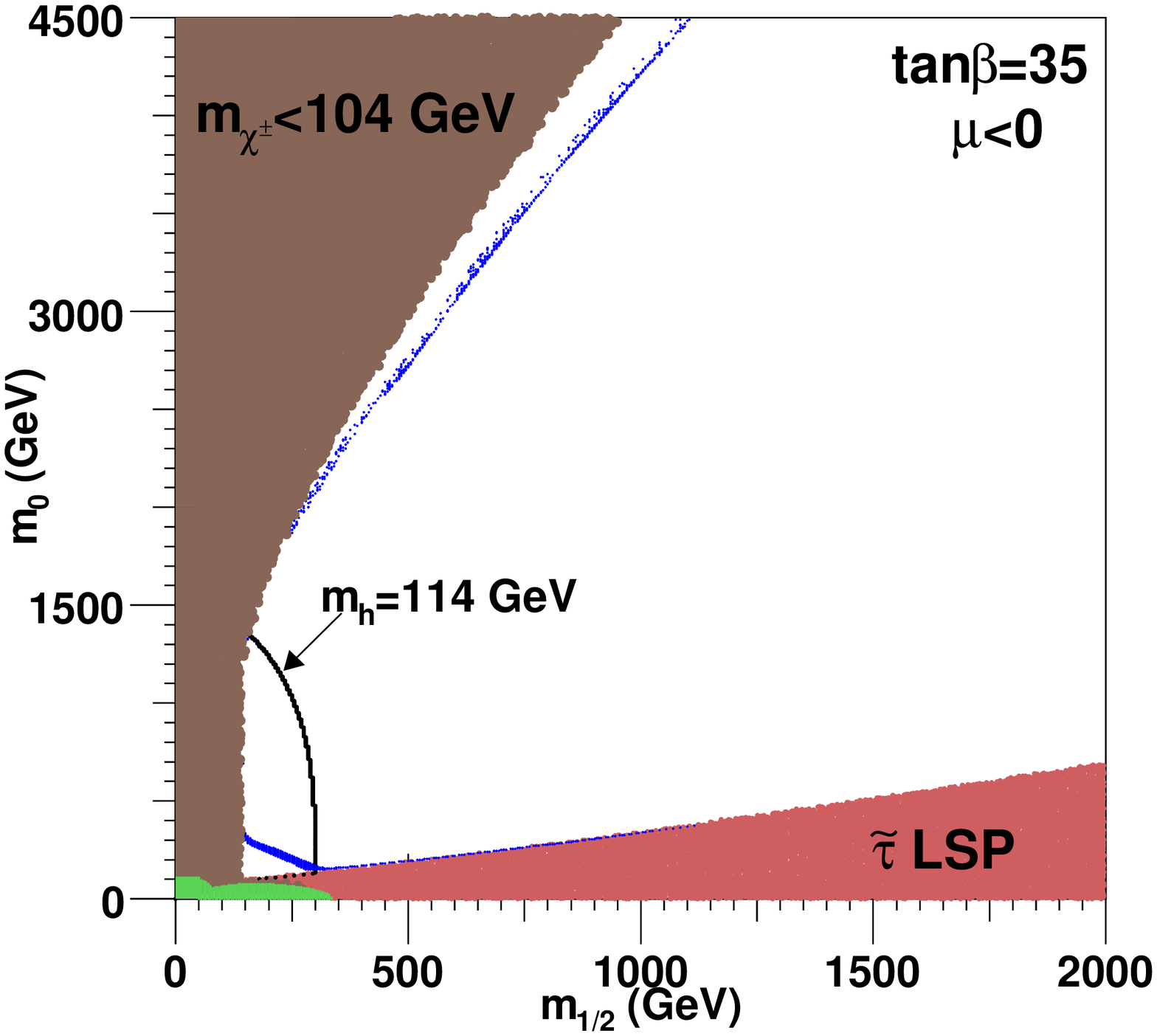}
\includegraphics[width=3.0in,angle=-90]{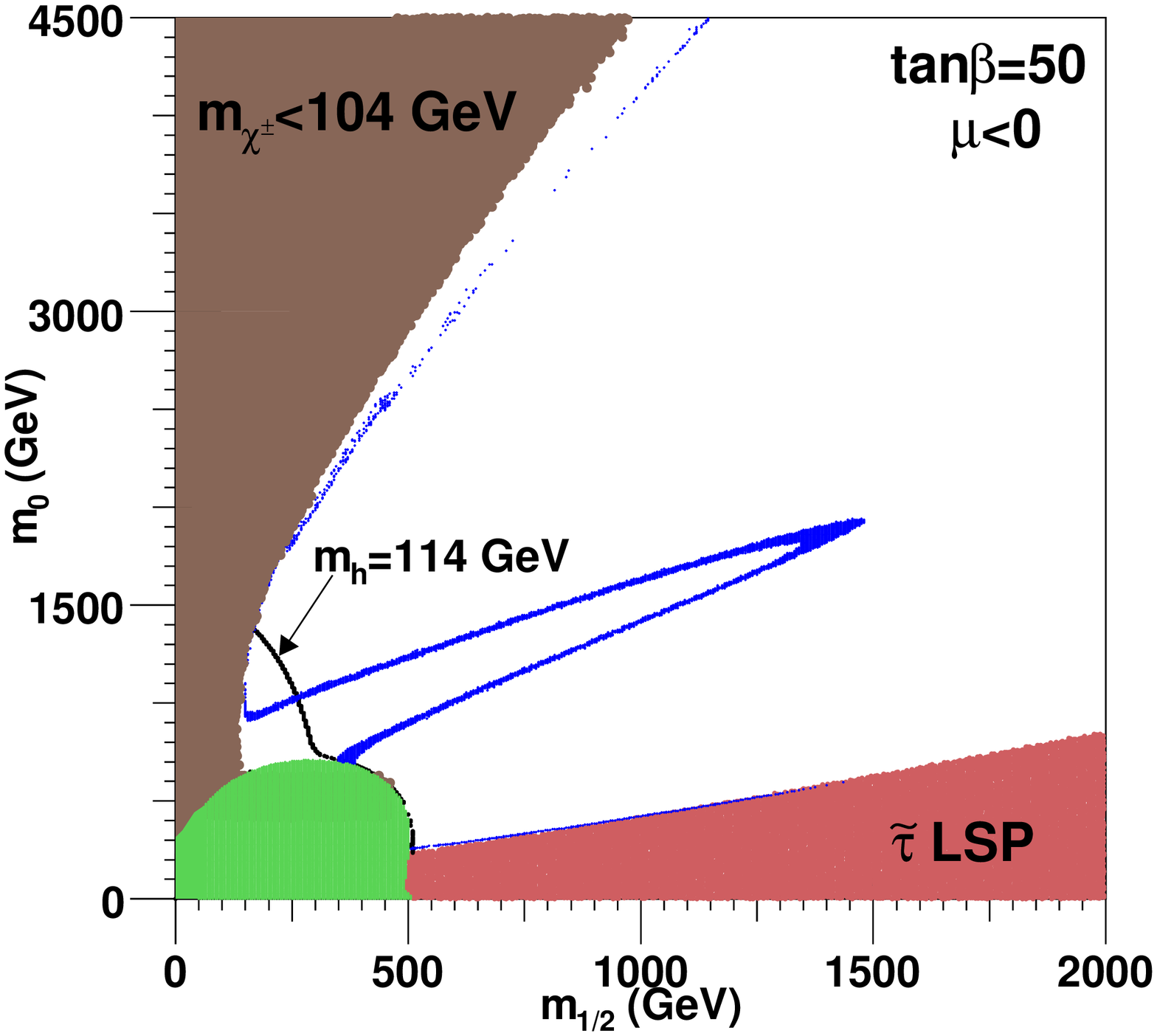}
\caption{As shown in Fig.~\ref{mzeromhalf}, but for $\mu < 0$. The lower left region of each frame is excluded on theoretical grounds.}
\label{mzeromhalfnegmu}
\end{figure}

In Figs.~\ref{mzeromhalf} and~\ref{mzeromhalfnegmu}, we show some of the phenomenological features of the CMSSM parameter space. In each frame, the narrow blue region predicts a thermal abundance of neutralinos which is within the cold dark matter density range determined by WMAP ($0.0913 < \Omega_{\chi^0} h^2 < 0.1285$, using 3$\sigma$ errors)~\cite{spergel}. The upper left region of each frame is excluded by the LEP chargino bound ($m_{\chi^{\pm}}> 104$ GeV)~\cite{pdg}. In the lower right region of each frame, the lightest supersymmetric particle is a stau, and thus does not provide a viable dark matter candidate. Also shown in each frame is the contour corresponding to the LEP Higgs mass bound ($m_h > 114$ GeV)~\cite{pdg} and lightly shaded regions corresponding to the parameter space preferred by measurements of the muon's magnetic moment  ($3.1\times 10^{-10} < {\delta a_{\mu}} < 55.9\times 10^{-10}$, using 3$\sigma$ errors~\cite{gm2SM}). We have used the DarkSUSY package to calculate these quantities~\cite{darksusy}.

Much of the parameter space shown in Figs.~\ref{mzeromhalf} and~\ref{mzeromhalfnegmu} appears to be ruled out or strongly disfavored by the constraints on the light Higgs mass and the muon's magnetic moment. In particular, the entire $m_0-m_{1/2}$ plane is ruled out by the LEP Higgs mass constraint in the case of $\tan \beta=3$. Similarly, neither of the frames with $\mu < 0$ have any points that are within the region preferred by the measurement of $\delta a_{\mu}$. Before we discard these regions of parameter space, however, a note of caution is called for. Departures from the assumptions implicit in the CMSSM can alter the nature of constraints such as that from the muon's magnetic moment and the Higgs mass bound considerably. For example, if instead of fixing the top trilinear coupling $A_t$ by its GUT-scale universal value $A_0$ by RGE running, we can maximize the light Higgs mass by selecting $A_t=2 m_{\tilde{q}} + \mu \cot \beta$, thereby reducing the impact of the LEP Higgs mass bound considerably without otherwise substantially changing the dark matter phenomenology. Similar departures from the assumptions of the CMSSM could also lead to substantial variations in the values of ${\delta a_{\mu}}$. For this reason, we will consider throughout this paper models which are (under the strict assumptions of the CMSSM) excluded by constraints on ${\delta a_{\mu}}$ and $m_h$.

\newpage
.
\newpage

From Figs.~\ref{mzeromhalf} and~\ref{mzeromhalfnegmu}, we see that only a small fraction of the parameter space predicts an abundance of neutralinos consistent with the measured density of dark matter. In particular, neutralino dark matter is expected to be overproduced relative to the observed dark matter abundance over the majority of the supersymmetric parameter space. The regions which do provide an acceptable dark matter abundance can be classified as follows:
\begin{itemize}
\item{The focus point region: For large values of $m_0$ and moderate or large values of $\tan \beta$, the lightest neutralino is a mixed bino-higgsino and, as a result, possesses large couplings which enable it to annihilate efficiently.}
\item{The stau coannihilation region: In the parameter space near the boundary of the $\tilde{\tau}$ LSP region, the lightest neutralino is nearly degenerate with the lightest stau. In these points, coannihilations between the lightest stau and lightest neutralino in the early universe lead to an acceptable density of neutralino dark matter.}
\item{The bulk region: The parameter space with light $m_0$ and light $m_{1/2}$ contains many light sparticles which, in some cases, enable the lightest neutralino to annihilate efficiently.}
\item{The $A$-funnel region: The parameter space with large $\tan \beta$ contains regions in which the lightest neutralino is able to annihilate efficiently through the CP-odd Higgs boson resonance $\chi^0 \chi^0 \rightarrow A \rightarrow f \bar{f}$. The corresponding region can be seen in the right frame of Fig.~\ref{mzeromhalfnegmu}.}
\end{itemize}

\begin{figure}[t]
\centering\leavevmode
\includegraphics[width=1.73in,angle=-90]{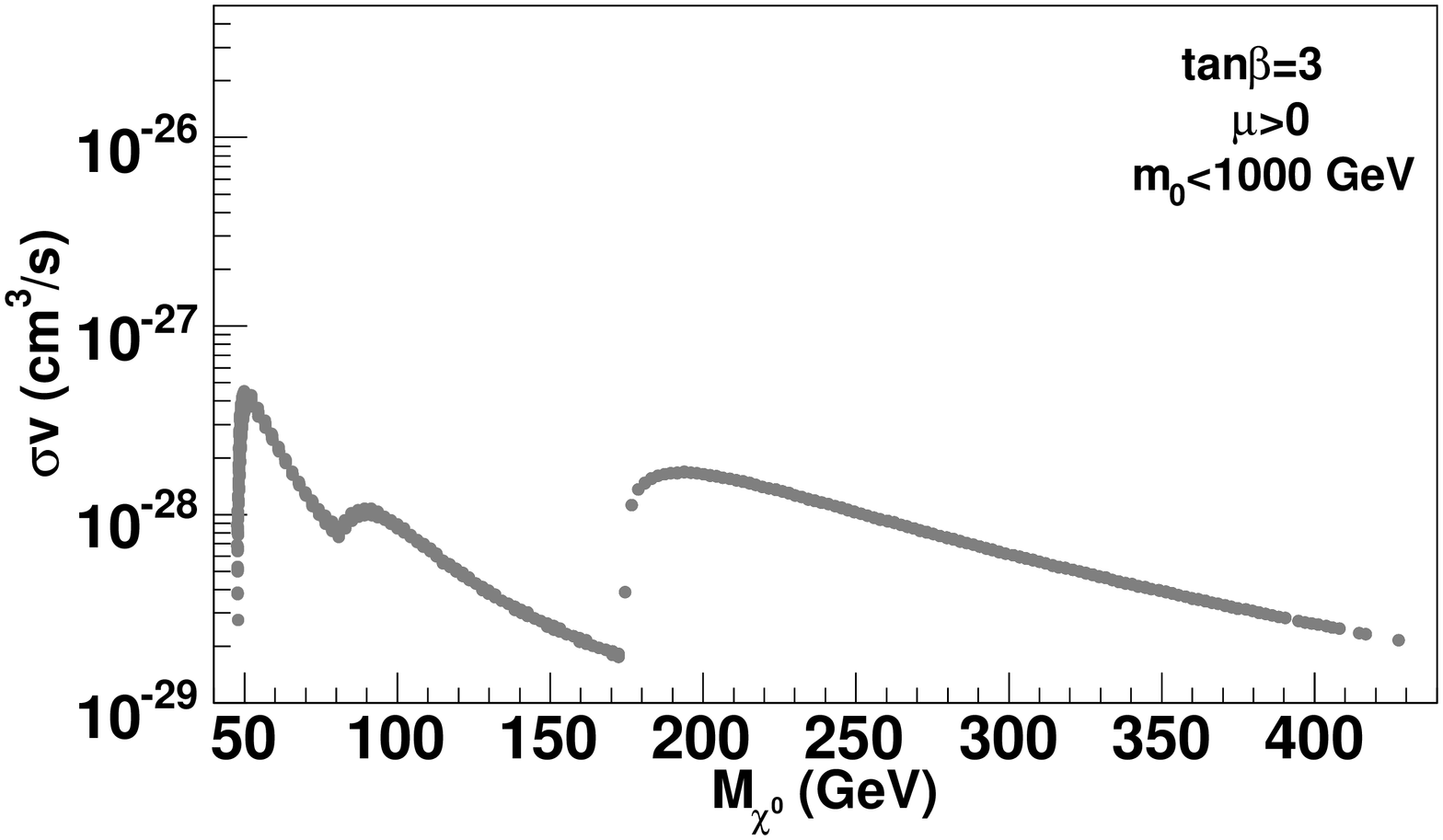}
\includegraphics[width=1.73in,angle=-90]{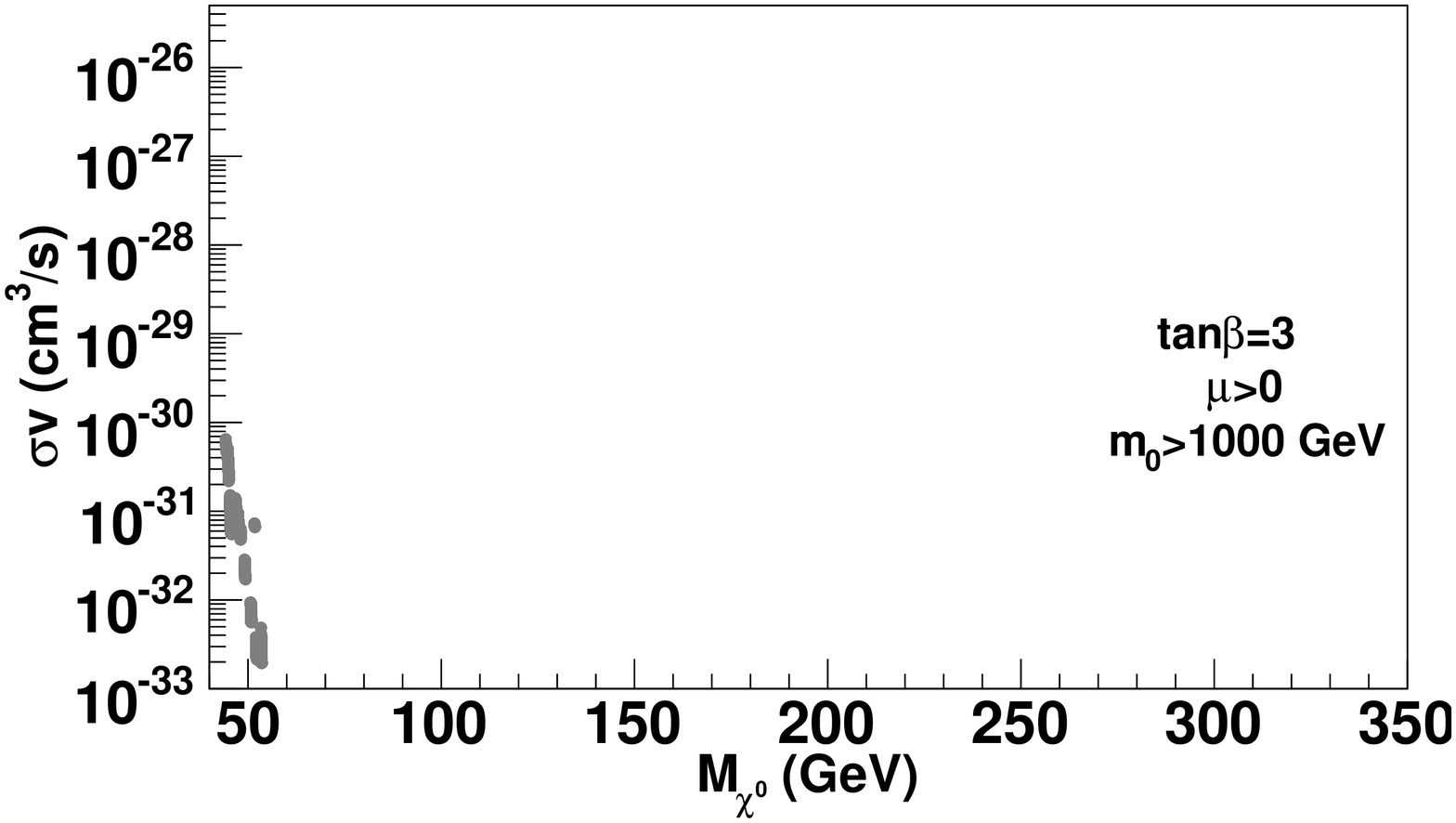}\\
\vspace{0.3cm}
\includegraphics[width=1.73in,angle=-90]{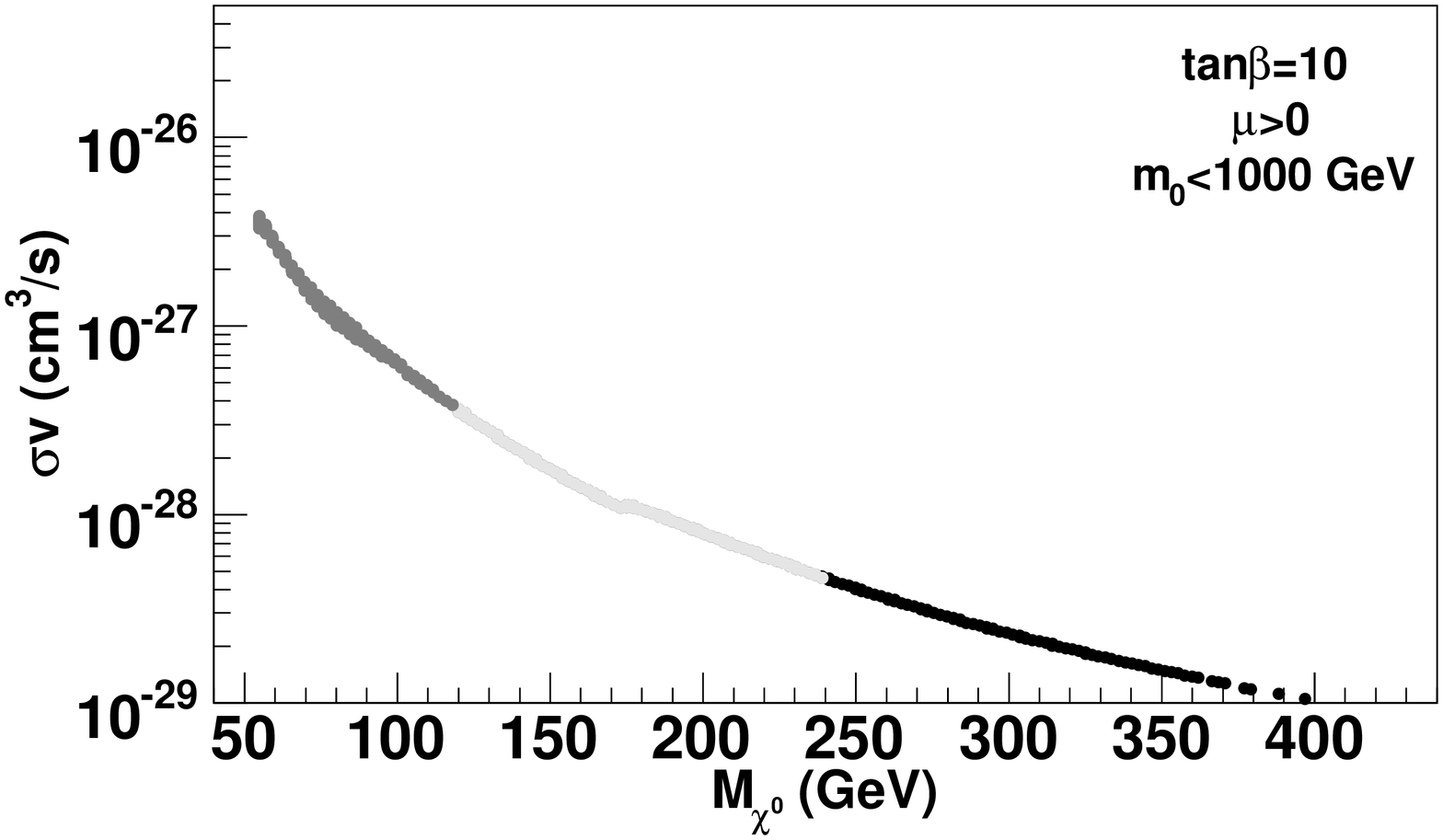}
\includegraphics[width=1.73in,angle=-90]{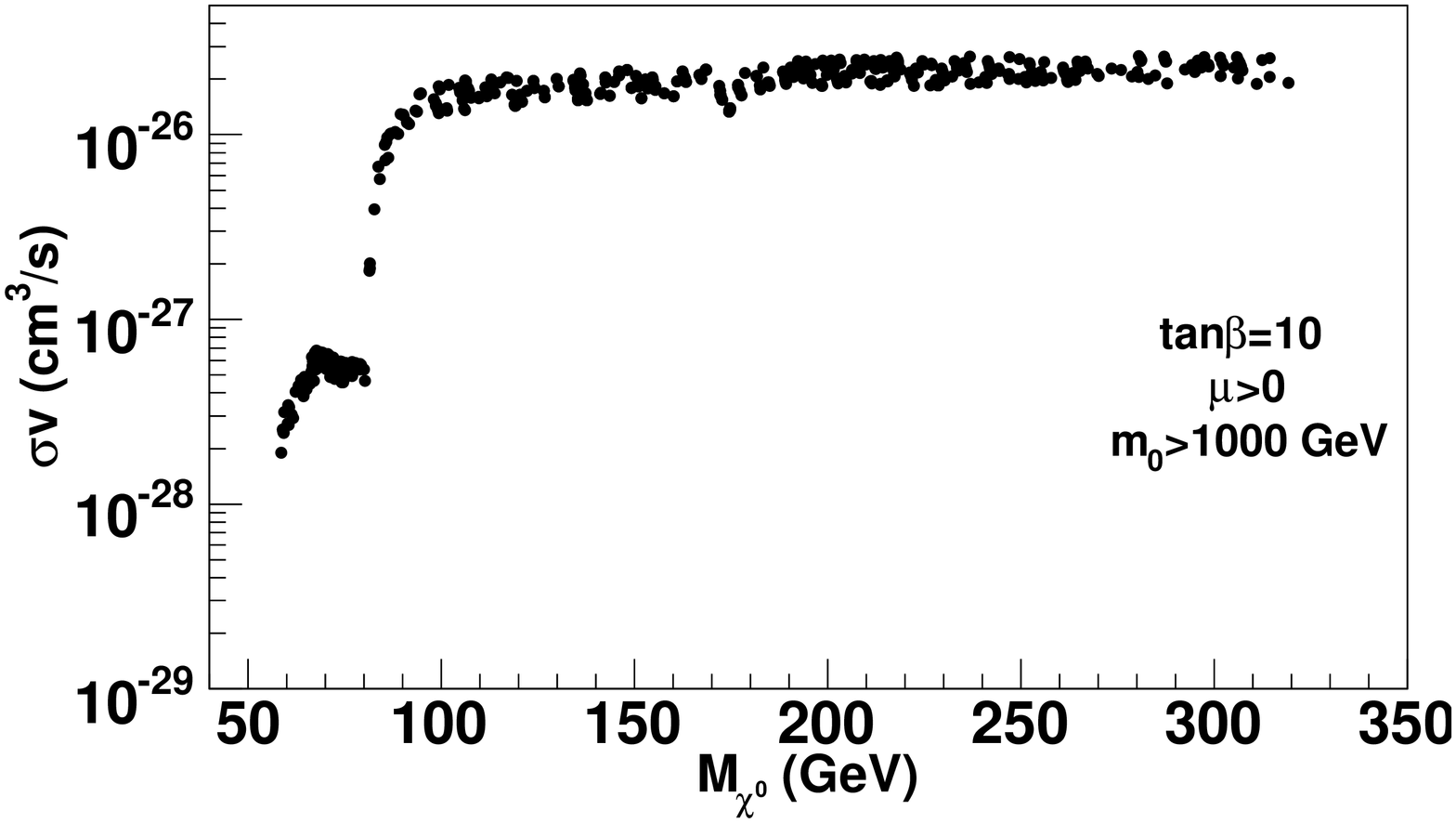}\\
\vspace{0.3cm}
\includegraphics[width=1.73in,angle=-90]{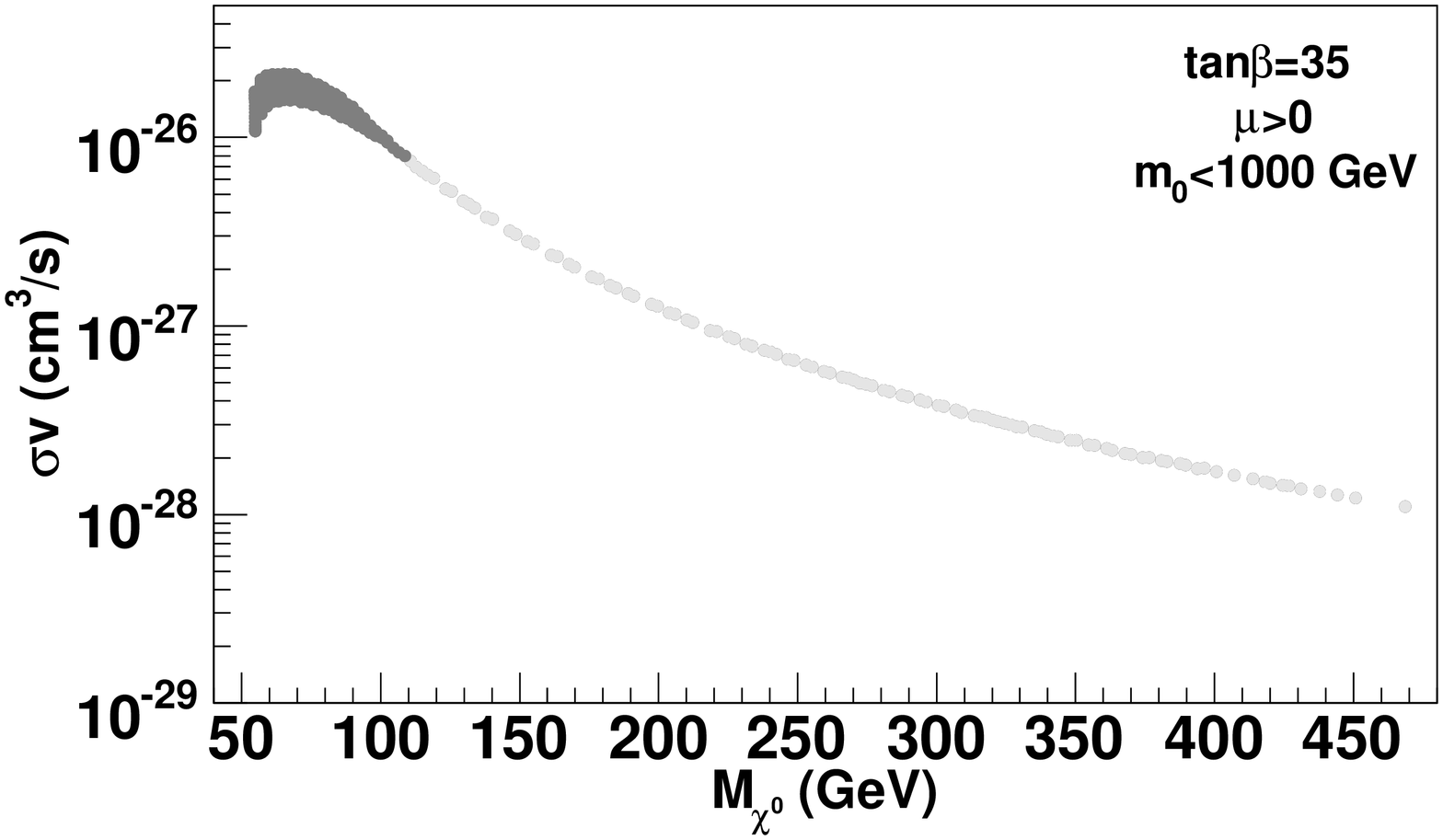}
\includegraphics[width=1.73in,angle=-90]{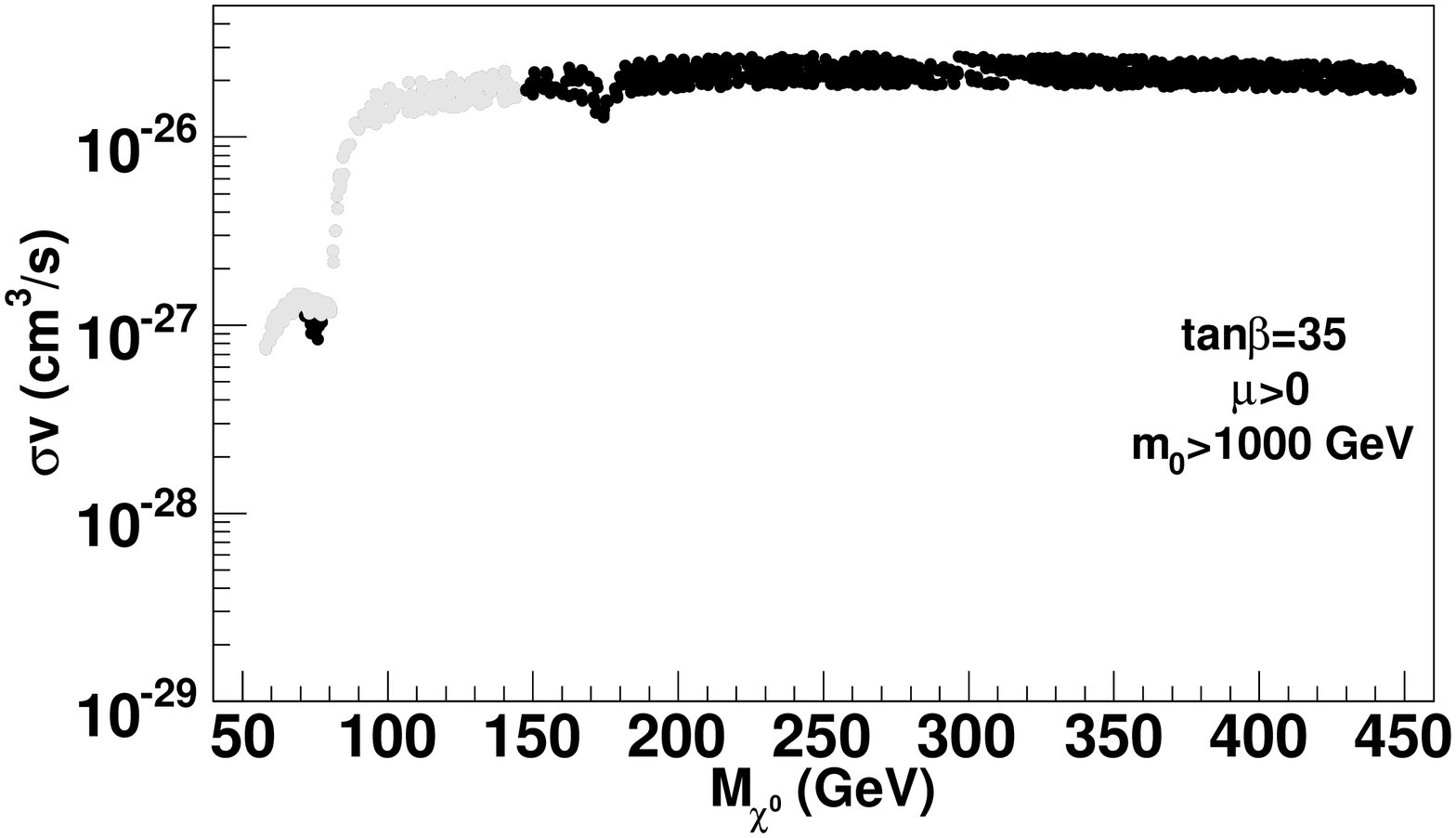}\\
\vspace{0.3cm}
\includegraphics[width=1.73in,angle=-90]{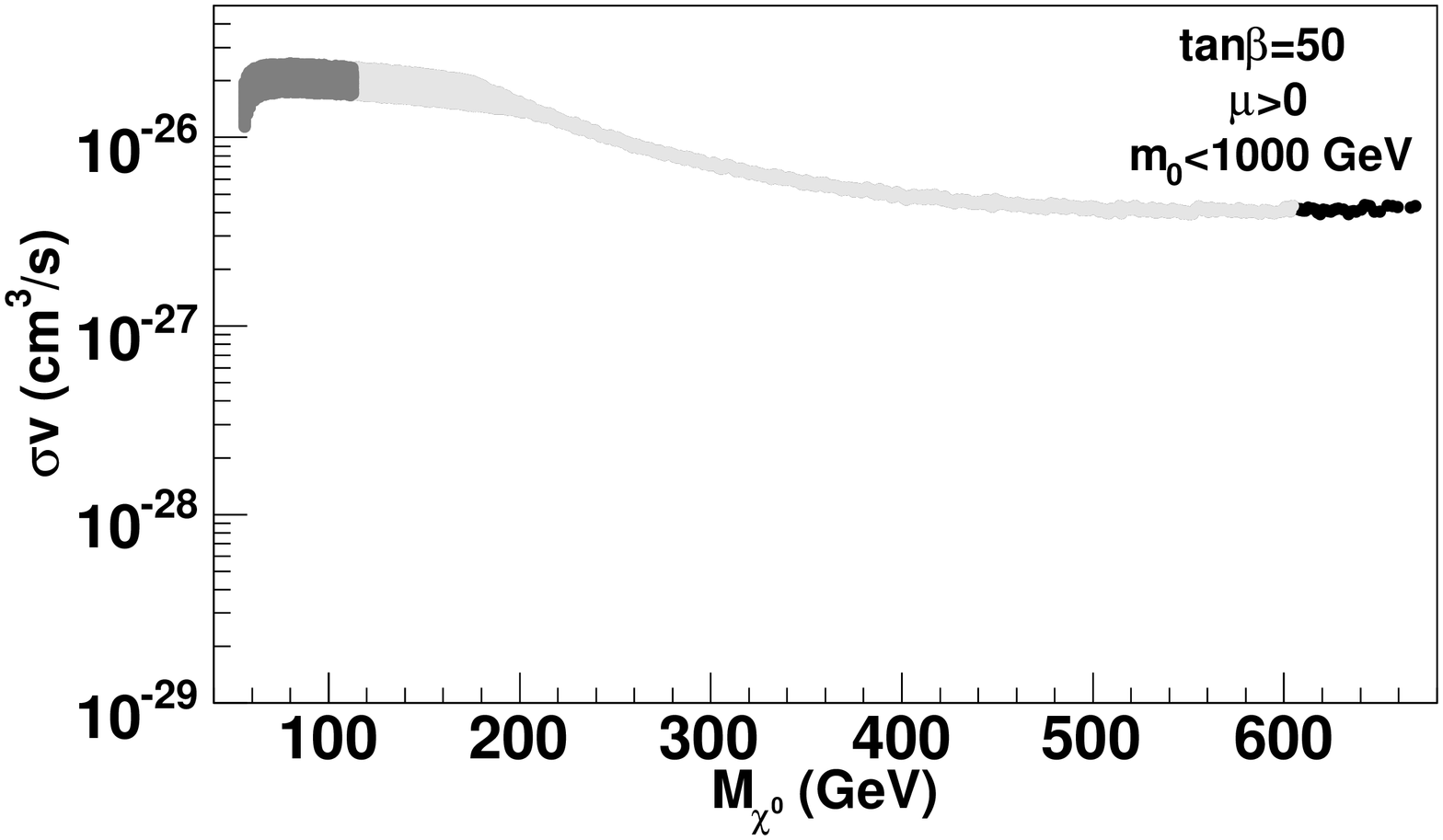}
\includegraphics[width=1.73in,angle=-90]{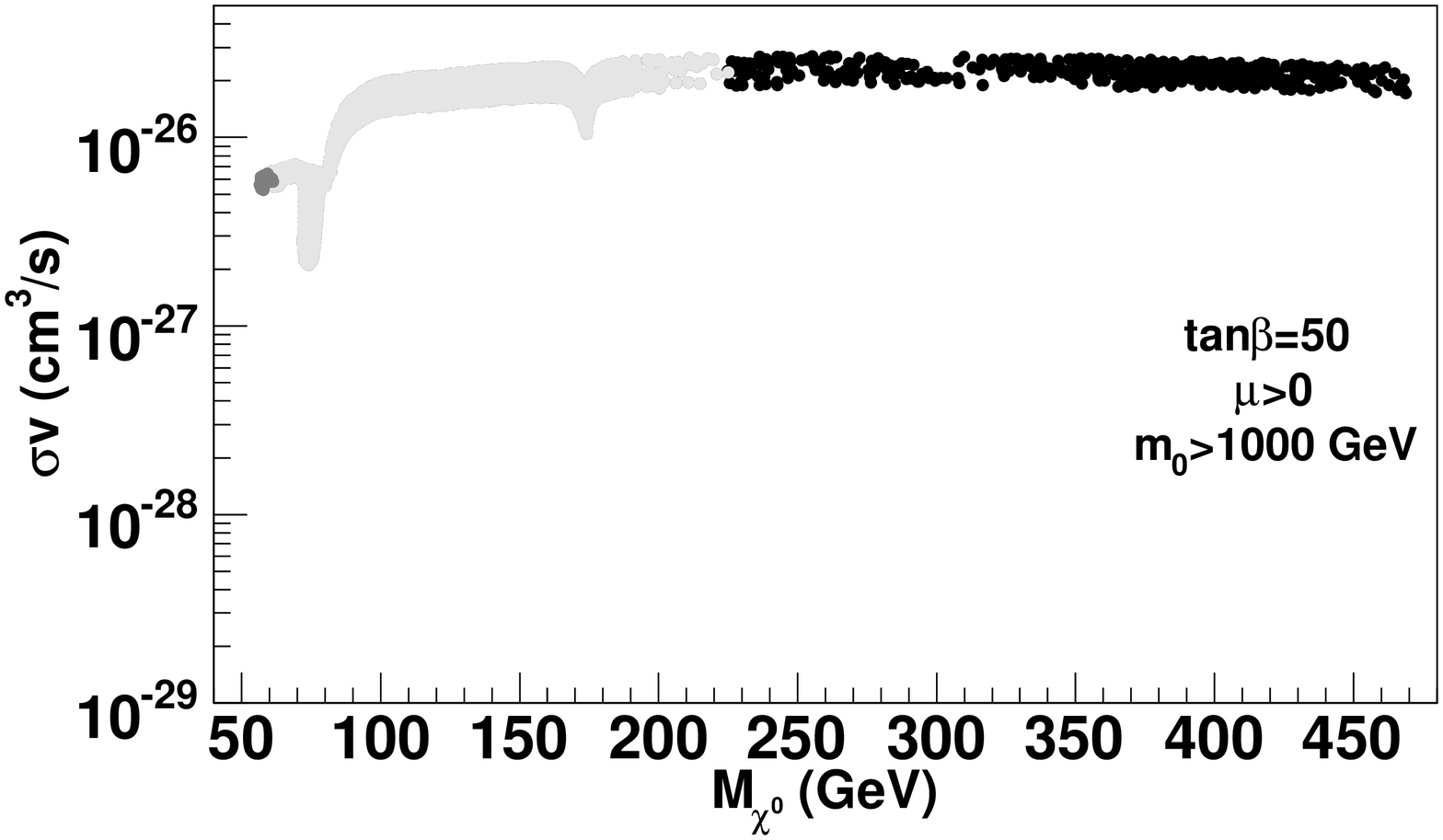}\\
\caption{The low-velocity neutralino annihilation cross section. Each point shown corresponds to a point in Fig.~\ref{mzeromhalf} that predicts a neutralino dark matter abundance consistent with the measured dark matter density, and that does not violate the LEP chargino mass bound. The dark gray regions are disfavored by the LEP Higgs mass bound. Of the remaining points, the light gray (black) regions are preferred (disfavored) by the measurements of the muon's magnetic moment. See text for more details.}
\label{crosssection}
\end{figure}

\begin{figure}[t]
\centering\leavevmode
\includegraphics[width=1.8in,angle=-90]{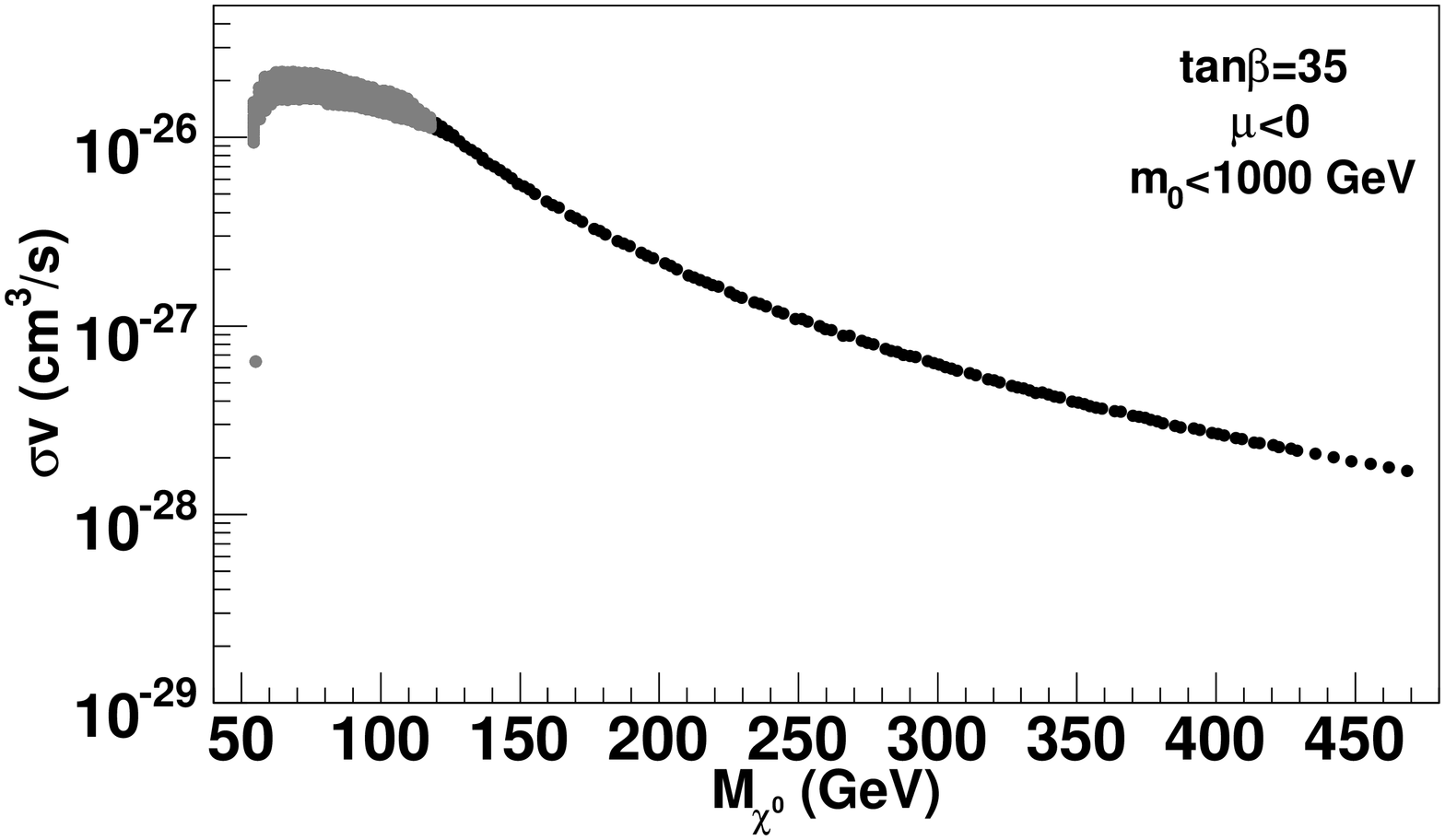}
\includegraphics[width=1.8in,angle=-90]{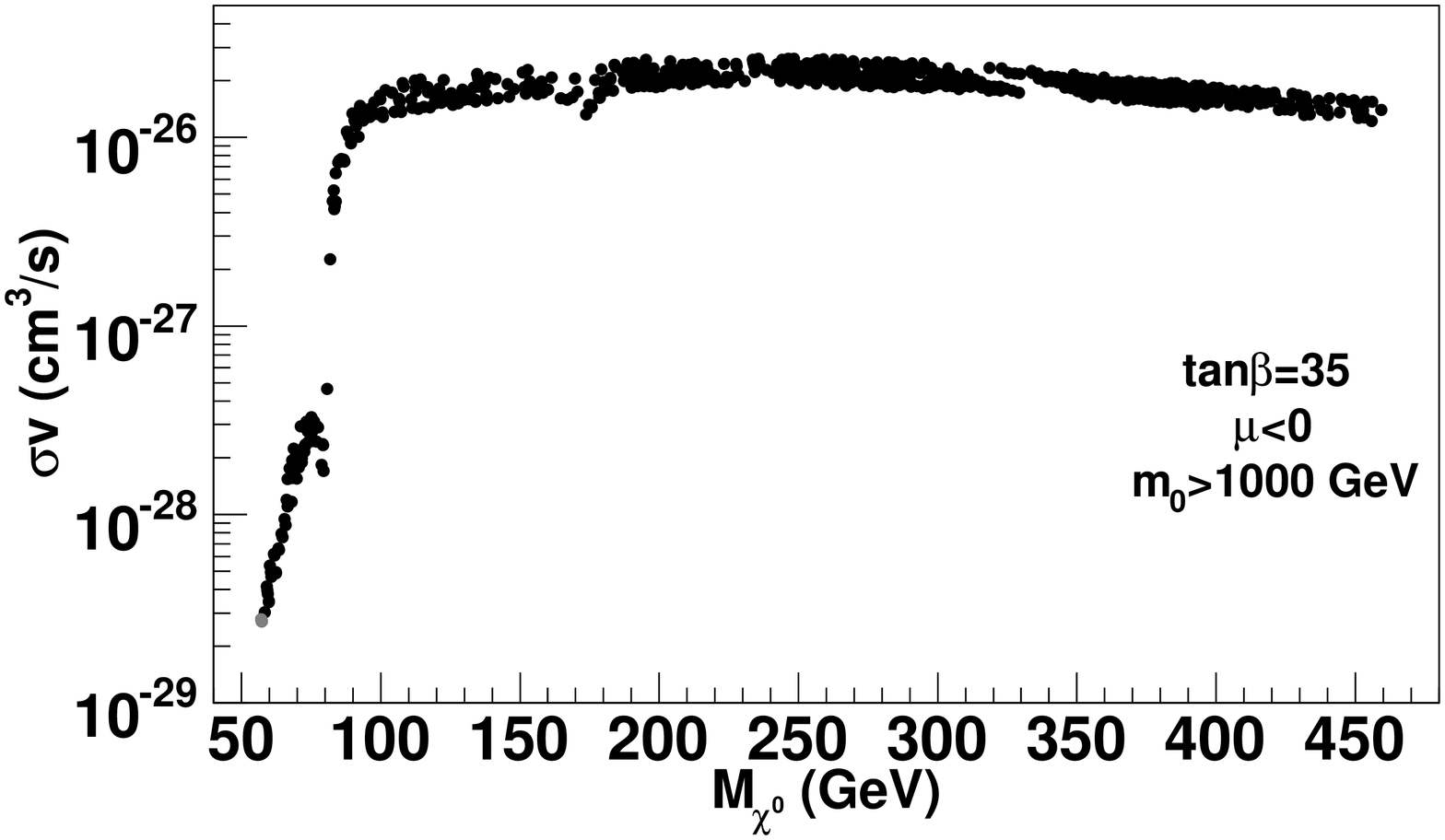}\\
\vspace{0.5cm}
\includegraphics[width=1.8in,angle=-90]{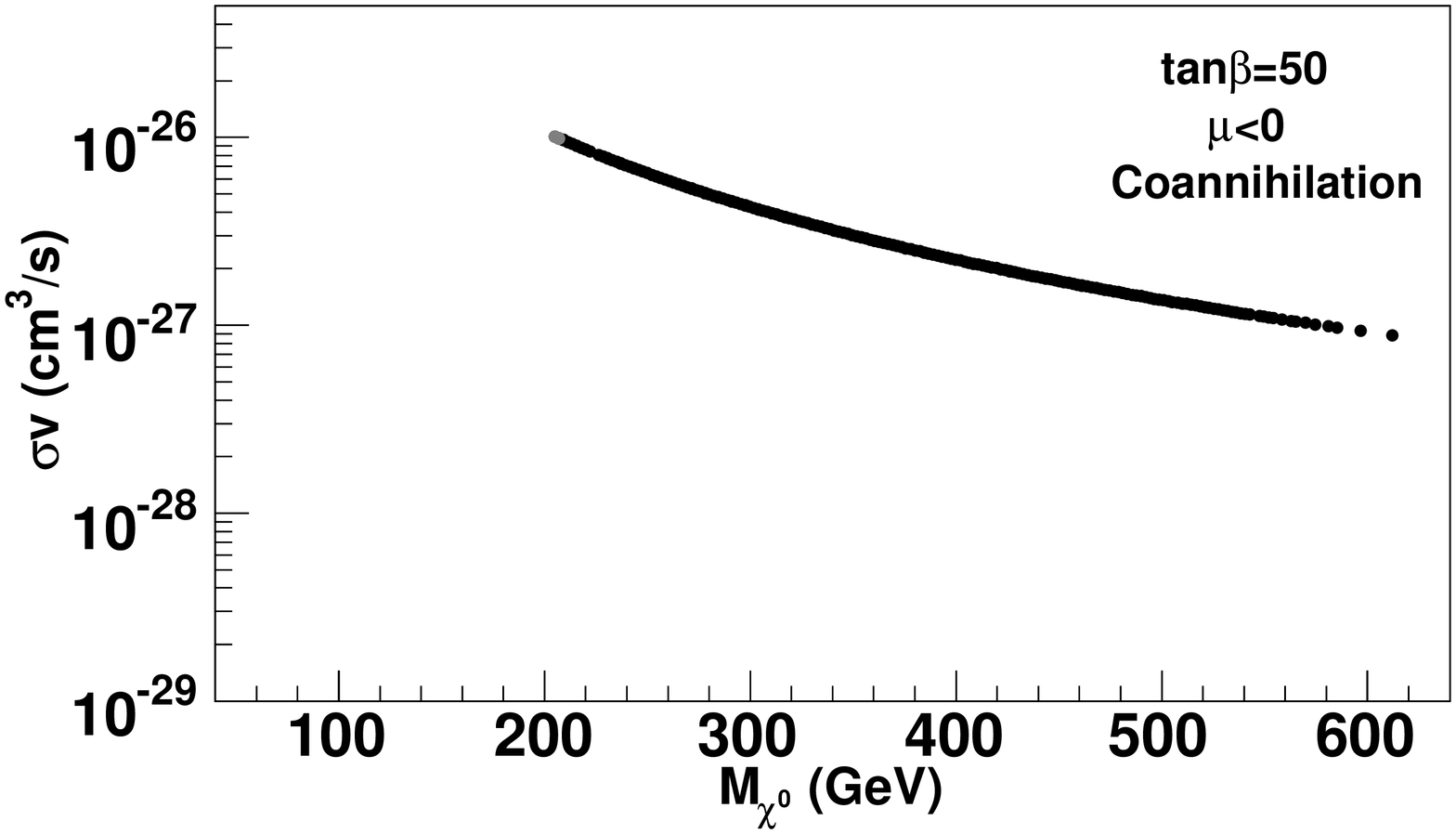}
\includegraphics[width=1.8in,angle=-90]{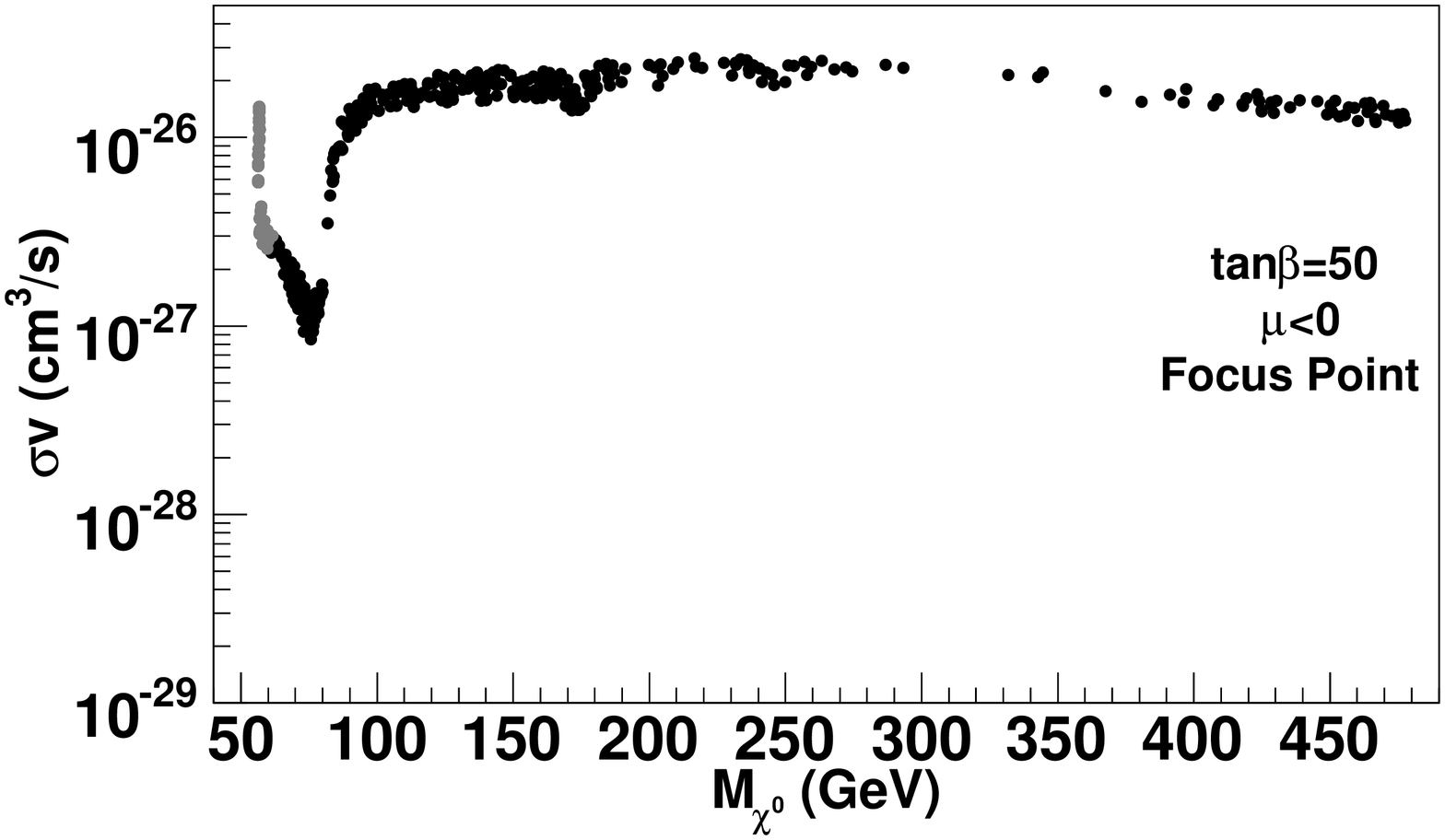}\\
\vspace{0.5cm}
\includegraphics[width=1.8in,angle=-90]{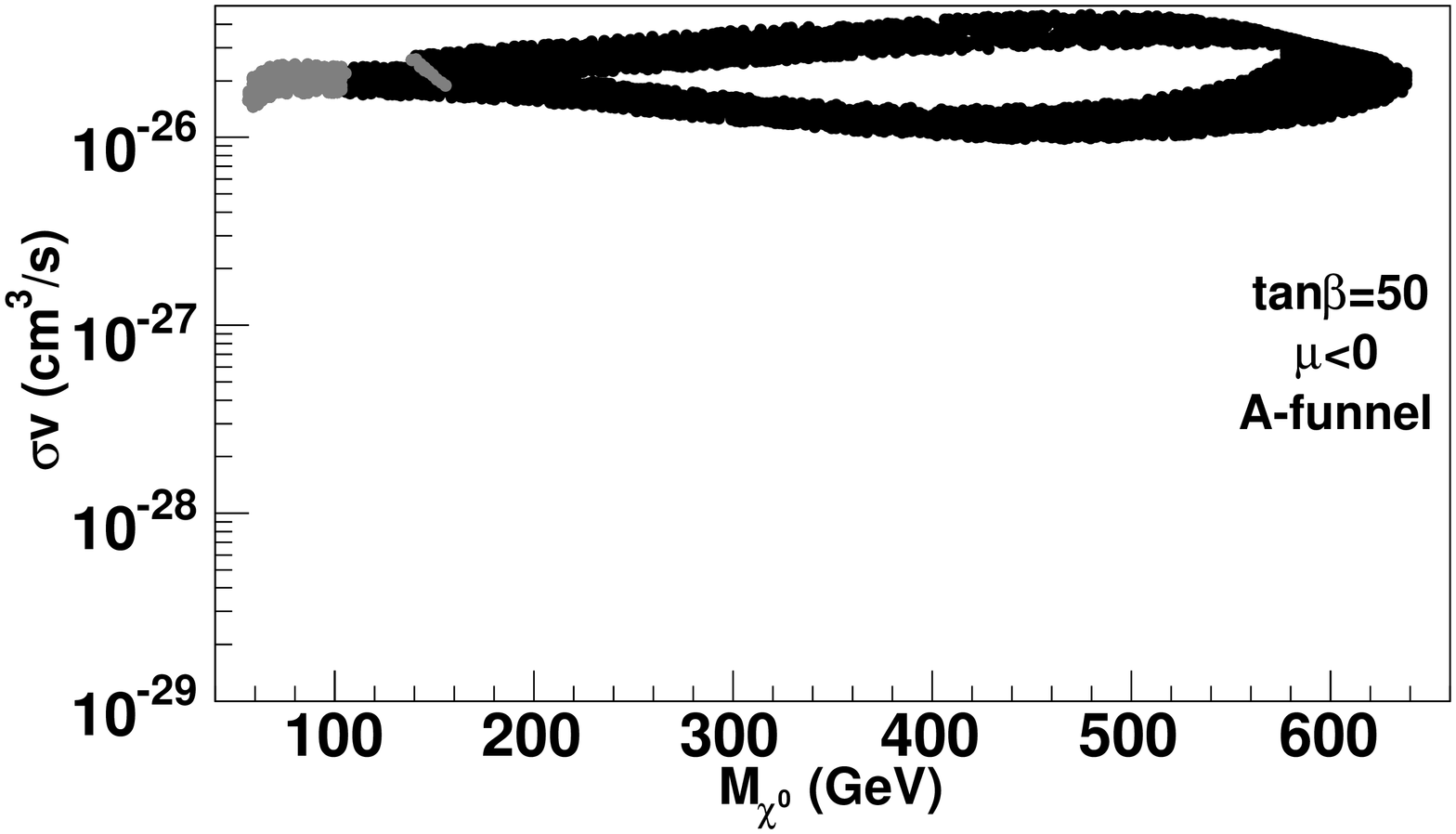}
\caption{As shown in Fig.~\ref{crosssection}, but for $\mu < 0$. See text for more details.}
\label{crosssectionnegmu}
\end{figure}

\begin{figure}[t]
\centering\leavevmode
\includegraphics[width=1.7in,angle=-90]{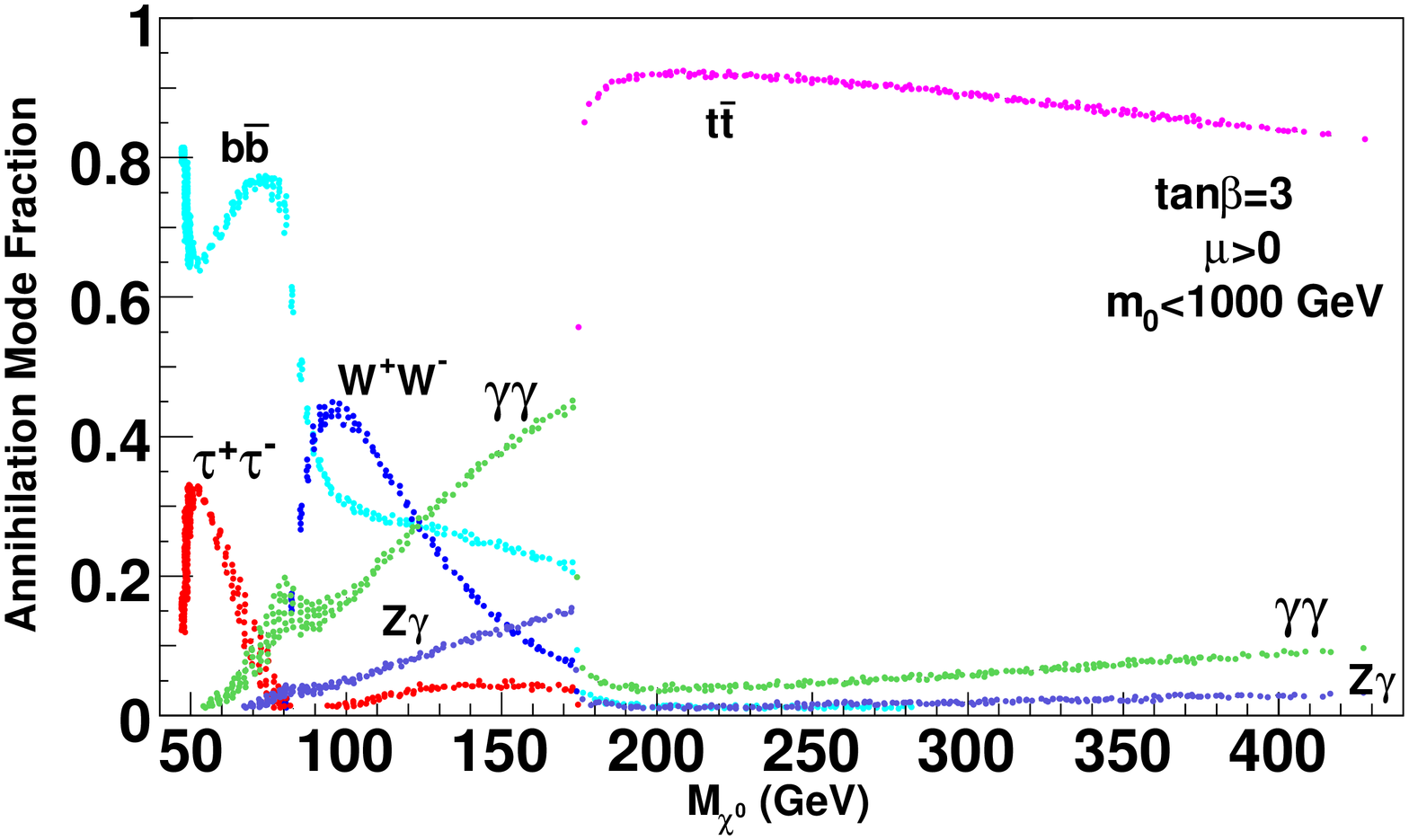}
\includegraphics[width=1.7in,angle=-90]{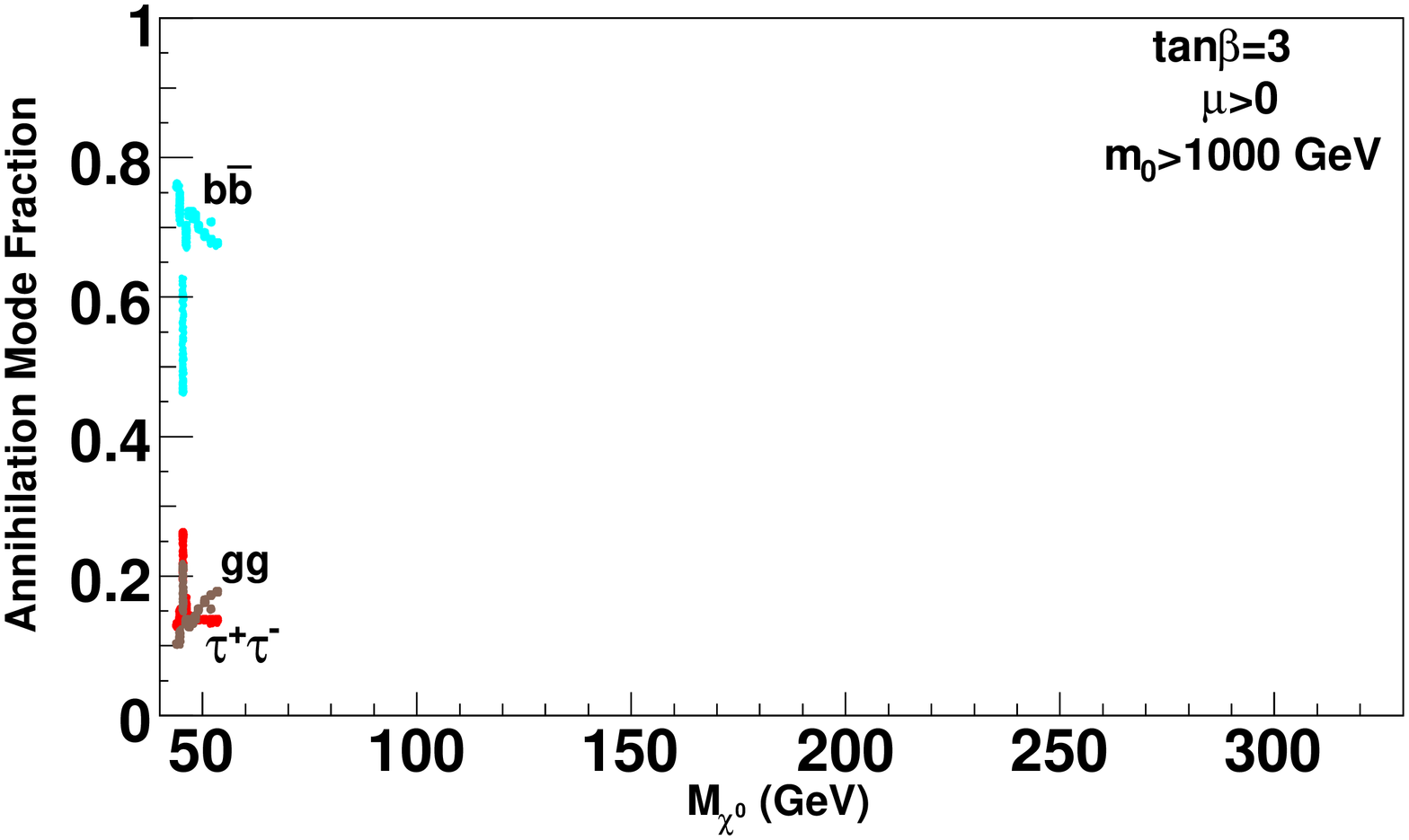}\\
\vspace{0.3cm}
\includegraphics[width=1.7in,angle=-90]{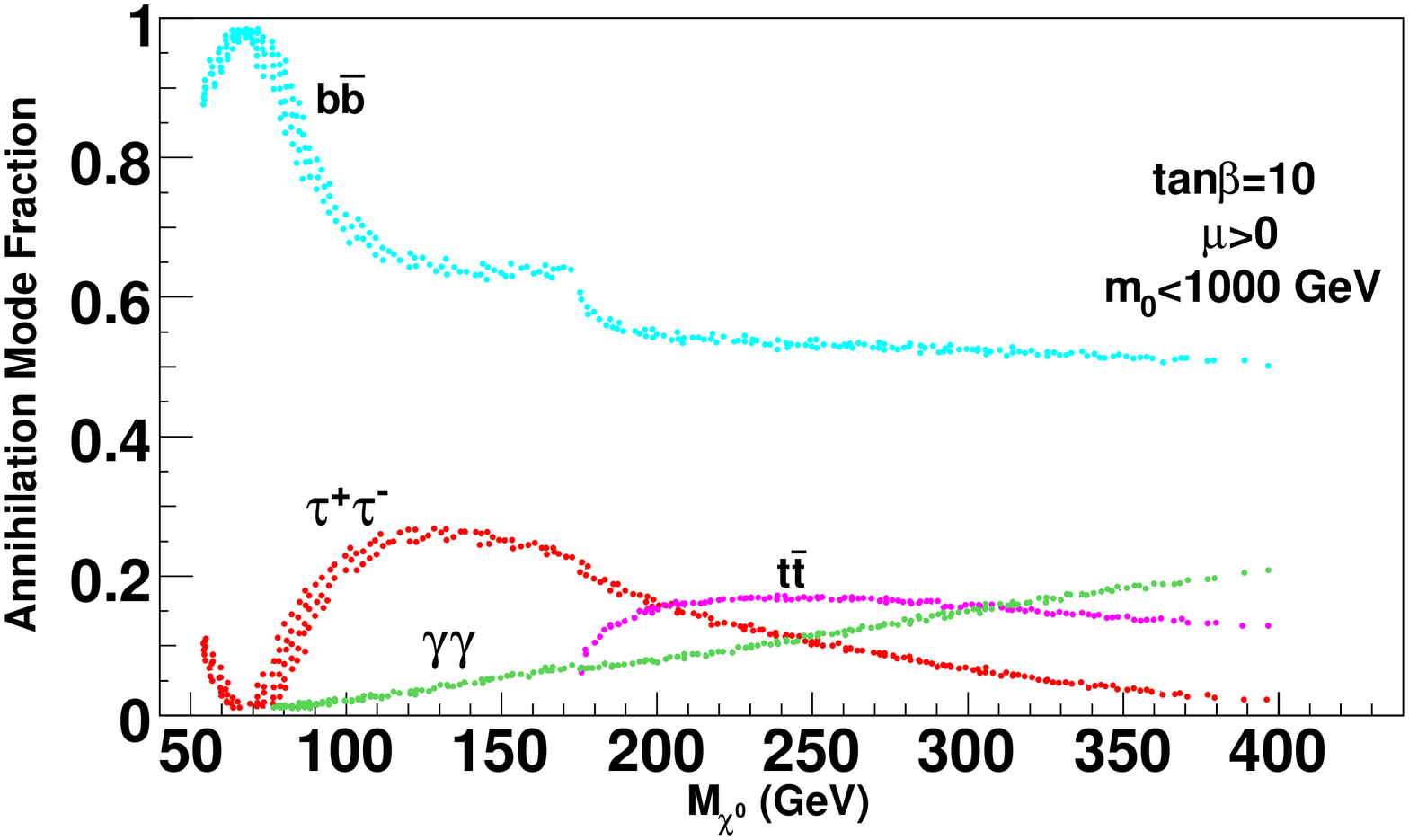}
\includegraphics[width=1.7in,angle=-90]{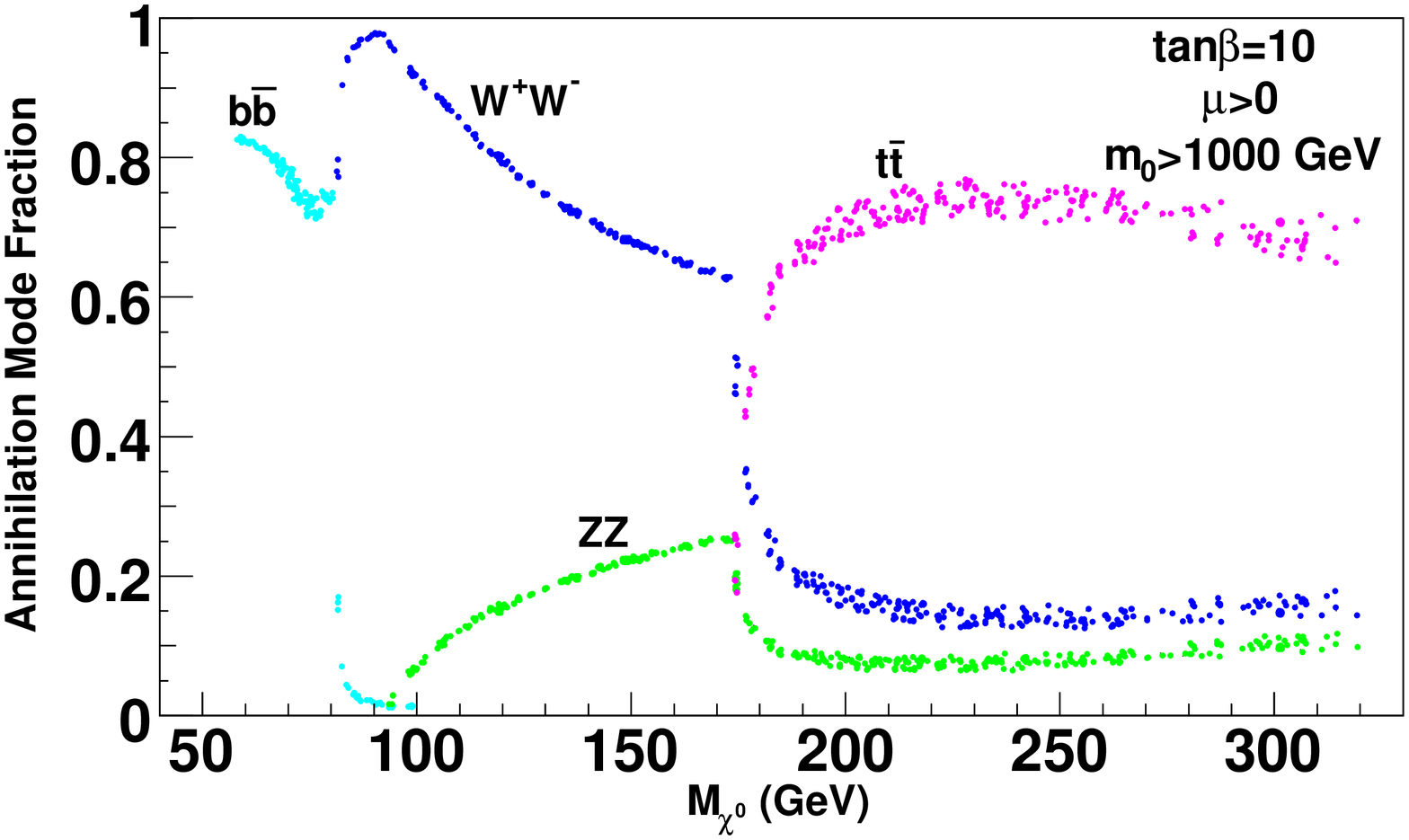}\\
\vspace{0.3cm}
\includegraphics[width=1.7in,angle=-90]{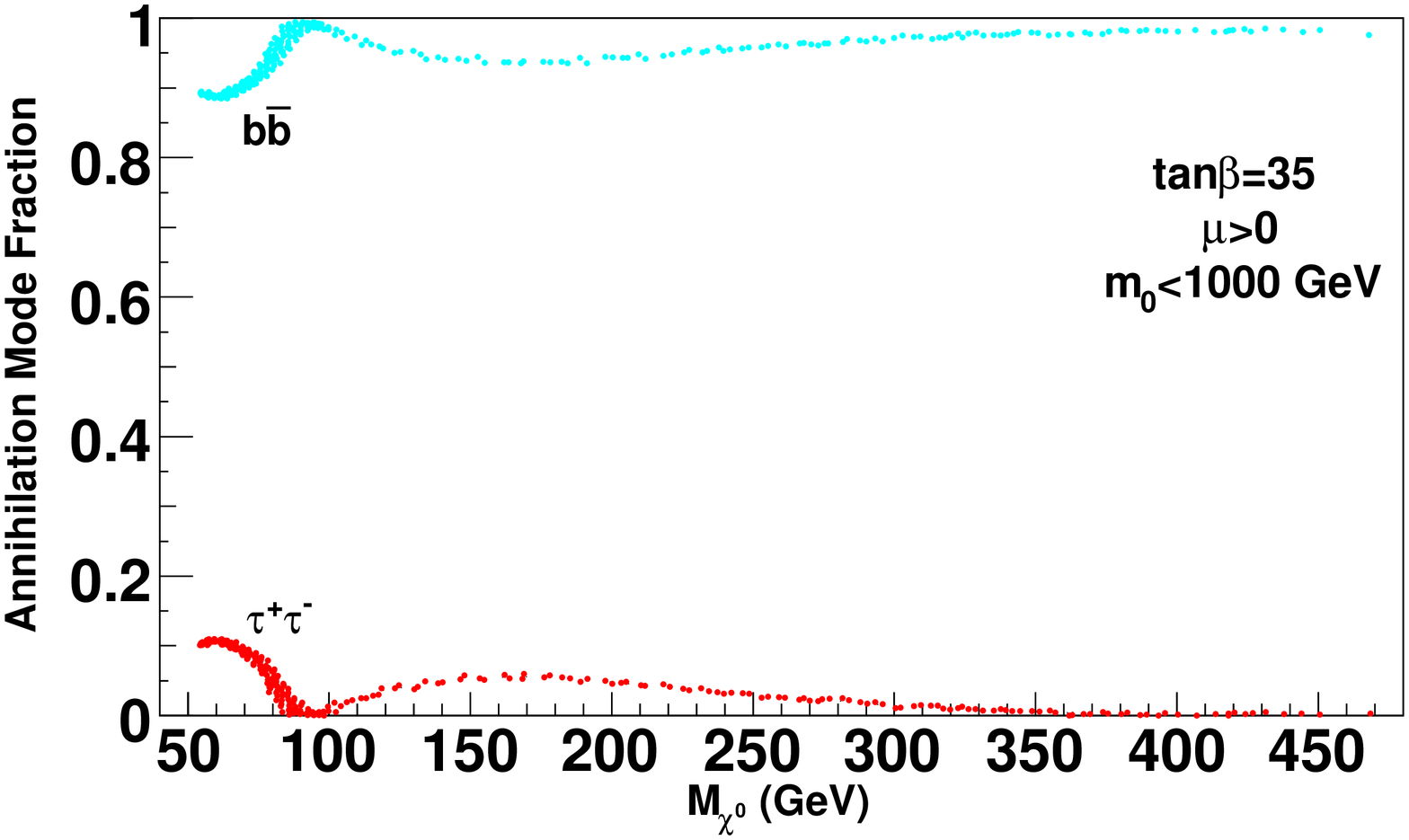}
\includegraphics[width=1.7in,angle=-90]{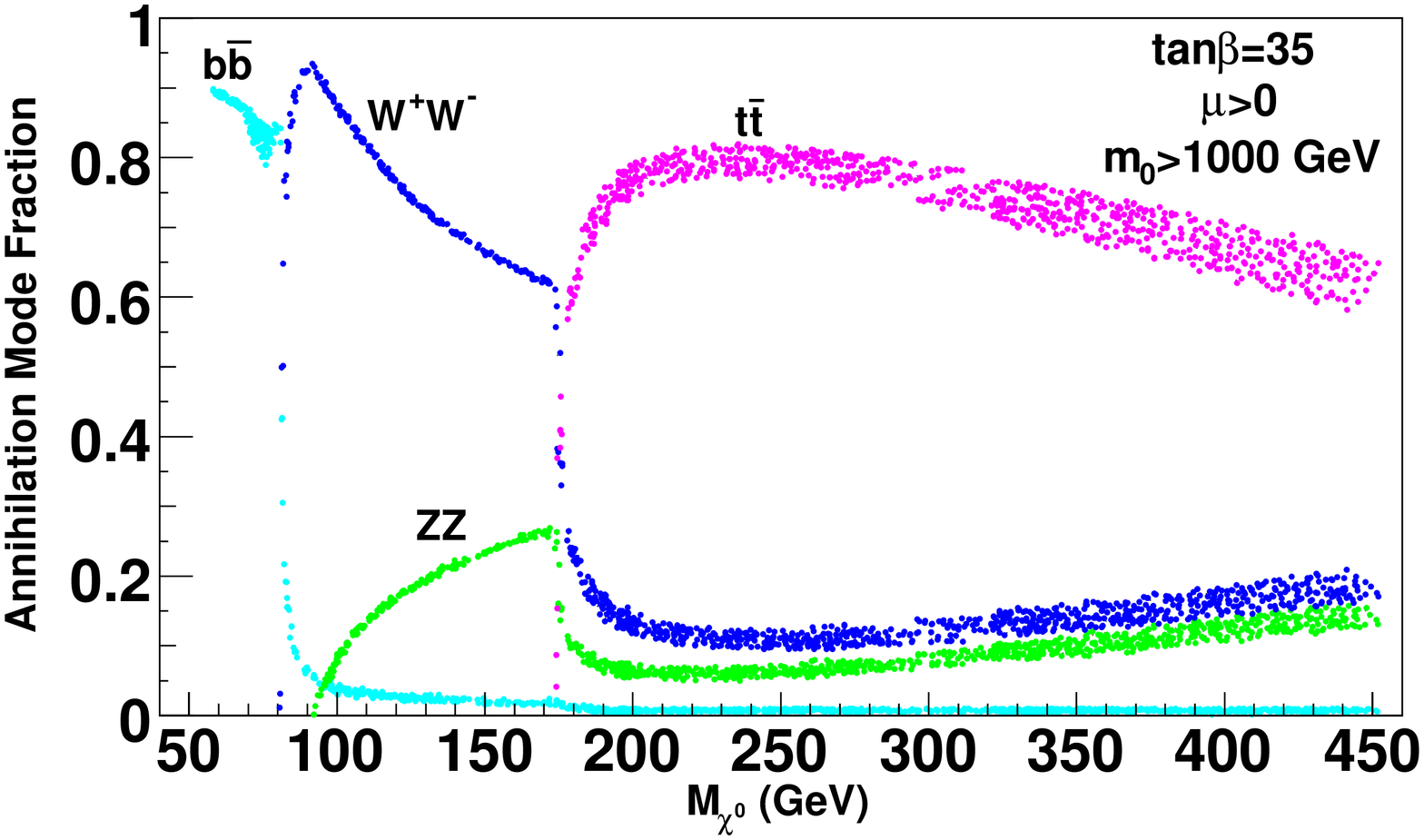}\\
\vspace{0.3cm}
\includegraphics[width=1.8in,angle=-90]{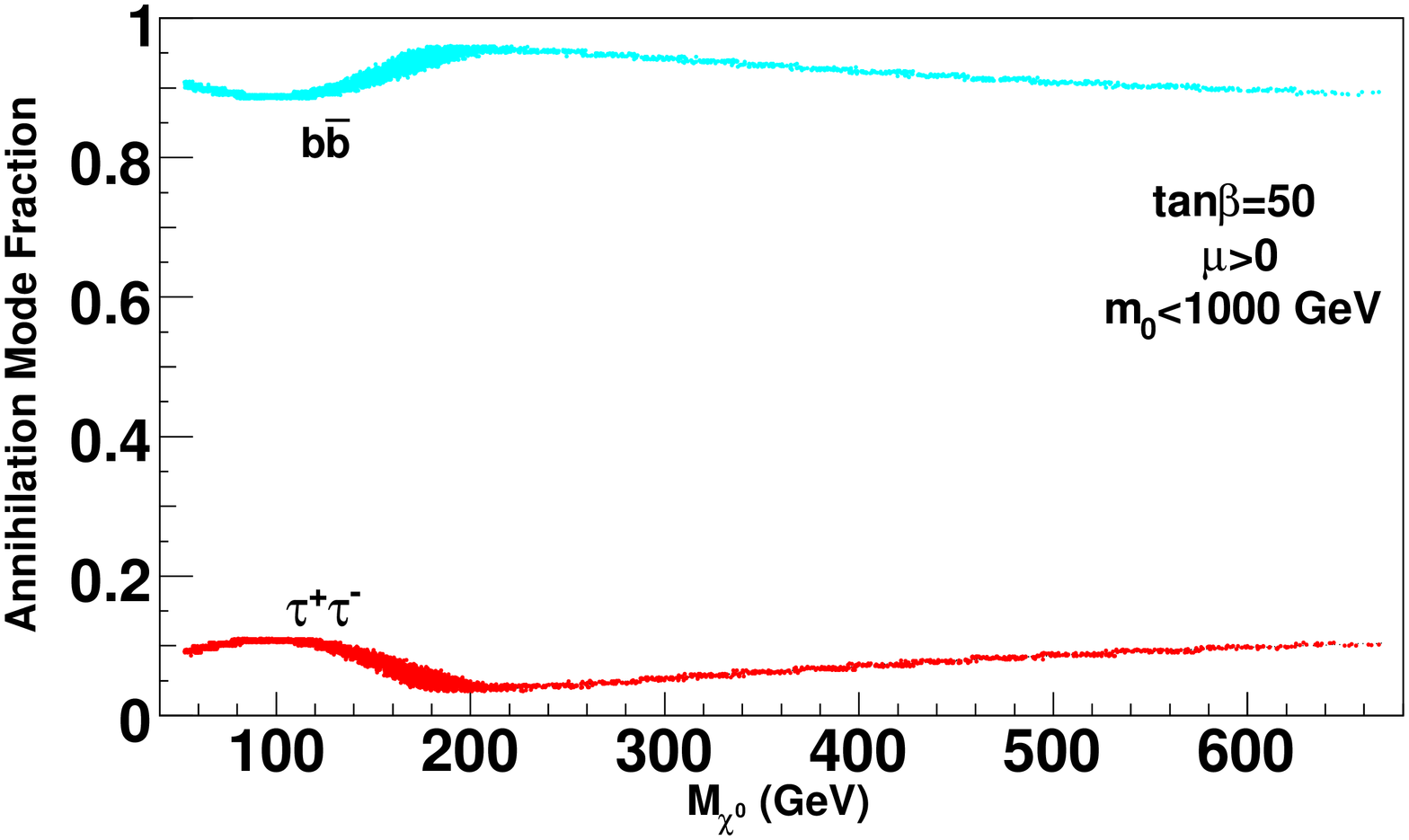}
\includegraphics[width=1.8in,angle=-90]{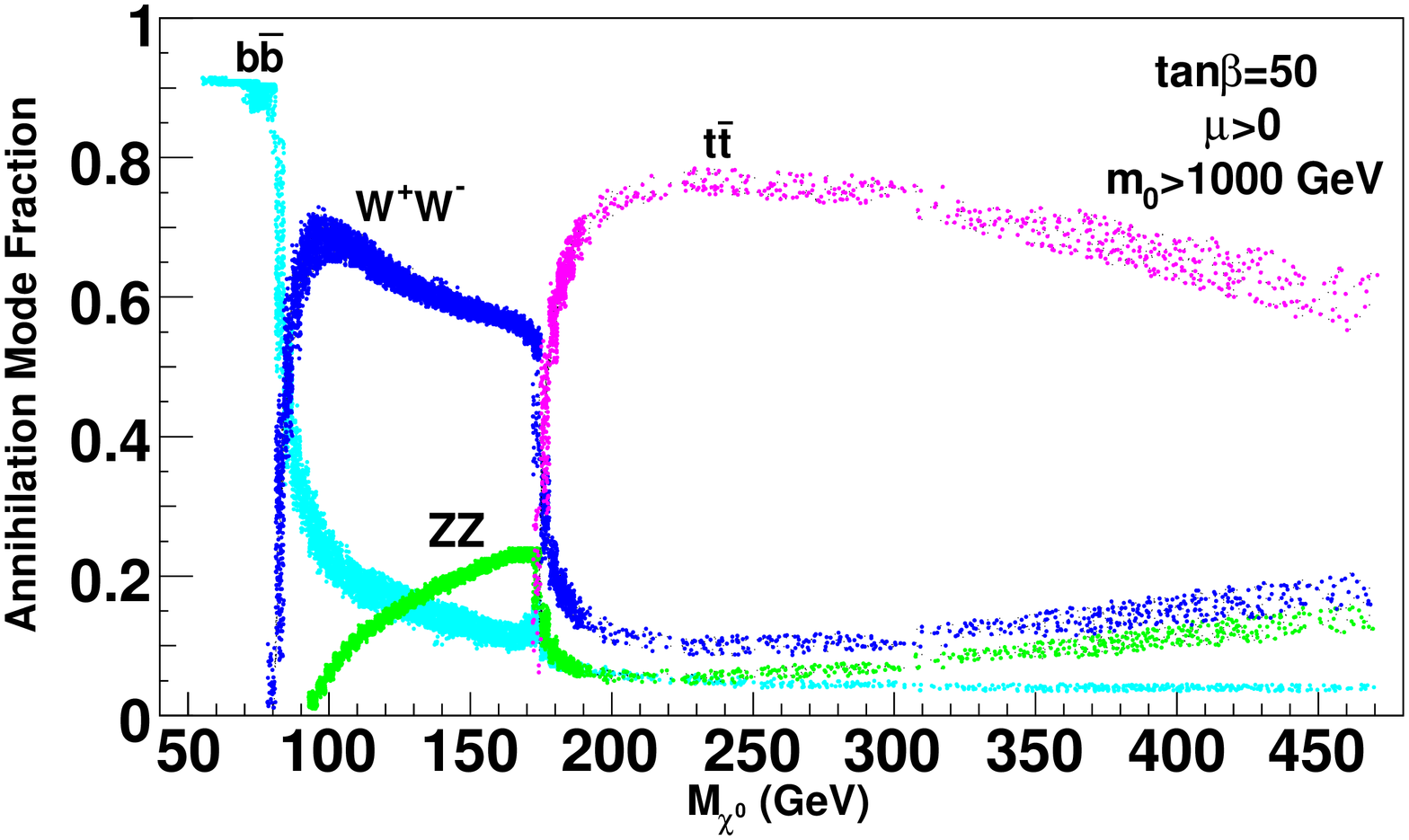}\\
\caption{The fraction of neutralino annihilations which produce various Standard Model final states. Each point shown corresponds to a point in Fig.~\ref{mzeromhalf} that predicts a neutralino dark matter abundance consistent with the measured dark matter density, and that does not violate the LEP chargino mass bound. We plot only the modes that contribute 12\% or more for at least one value of the neutralino mass in a given frame.}
\label{modes}
\end{figure}

\begin{figure}[t]
\centering\leavevmode
\includegraphics[width=1.8in,angle=-90]{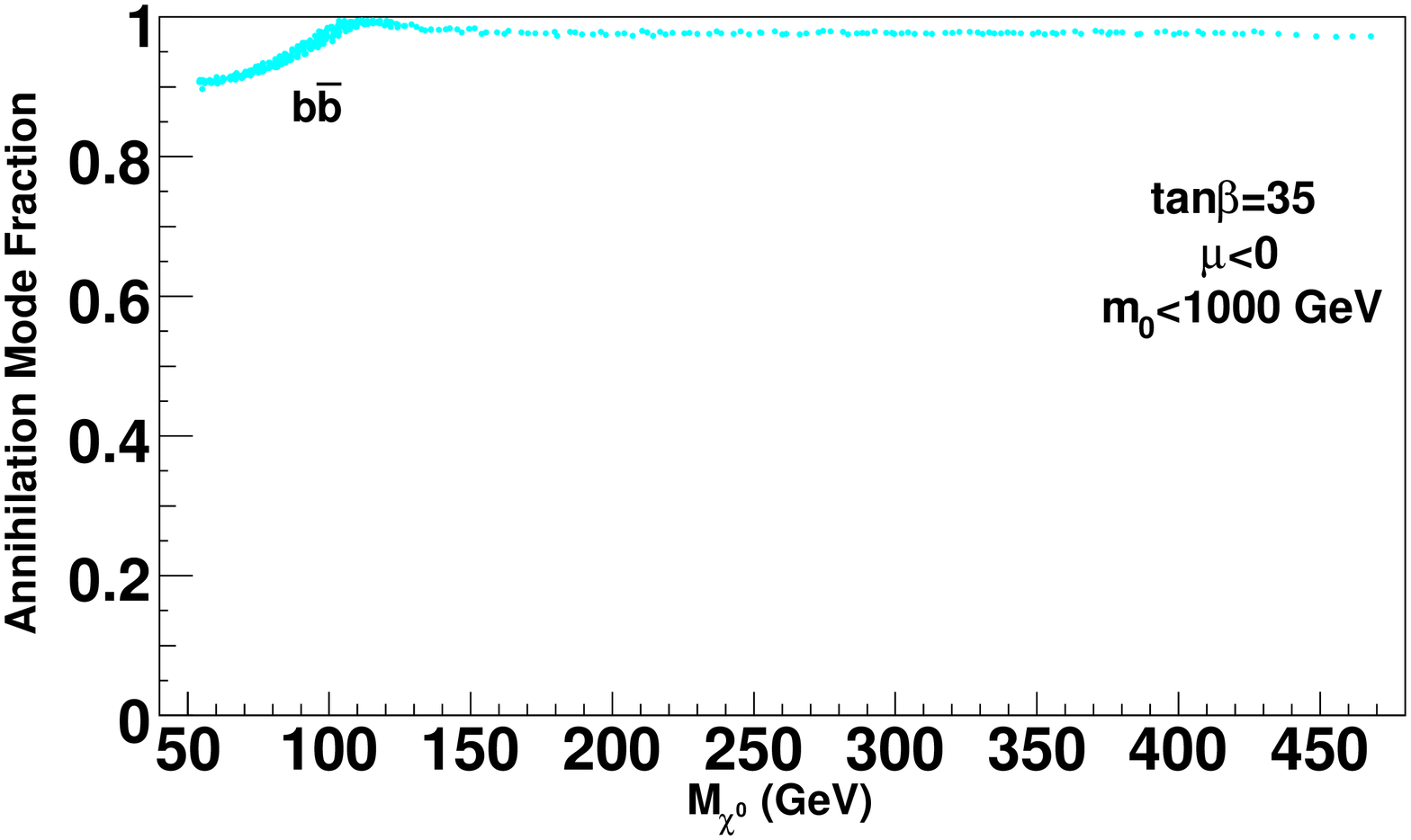}
\includegraphics[width=1.8in,angle=-90]{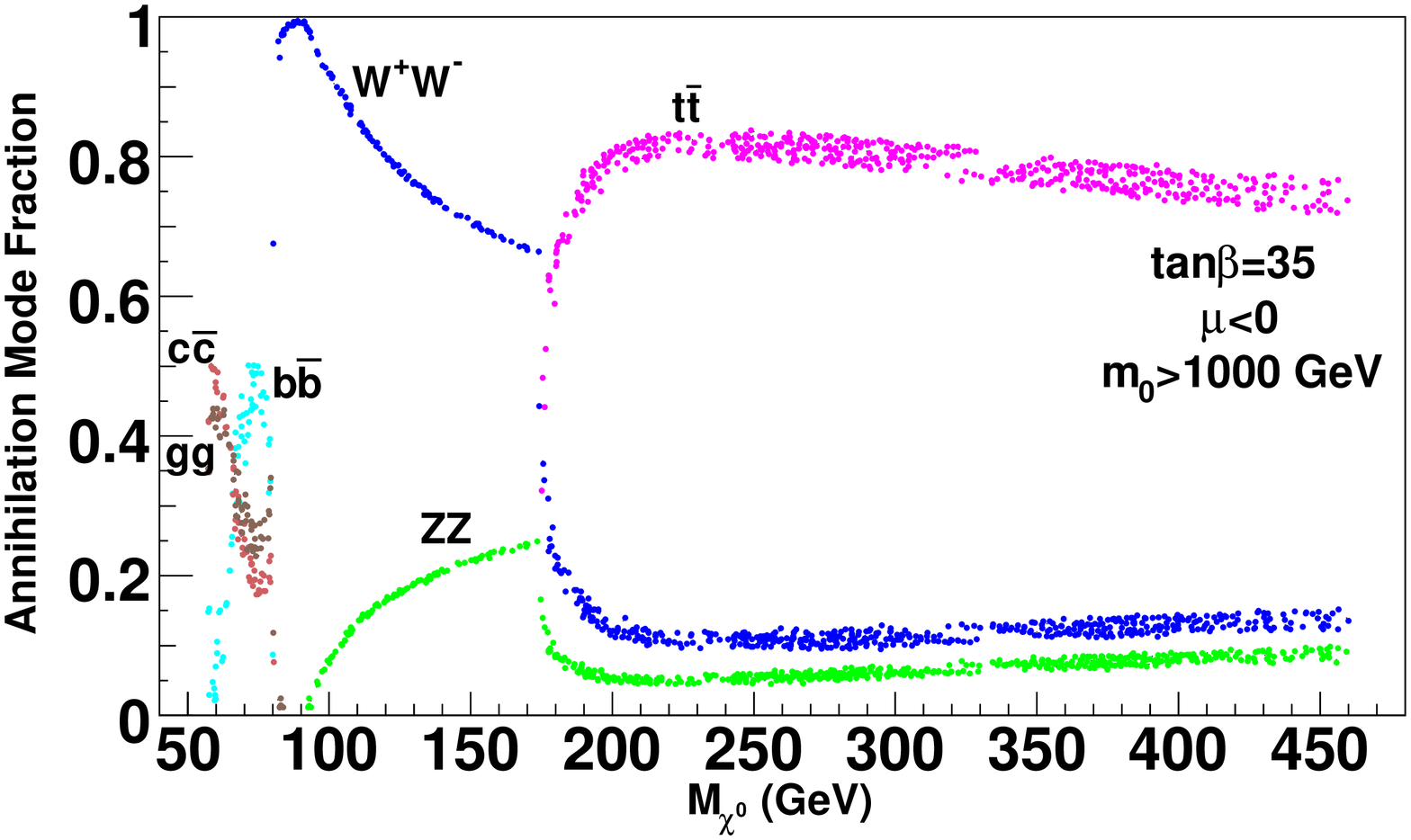}\\
\vspace{0.5cm}
\includegraphics[width=1.8in,angle=-90]{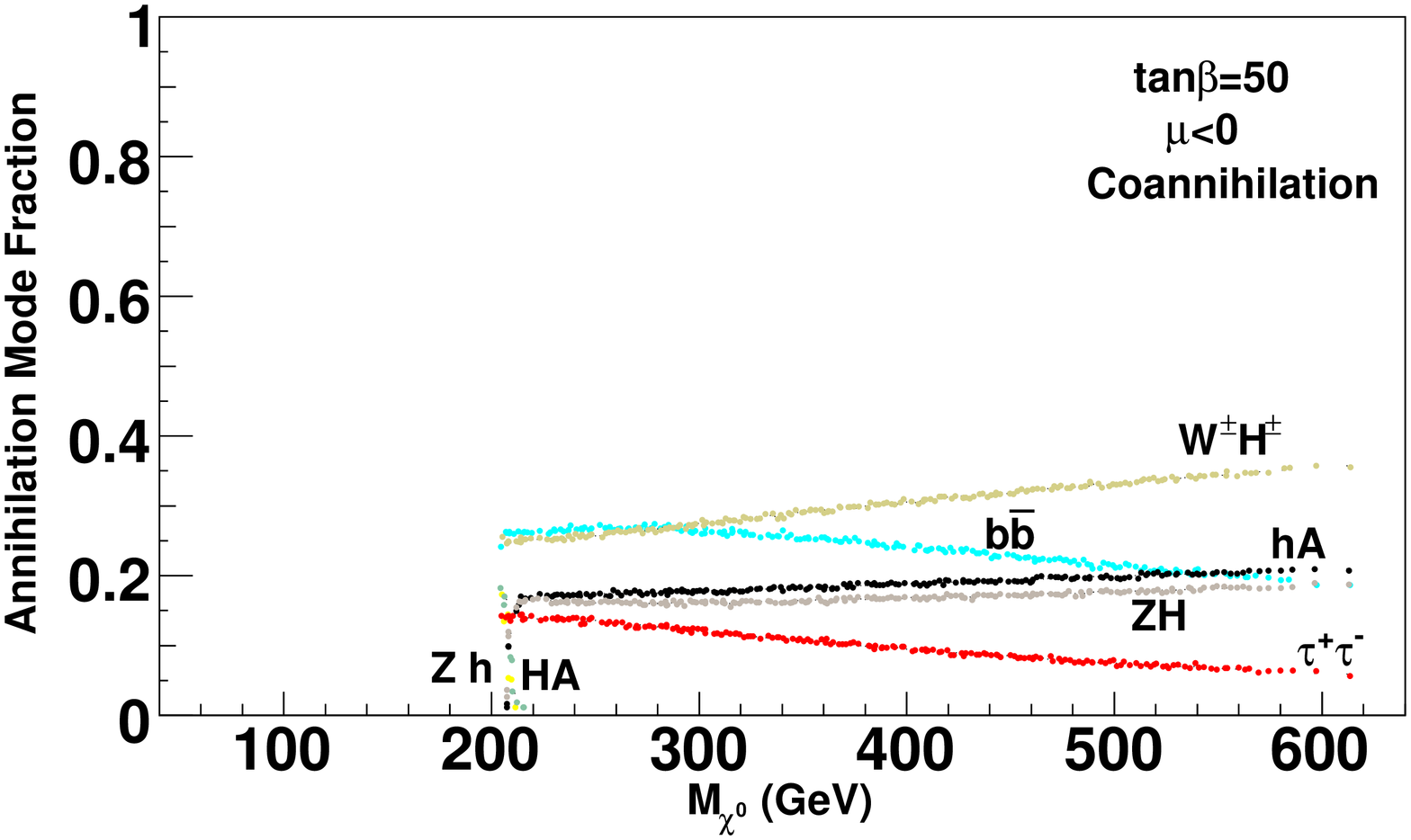}
\includegraphics[width=1.8in,angle=-90]{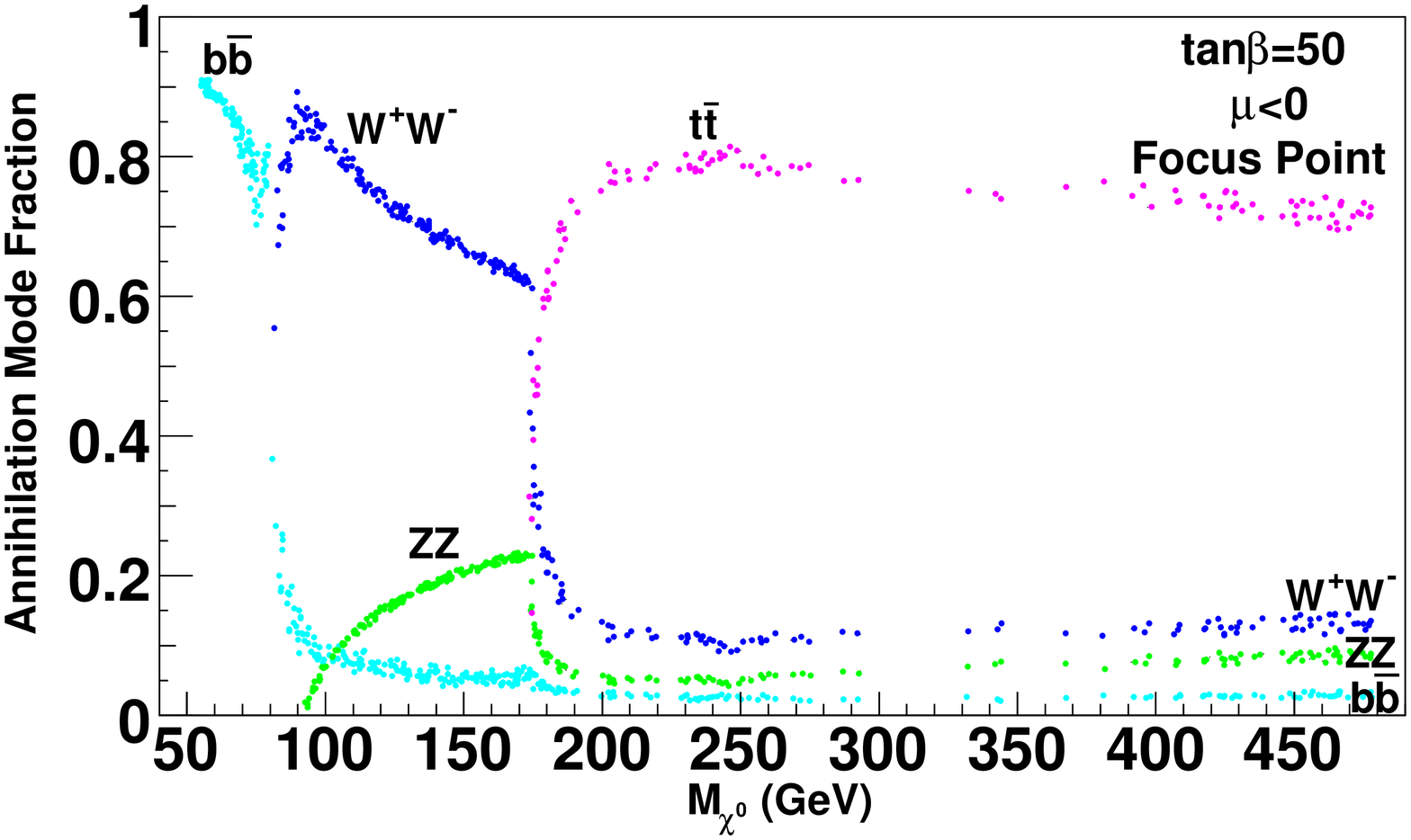}\\
\vspace{0.5cm}
\includegraphics[width=1.8in,angle=-90]{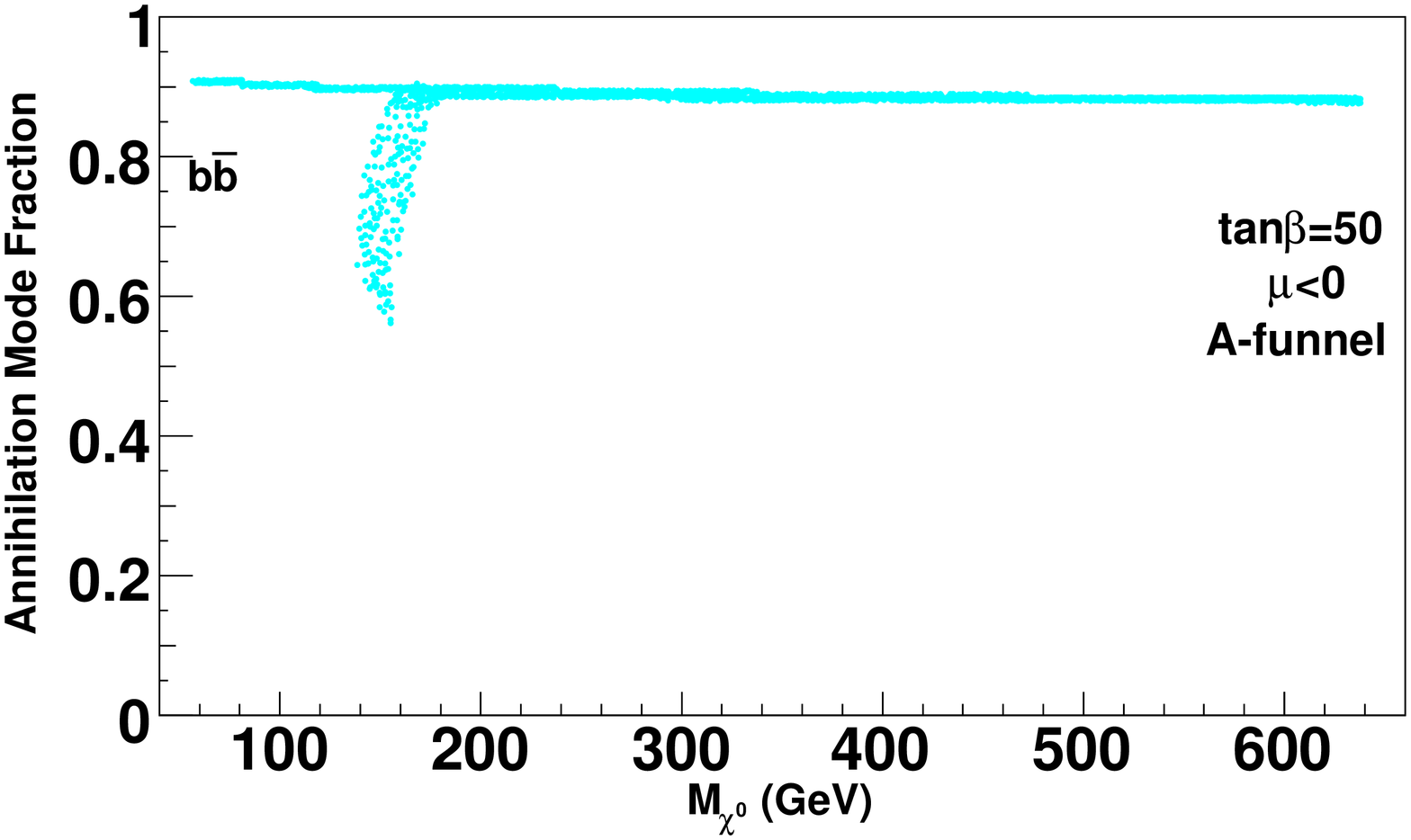}
\caption{As shown in Fig.~\ref{modes}, but for $\mu < 0$.}
\label{modesnegmu}
\end{figure}

Focusing on the regions of parameter space which lead to an acceptable abundance of neutralino dark matter, we plot the neutralino annihilation cross section in Figs.~\ref{crosssection} and~\ref{crosssectionnegmu}, and the fraction of neutralino annihilations which go to various Standard Model final states in Figs.~\ref{modes} and~\ref{modesnegmu}, each in the low-velocity limit, as is appropriate for the WMAP Haze and indirect detection in general.

In Figs.~\ref{crosssection} and~\ref{crosssectionnegmu}, we see that, although the parameter space corresponding to the stau-coannihilation region predicts annihilation cross sections well below the range preferred in Fig.~\ref{sigma}, much of the focus point, bulk (with large $\tan \beta$), and $A$-funnel regions naturally predict a cross section very close to that required to produce the WMAP Haze. 

In the left frames of Figs.~\ref{modes} and~\ref{modesnegmu} ($m_0 < 1000$ GeV, corresponding to the stau-coannihilation and bulk regions), we see that most of the annihilations produce $b\bar{b}$ with a smaller contribution from $\tau^+ \tau^-$ and in some cases $t \bar{t}$. In the right frames ($m_0 > 1000$ GeV, corresponding to the focus point region), we see that neutralino annihilations produce mostly heavy quarks, gauge bosons, or a combination thereof. In the $A$-funnel region, neutralino annihilations proceed largely to $b\bar{b}$.

\newpage

\newpage
.

\newpage
.

\newpage
.

\newpage

.

\newpage
.

\section{Regions of Supersymmetric Parameter Space Consistent With Being the Source of the WMAP Haze}

\begin{figure}[t]
\centering\leavevmode
\includegraphics[width=1.6in,angle=-90]{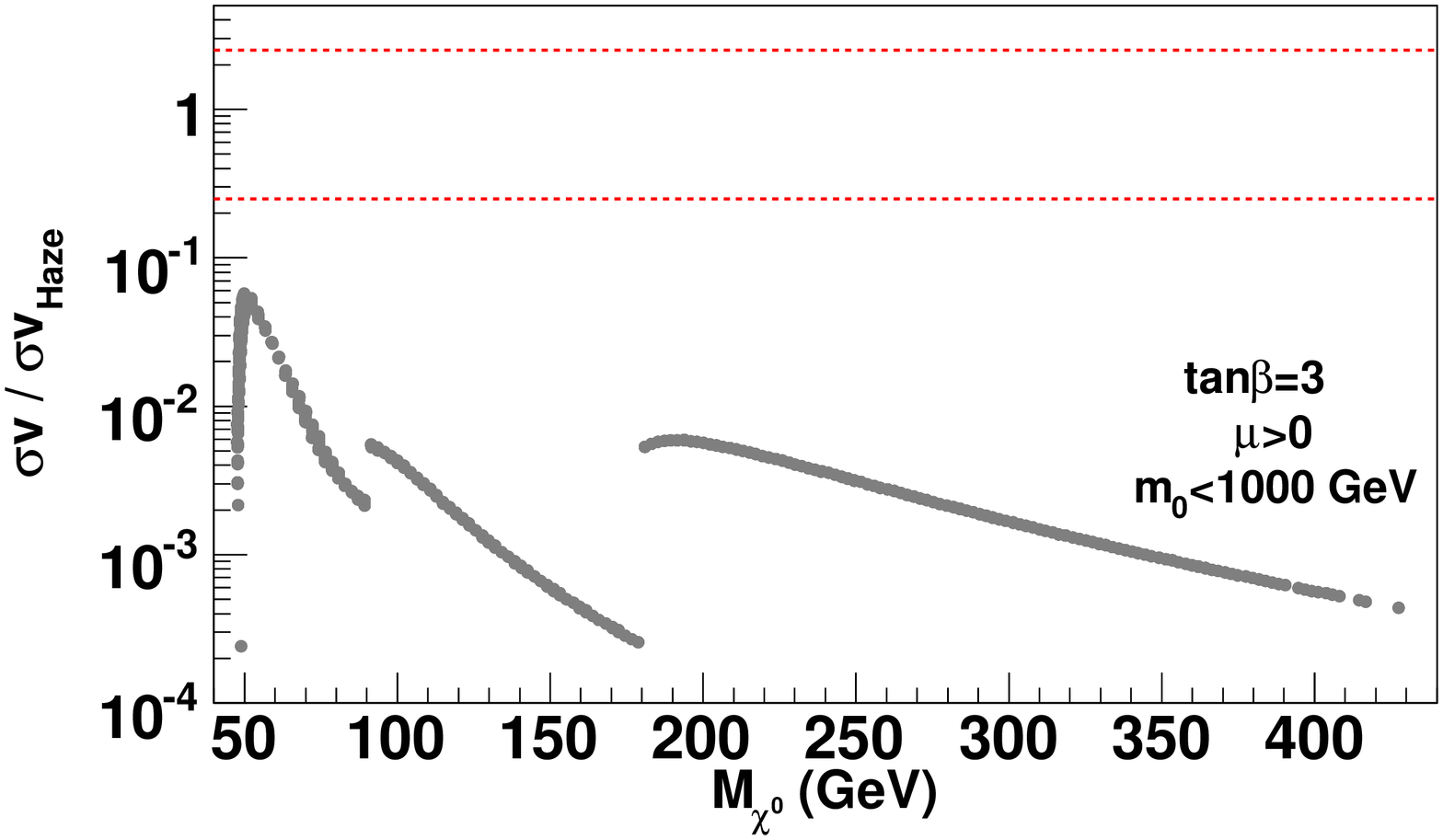}
\includegraphics[width=1.6in,angle=-90]{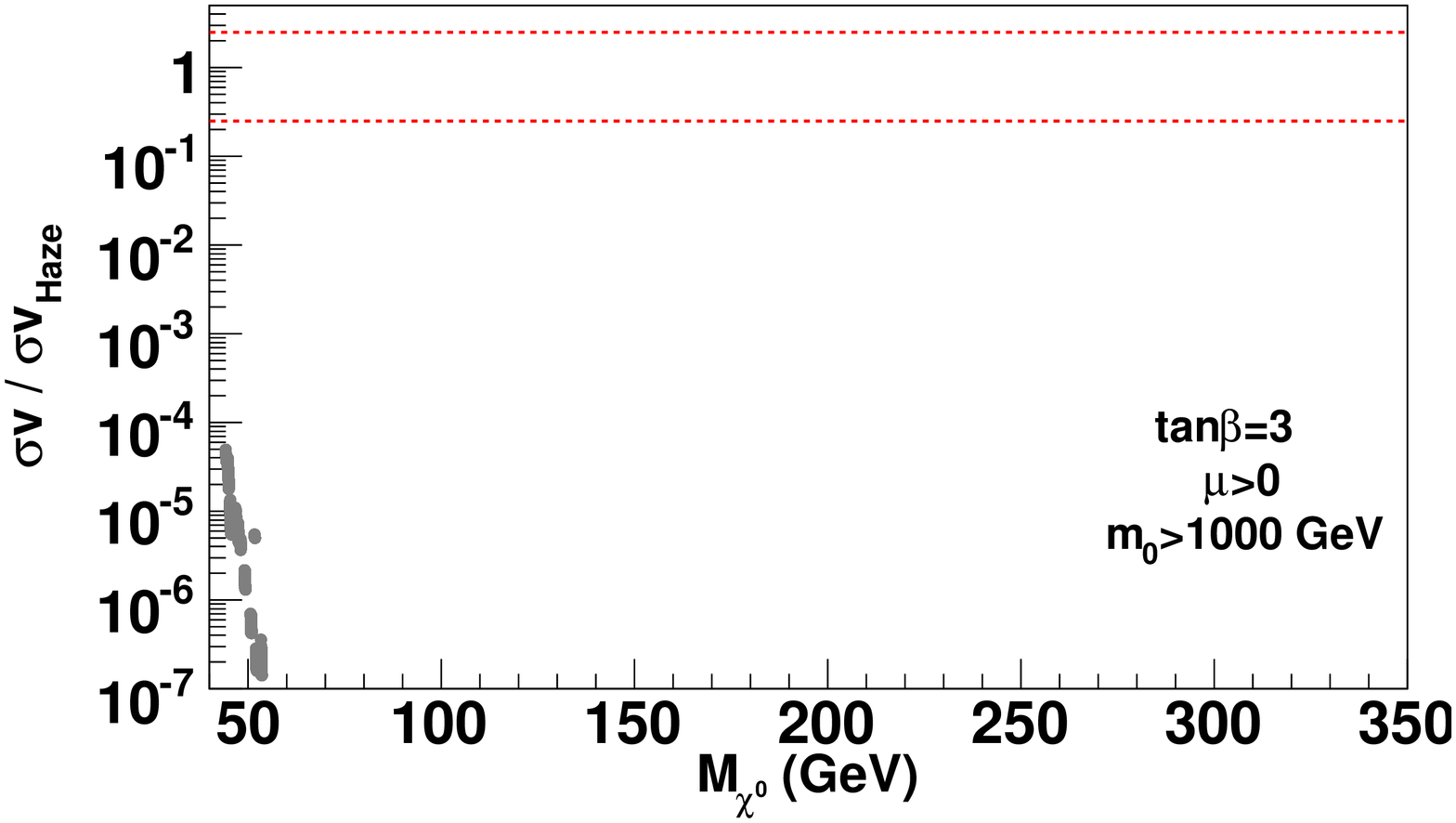}\\
\vspace{0.3cm}
\includegraphics[width=1.6in,angle=-90]{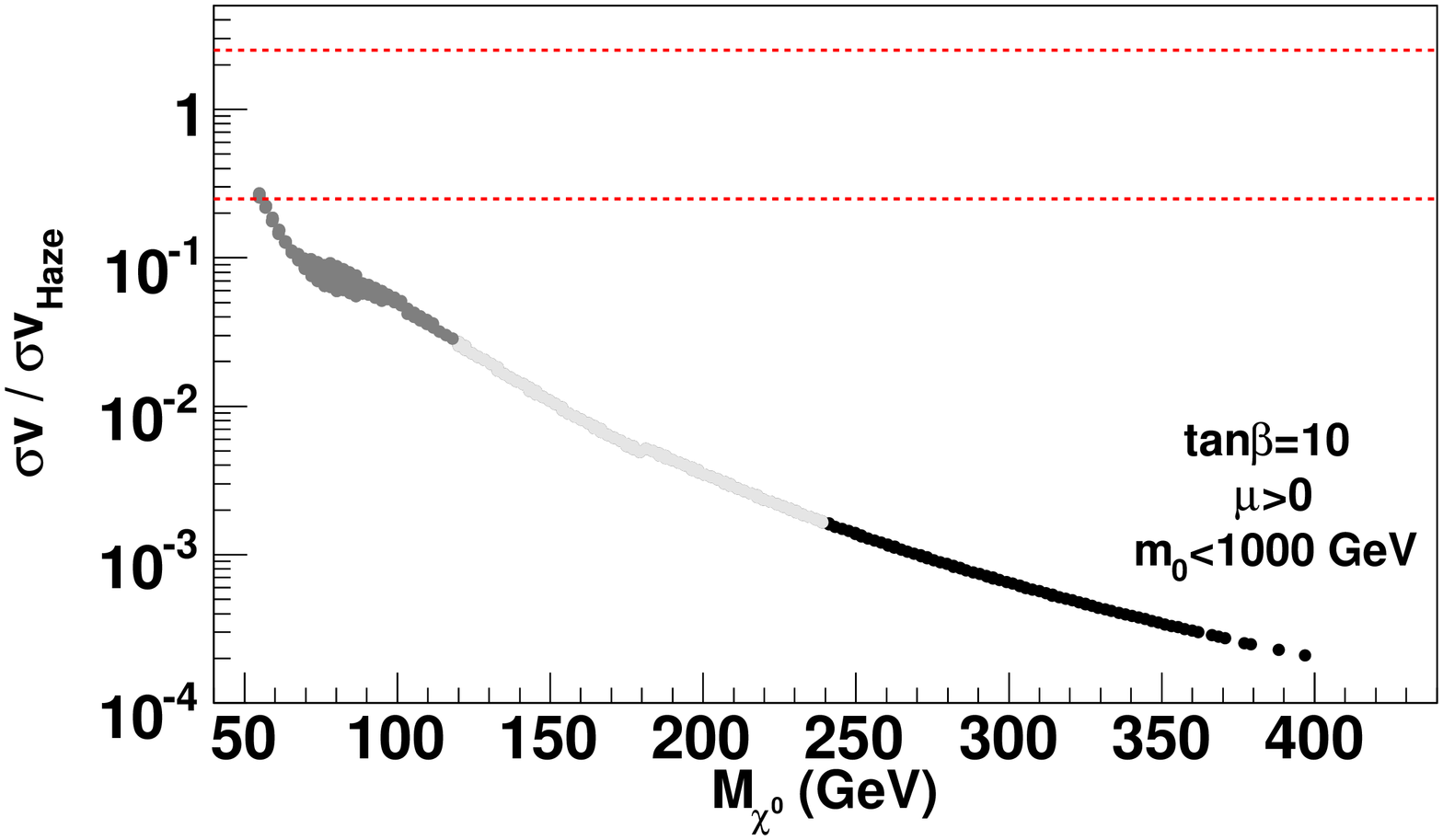}
\includegraphics[width=1.6in,angle=-90]{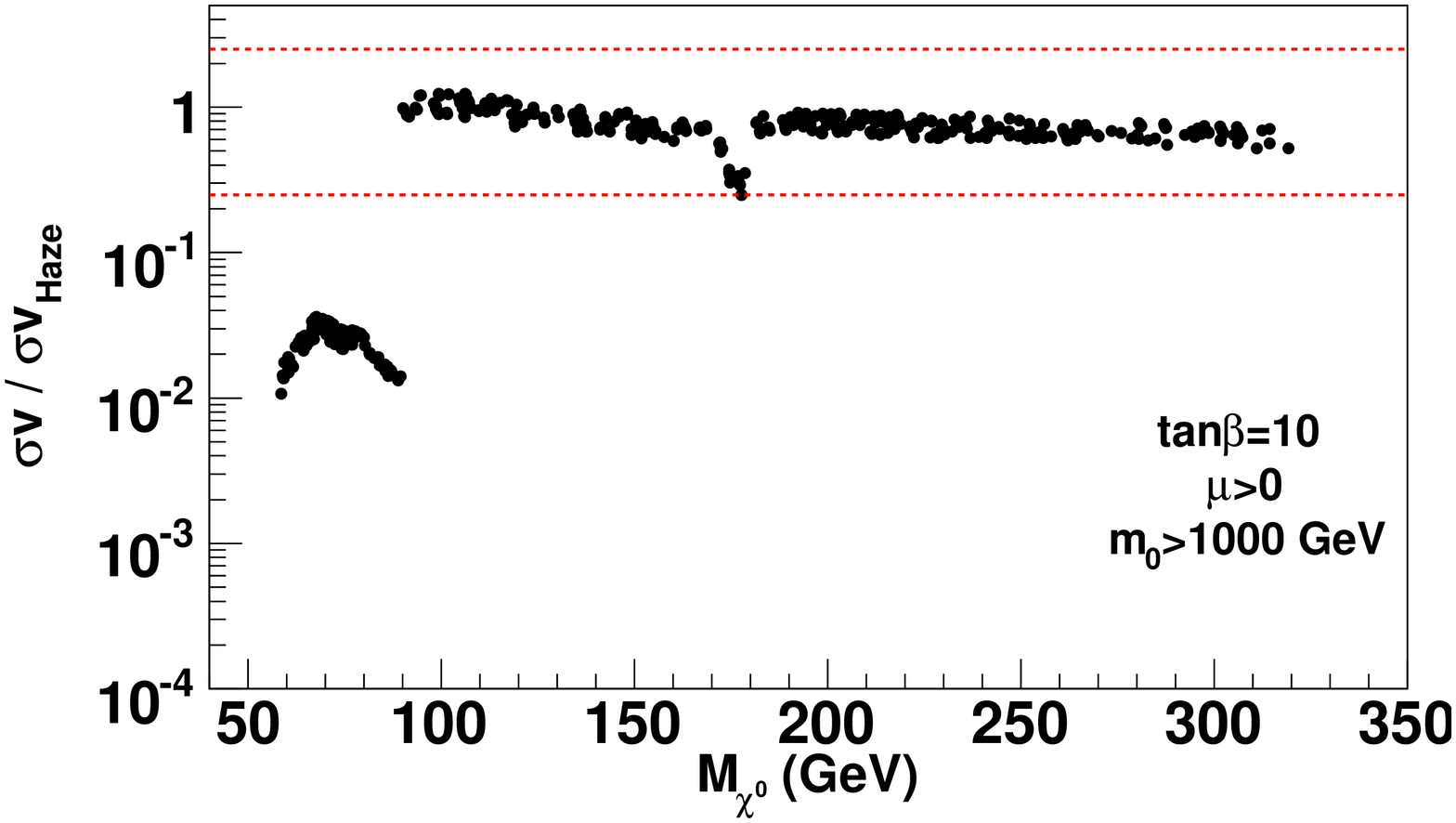}\\
\vspace{0.3cm}
\includegraphics[width=1.6in,angle=-90]{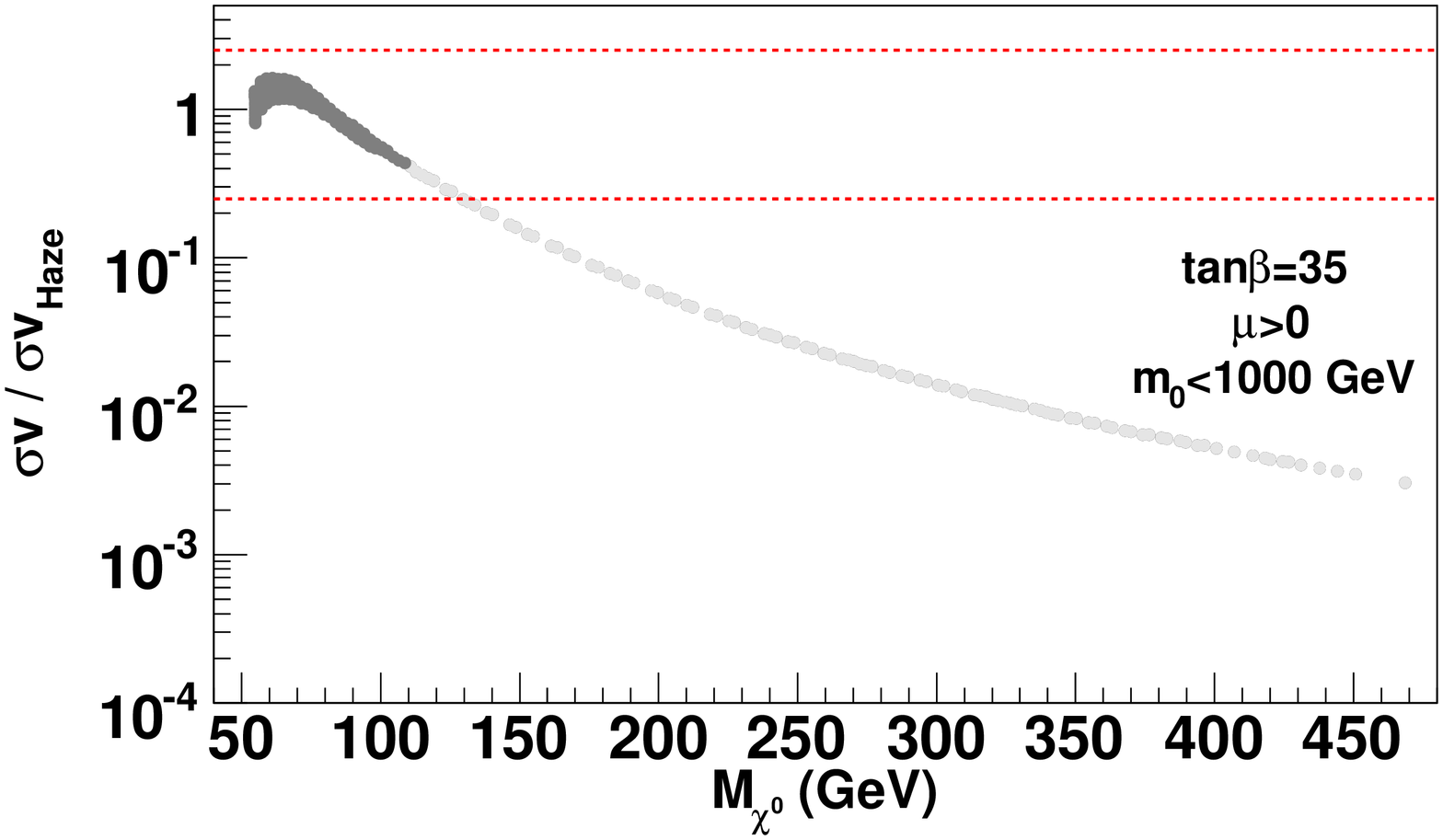}
\includegraphics[width=1.6in,angle=-90]{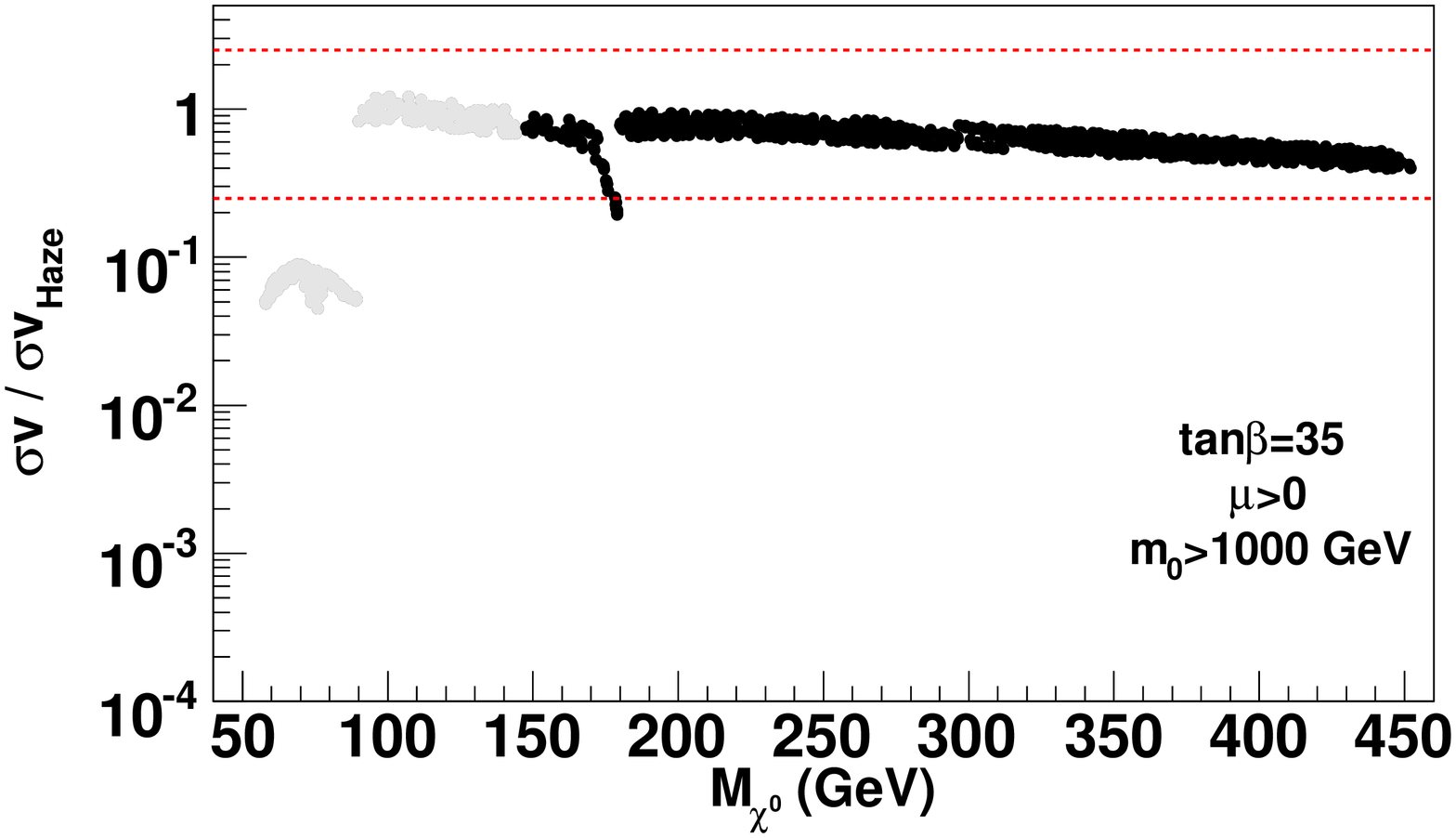}\\
\vspace{0.3cm}
\includegraphics[width=1.6in,angle=-90]{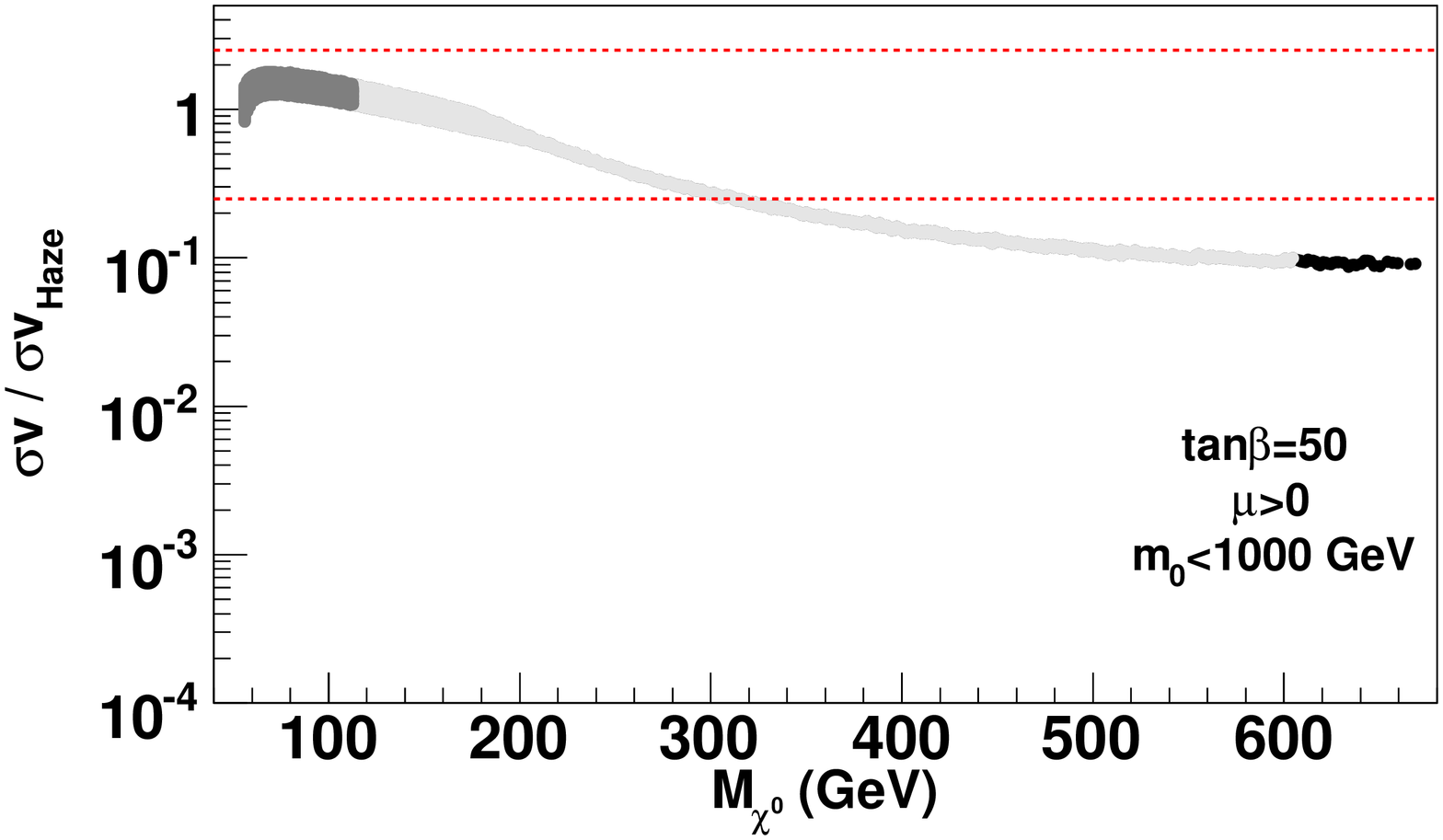}
\includegraphics[width=1.6in,angle=-90]{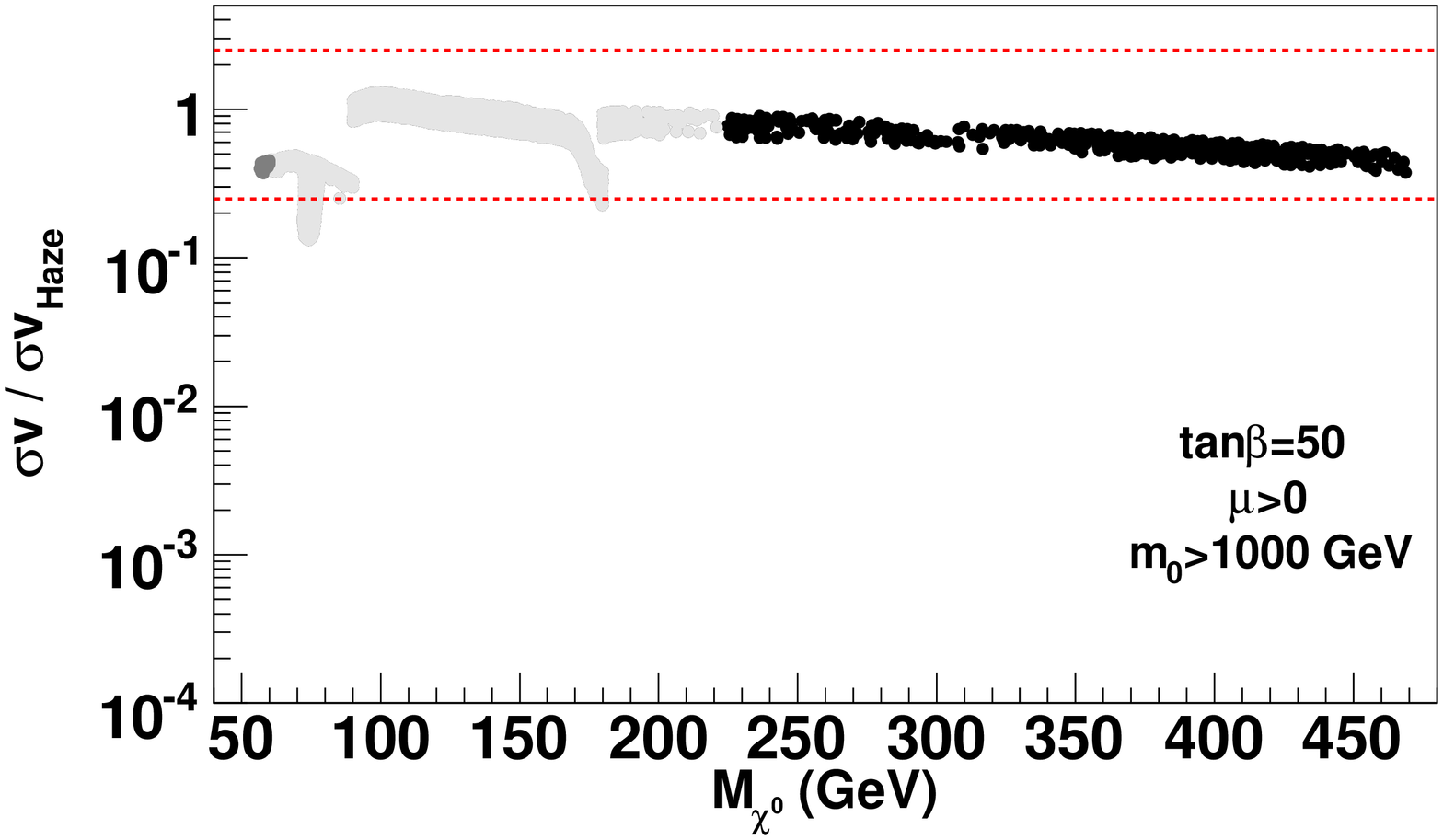}\\
\caption{The ratio of the low-velocity neutralino annihilation cross section to the value required to generate the observed intensity of the WMAP Haze. Each point shown corresponds to a point in Fig.~\ref{mzeromhalf} that predicts a neutralino dark matter abundance consistent with the measured dark matter density, and that does not violate the LEP chargino mass bound. The dark gray regions are disfavored by the LEP Higgs mass bound. Of the remaining points, the light gray (black) regions are preferred (disfavored) by the measurements of the muon's magnetic moment. The ratios shown here were calculated using a ratio of magnetic field energy density to magnetic field plus radiation field density of $U_B/(U_B+U_{\rm rad})$=0.25. The horizontal lines shown correspond to the range of cross sections that lead to the observed intensity for $U_B/(U_B+U_{\rm rad})$ within the range of 0.1-1.0. See text for more details.}
\label{ratio}
\end{figure}

\begin{figure}[t]
\centering\leavevmode
\includegraphics[width=1.8in,angle=-90]{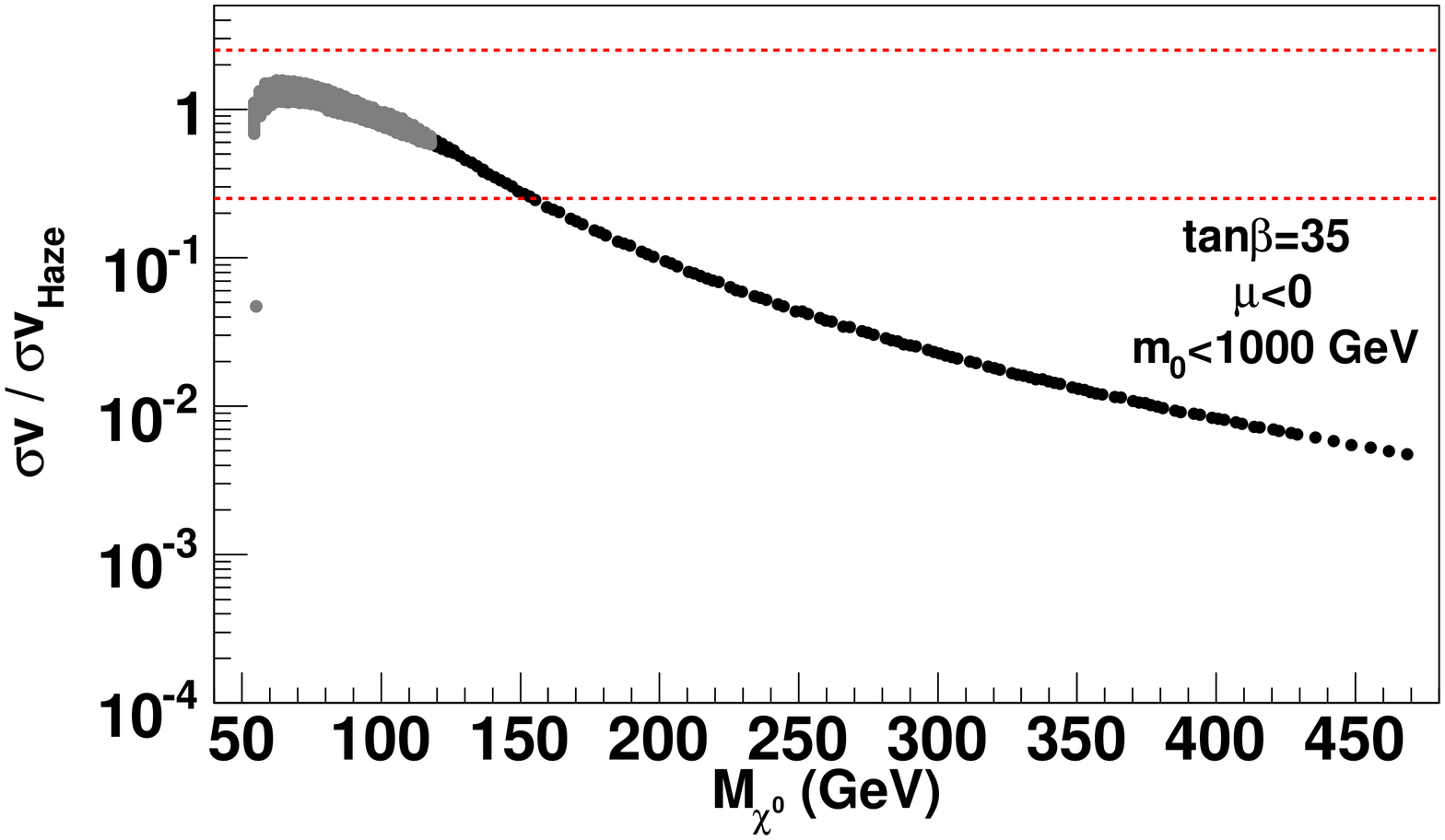}
\includegraphics[width=1.8in,angle=-90]{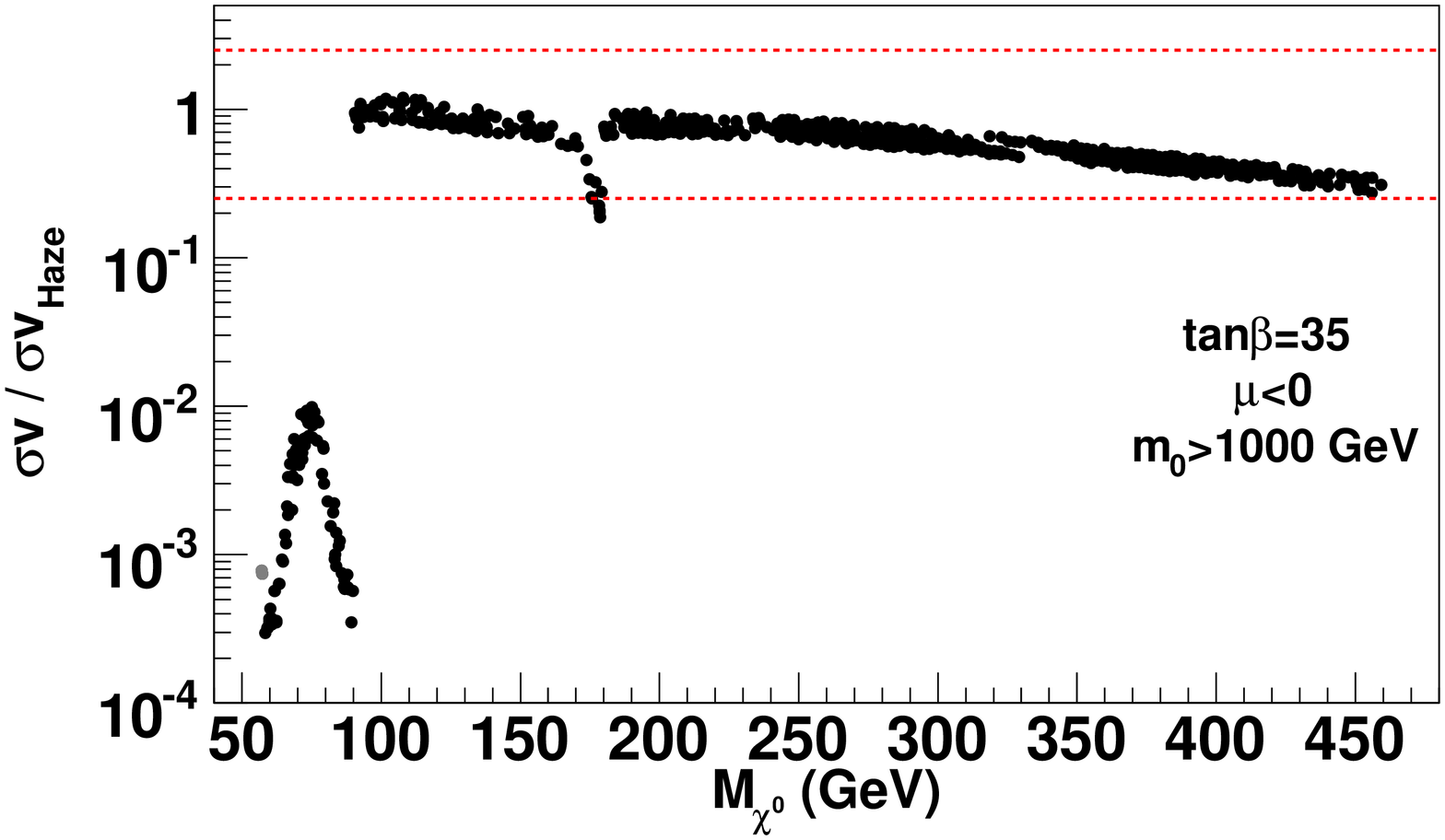}\\
\vspace{0.5cm}
\includegraphics[width=1.8in,angle=-90]{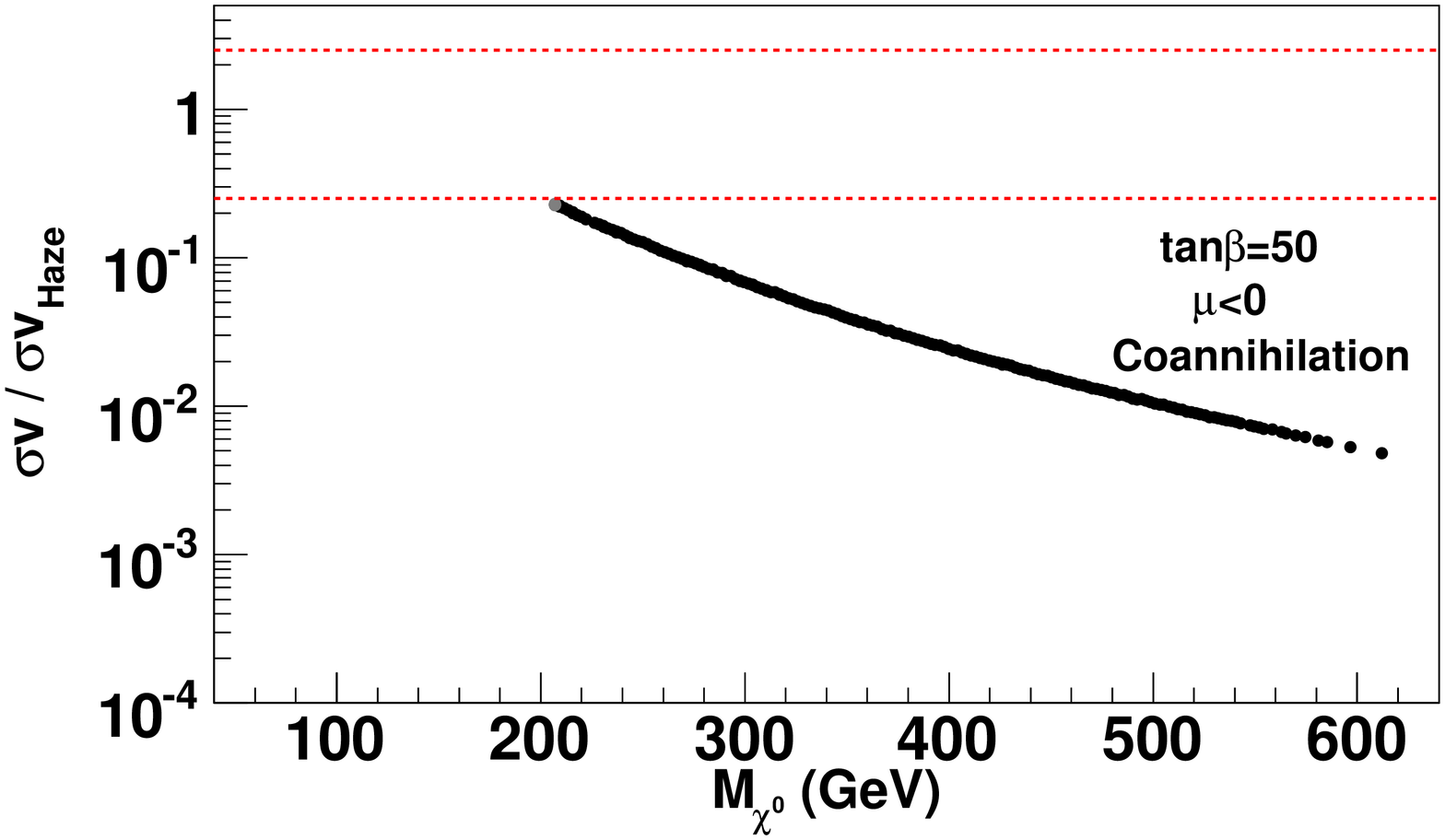}
\includegraphics[width=1.8in,angle=-90]{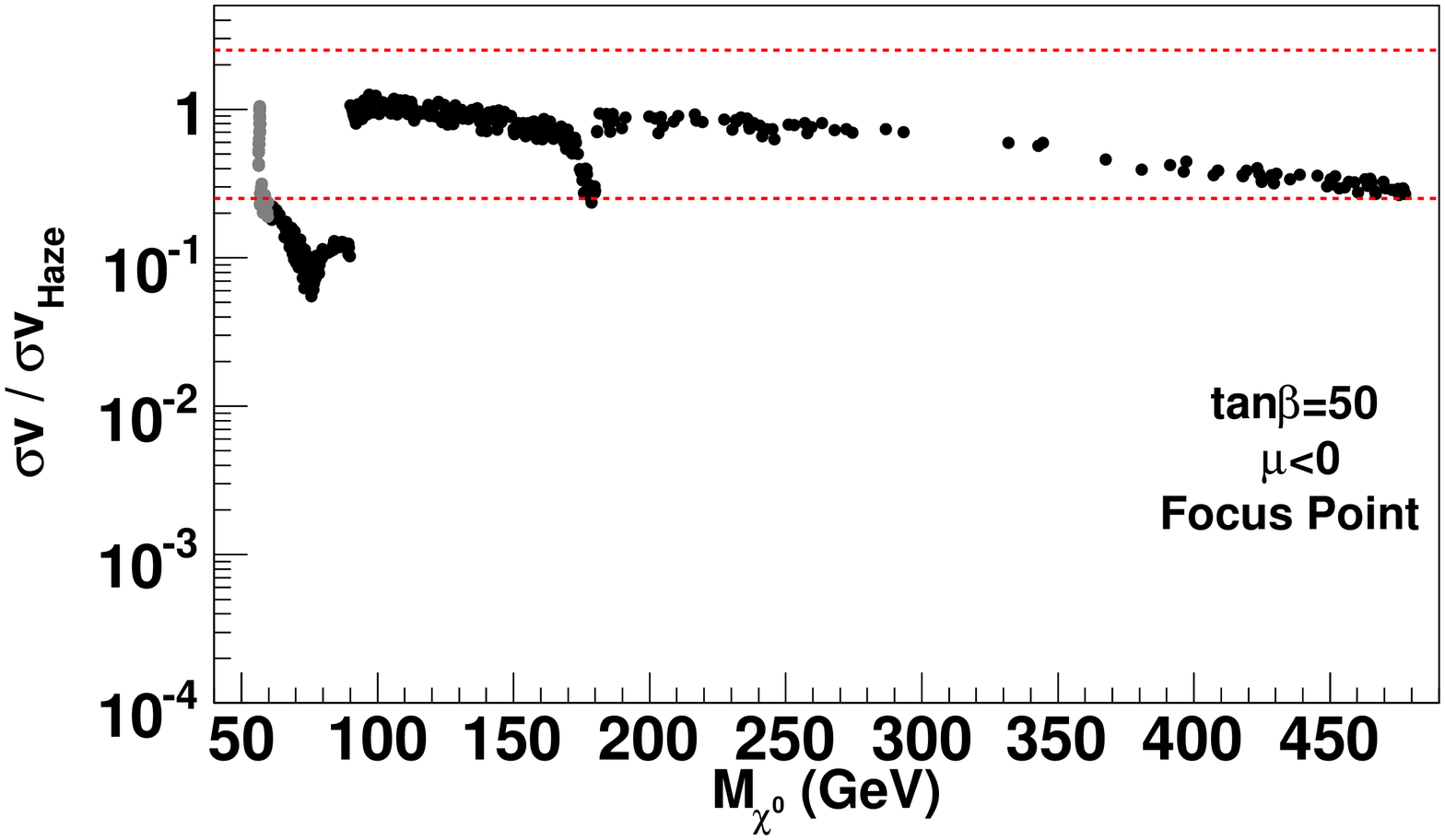}\\
\vspace{0.5cm}
\includegraphics[width=1.8in,angle=-90]{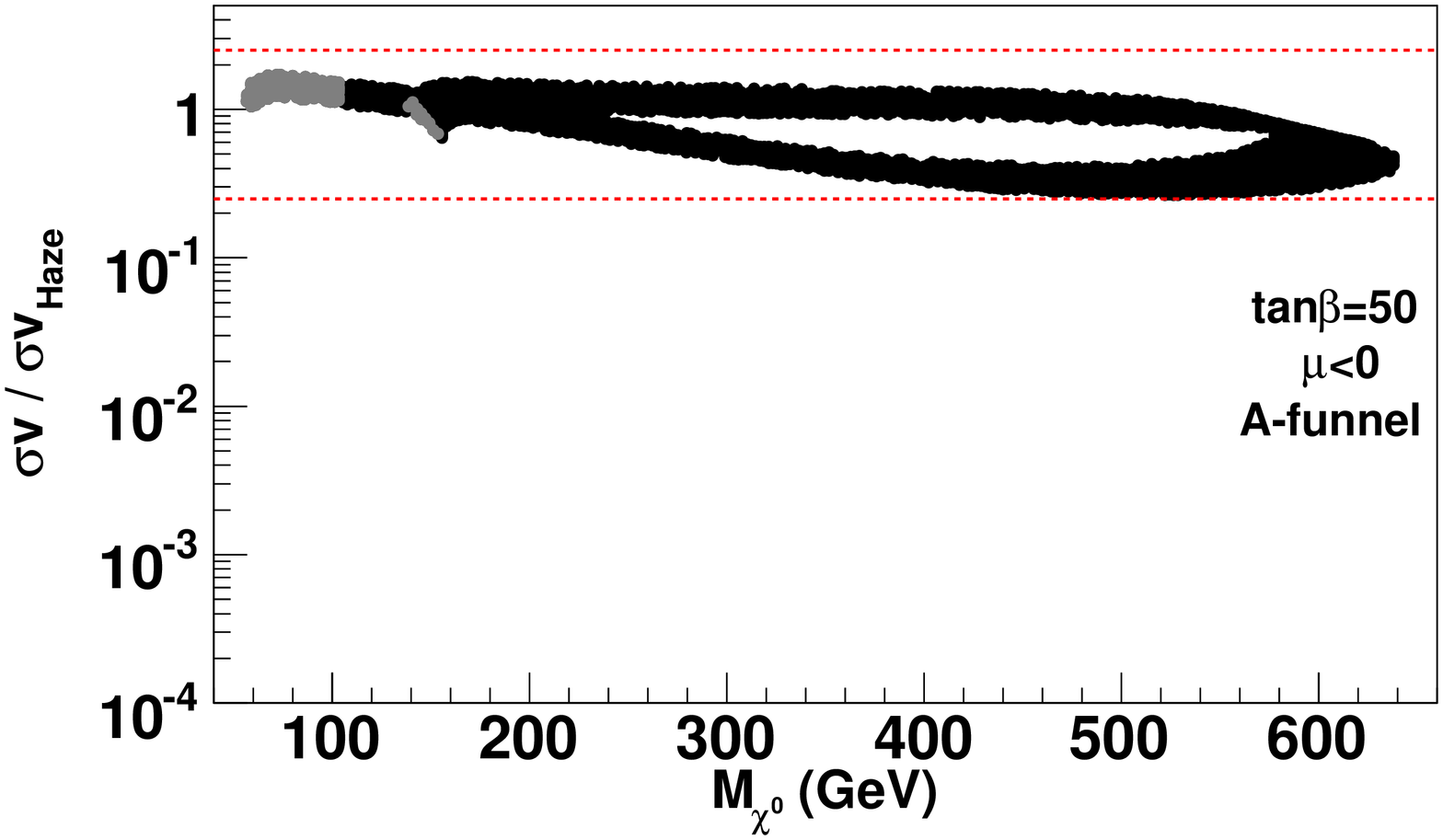}
\caption{As shown in Fig.~\ref{ratio}, but for $\mu < 0$.}
\label{rationegmu}
\end{figure}

We will now proceed to combine the results obtained in the previous sections in order to determine which regions of supersymmetric parameter space contain a neutralino with the characteristics required to generate the observed properties of the WMAP Haze. In Figs.~\ref{ratio} and~\ref{rationegmu}, we combine the information shown in Fig.~\ref{sigma} and Figs.~\ref{crosssection}-\ref{crosssectionnegmu} to determine which regions predict the required synchrotron intensity. For our default choice of $U_B/(U_B+U_{\rm rad})=0.25$, a value of $\sigma v/\sigma v_{\rm Haze}$ of approximately 1 is required to correctly normalize the flux of synchrotron to the intensity observed by WMAP. The horizontal lines shown in the figures reflect the range of $U_B/(U_B+U_{\rm rad})=0.1$ to 1.0. We consider any model that lies within these contours to adequately reproduce the intensity (although not necessarily the spectral shape) of the WMAP Haze. From these figures, we conclude the following:
\begin{itemize}
\item{The focus point region: The majority of the parameter space in the focus region is predicted to naturally generate a signal consistent with the observed properties of the WMAP Haze. In particular, the entire focus point parameter space with $M_{\chi^0} > M_W$ produces a signal in agreement with the spectrum and intensity of the WMAP Haze.}

\item{The stau coannnihilation region: Neutralinos in the stau-coannihilation region consistently under-produce the intensity of synchrotron emission relative to the intensity of the WMAP Haze.}

\item{The bulk region: Although, in much of the bulk region, the neutralino is light and annihilates largely to $b\bar{b}$, leading to a spectrum too soft to accommodate the WMAP Haze, we find that in the parameter space near $\tan \beta \sim 50$, $\mu >0$, and $M_{\chi^0} \sim 125-300$ GeV, the synchrotron emission is consistent with the observed properties of the WMAP Haze.}

\item{The $A$-funnel region:  Much like the focus point, the $A$-funnel region consistently predicts a neutralino annihilation cross section near the value required to normalize the WMAP Haze. So long as $M_{\chi^0} \gsim 125$ GeV, $A$-funnel neutralinos are consistent with being the source of this signal.}
\end{itemize}

We will next go on to study the prospects for the direct and indirect detection of neutralinos within these parameter ranges suitable for generating the WMAP Haze.

\section{Prospects for Direct and Indirect Detection}

If annihilating neutralinos are in fact the source of the observed anomalous emission known as the WMAP Haze, then we can use this information to better evaluate the prospects for future direct and indirect detection experiments to detect dark matter. In this section, we will make such evaluations, focusing on future direct detection experiments and on next generation neutrino telescopes such as IceCube.

Direct detection experiments attempt to observe the elastic scattering of neutralinos or other WIMPs with target nuclei in their detector.

The spin-independent neutralino-nucleon elastic scattering cross section is given by:
\begin{equation}
f_{p,n}=\sum_{q=u,d,s} f^{(p,n)}_{T_q} a_q \frac{m_{p,n}}{m_q} + \frac{2}{27} f^{(p,n)}_{TG} \sum_{q=c,b,t} a_q  \frac{m_{p,n}}{m_q},
\label{feqn}
\end{equation}
which, in turn, is built up from the neutralino-quark couplings, $a_q$. The quantity $f^{(p,n)}_{T_q}$ and $f^{(p,n)}_{TG}$ denote the quark and gluon content of the nucleon, respectively. The first term in Eq.~\ref{feqn} accounts for interactions with the quarks in the target nuclei. This can occur through either $t$-channel CP-even Higgs exchange, or $s$-channel squark exchange. The second term corresponds to interactions with the gluons in the target through a quark/squark loop diagram.

The couplings $a_q$ depend on many of the features of the supersymmetric spectrum. In particular, the resulting cross section tends to scale with the product of the gaugino and higgsino fractions of the lightest neutralino, thus leading to potentially large cross sections for neutralinos in the focus point region. Scenarios with light $H$ or light squarks and large $\tan \beta$ can also contain neutralinos with an enhanced elastic scattering cross sections with nuclei~\cite{jungman}.

\newpage
.
\newpage
.
\newpage

\begin{figure}[t]
\centering\leavevmode
\includegraphics[width=3.0in,angle=-90]{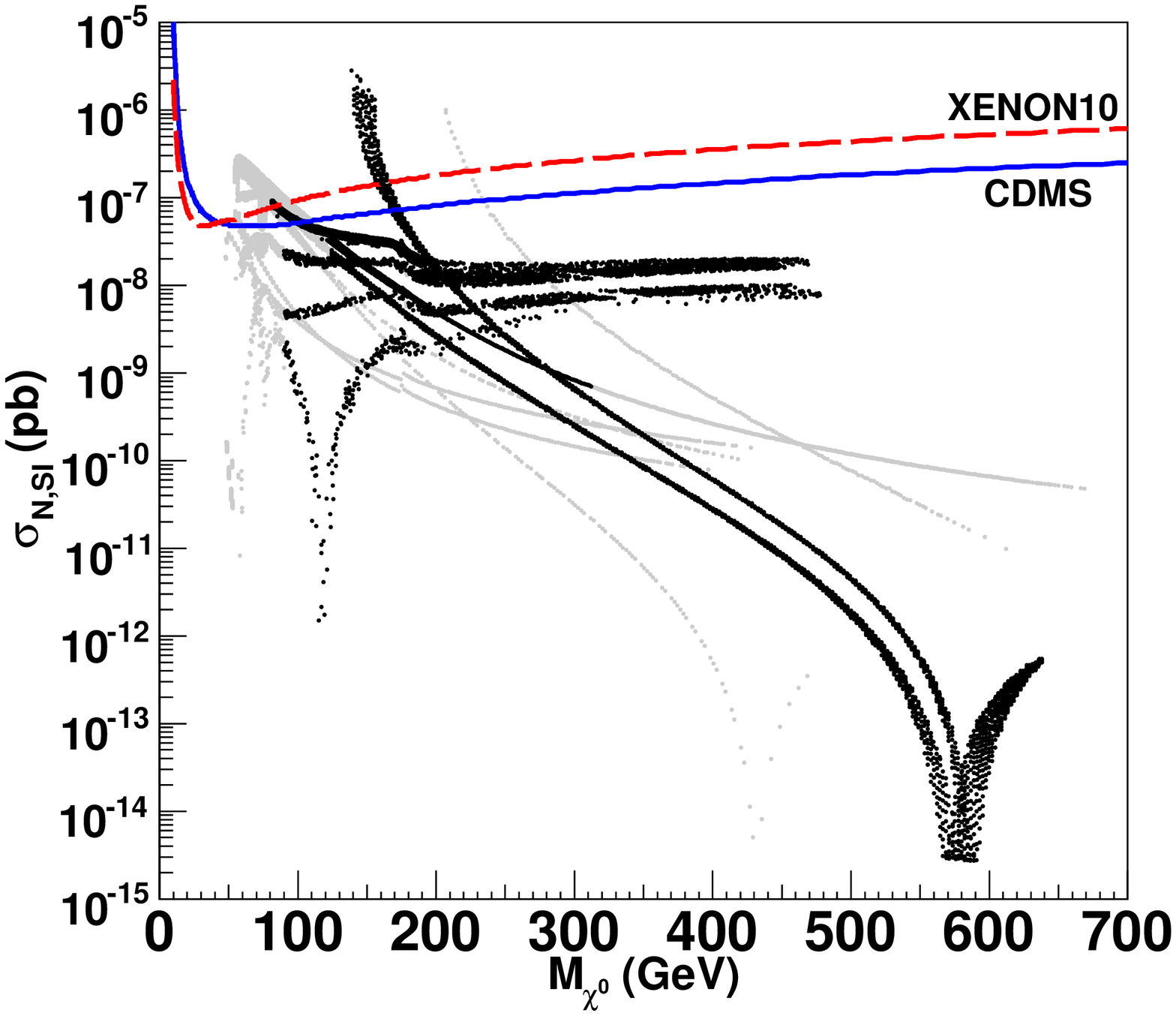}
\includegraphics[width=3.0in,angle=-90]{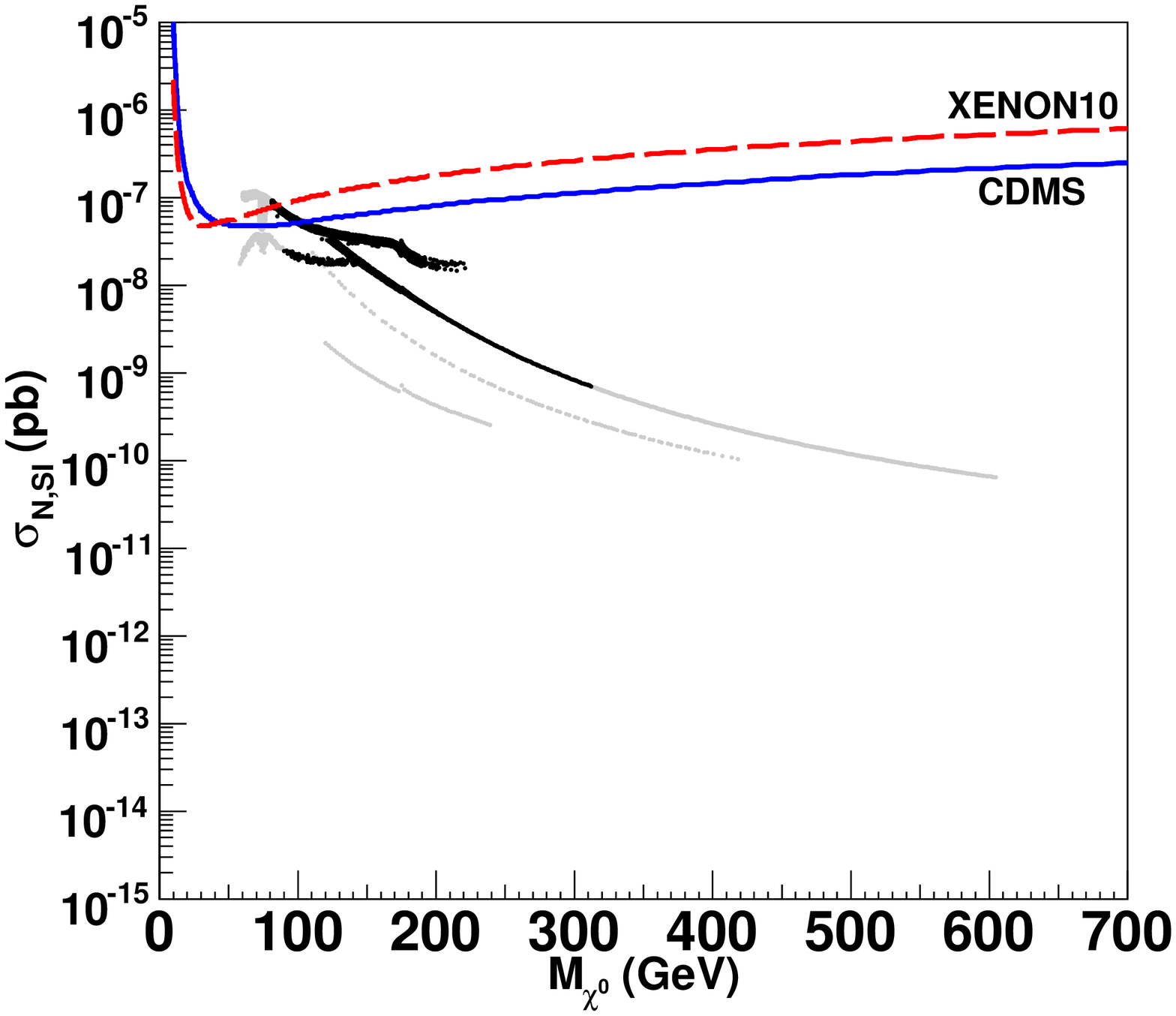}
\caption{The prospects for the direct detection of neutralino dark matter. Each point shown predicts a neutralino relic abundance consistent with the measured dark matter density. In the left (right) frame, we do not (do) take into account constraints from the muon's magnetic moment and the mass of the lightest Higgs boson. In each frame, the dark regions represent parameter space in which the observed properties of the WMAP Haze can be generated, whereas the lighter regions are inconsistent as a source of this emission.}
\label{direct}
\end{figure}

In Fig.~\ref{direct}, we show the spin-independent elastic scattering cross sections of neutralinos over the CMSSM parameter space and compare them to the current constraints from the direct detection experiments CDMS~\cite{cdms} and XENON~\cite{xenon}. In each frame, we show only those points that predict a neutralino relic abundance consistent with the measured dark matter density. In the left (right) frame, we do not (do) take into account constraints from the muon's magnetic moment and the mass of the lightest Higgs boson. In each frame, the dark regions represent parameter space in which the observed properties (both the spectrum and intensity) of the WMAP Haze can be generated, whereas the lighter regions are inconsistent as a source of this emission.

\begin{figure}[t]
\centering\leavevmode
\includegraphics[width=3.0in,angle=-90]{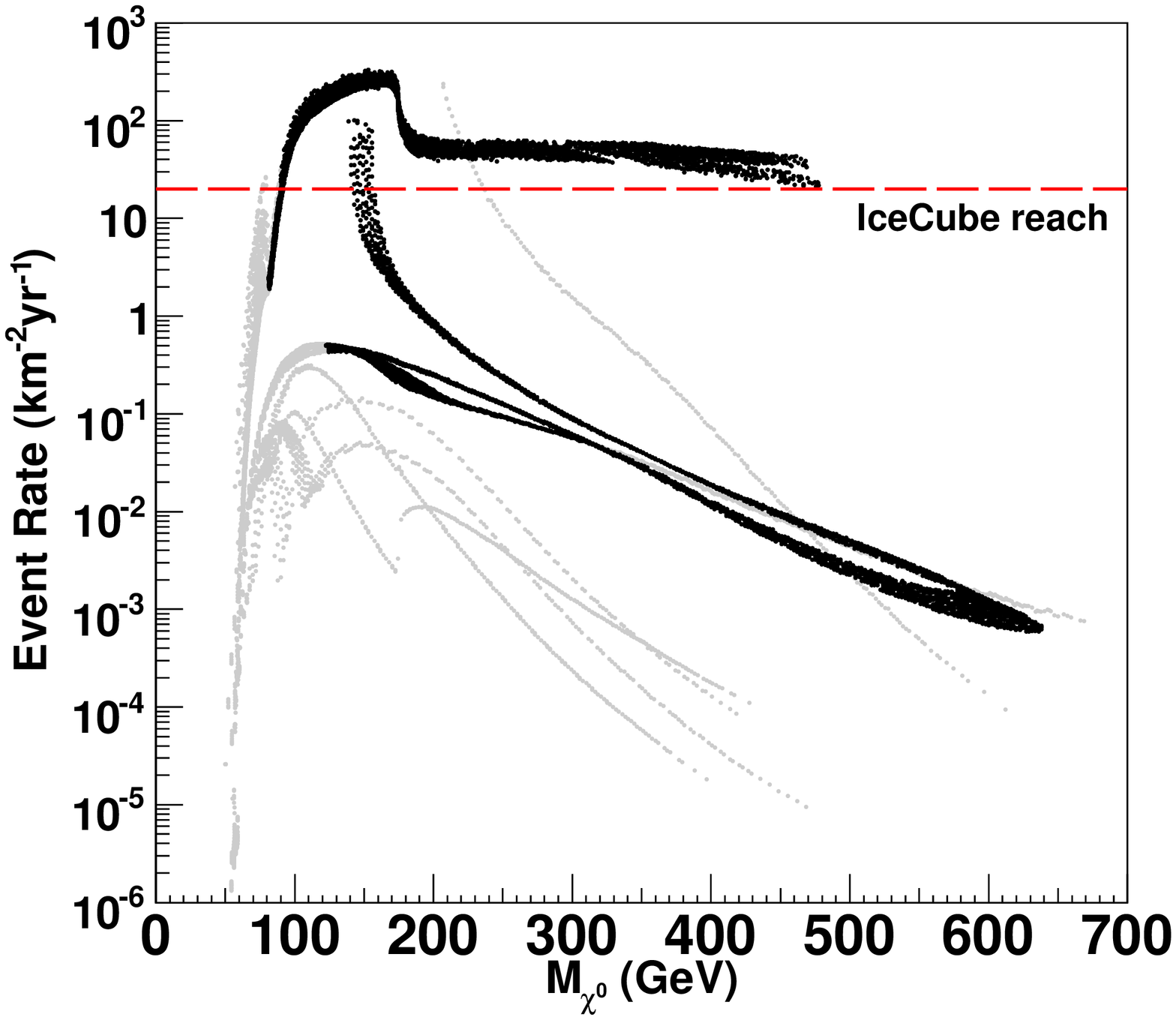}
\includegraphics[width=3.0in,angle=-90]{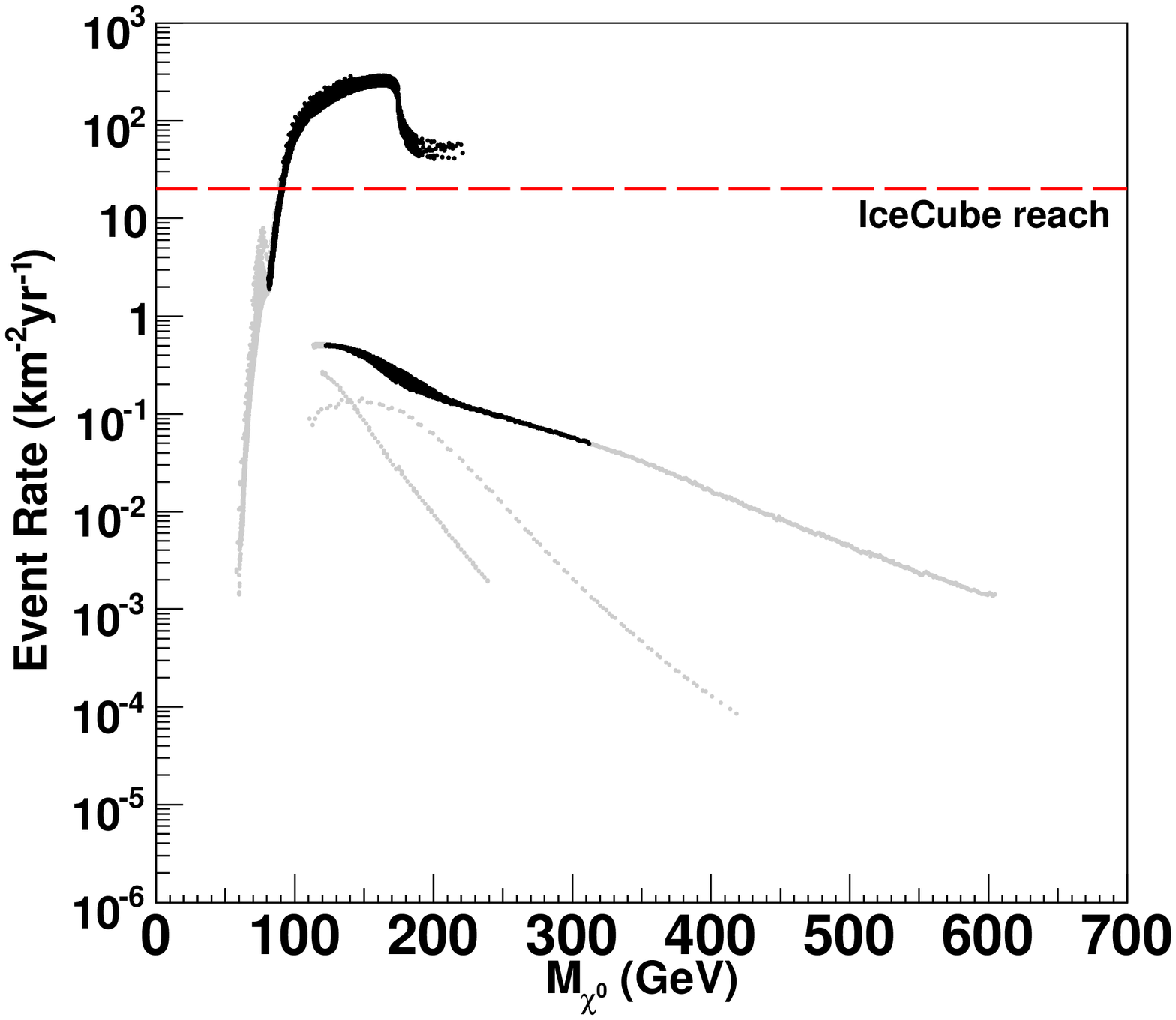}
\caption{The prospects for the indirect detection of neutralino dark matter by observing high energy neutrinos from neutralino annihilations in the Sun. Each point shown predicts a neutralino relic abundance consistent with the measured dark matter density. In the left (right) frame, we do not (do) take into account constraints from the muon's magnetic moment and the mass of the lightest Higgs boson. In each frame, the dark regions represent parameter space in which the observed properties of the WMAP Haze can be generated, whereas the lighter regions are inconsistent as a source of this emission. In calculating the event rate, we have imposed a muon energy threshold of 50 GeV.}
\label{neutrino50}
\end{figure}

Neutralinos which scatter with nuclei in the Sun can become gravitationally bound, leading them to accumulate in the core where they eventually annihilate. The neutrinos produced in these annihilations are potentially observable~\cite{neutrino}. In Fig.~\ref{neutrino50}, we show the prospects for a kilometer-scale high energy neutrino telescope such as IceCube~\cite{icecube} to detect the neutrinos from neutralino annihilation. As in Fig.~\ref{direct}, the dark regions denote parameter space in which the spectrum and intensity of the synchrotron emission is consistent with the observed properties of the WMAP Haze, whereas the lighter regions it is not.

From Figs.~\ref{direct} and~\ref{neutrino50} we can see that the regions of supersymmetric parameter space which are capable of generating the observed properties of the WMAP Haze also lead to somewhat favorable prospects for other direct and indirect detection efforts, especially if constraints on the Higgs mass and muon magnetic moment are taken into account. In particular, from the right frame of Fig.~\ref{direct}, we see that much of the dark region lies within an order of magnitude of current direct detection constraints. Only the $\tan \beta=50$, $\mu > 0$, $m_0 <1000$ GeV (bulk) region is consistent with being the source of the WMAP Haze while also being beyond the very near future reach of such experiments. The rest of the WMAP Haze favored parameter space falls in the focus point, which is ideally suited for direct detection.

The prospects for IceCube are also encouraging. Focus point neutralinos heavier than $\sim 100$ GeV are expected to be within the reach of next generation high energy neutrino telescopes, such as IceCube. 

We have not here discussed other indirect detection channels, such as those using gamma-rays, or charged cosmic rays. As these techniques rely on the neutralino's annihilation cross section, rather than their elastic scattering cross sections, theirs prospects are strengthened by the requirement that $\sigma v \sim 3 \times 10^{-26}$ cm$^3$/s (see Fig.~\ref{sigma}), but relatively little can be said beyond this.

\section{Discussion and Conclusions}

It has previously been shown that the anomalous emission from the inner Milky Way known as the WMAP Haze can be generated by a WIMP with a mass within the range of 80 GeV to several TeV and an annihilation cross section near the value predicted for an $s$-wave annihilating thermal relic. In this paper, we have studied the possibility that annihilating neutralinos are the source of this signal. Confining our study to the Constrained Minimal Supersymmetric Standard Model (CMSSM), we find that a large fraction of the phenomenologically viable parameter space naturally leads to an annihilation cross section and spectrum of annihilation products consistent with the observed properties of the WMAP Haze (both the spectrum and intensity). In particular, the focus point, $A$-funnel, and high $\tan \beta$ bulk regions of the CMSSM parameter space are each well suited for generating this anomalous signal. We find that neutralinos in the stau coannihilation, or low $\tan \beta$ bulk region, in contrast, generate a spectrum of synchrotron emission which is either too faint, too soft, or both, to account for the WMAP Haze. Within this context, it is interesting to note that the focus point region of the CMSSM is also favored by global fits to indirect data~\cite{allanach}.

If the WMAP Haze is in fact generated by annihilating neutralinos, then the prospects for their direct and indirect detection are each promising. In particular, direct detection experiments are currently quite likely to be within 1-2 orders of magnitude of the sensitivity required to detect such a particle. The prospects for next generation neutrino telescopes such as IceCube are also quite encouraging, especially in the focus point region of parameter space.

\bigskip

We would like to thank Greg Dobler and Doug Finkbeiner for very helpful discussions. This work has been supported by the US Department of Energy and by NASA grant NAG5-10842. GC has been supported by the Fermilab Summer Internships in Science and Technology Program.

\end{document}